\documentclass[fleqn,usenatbib]{mnras}

\usepackage{newtxtext,newtxmath}

\usepackage{blindtext}

\usepackage[T1]{fontenc}
\usepackage{tikz}
\usepackage{subcaption}
\usetikzlibrary{matrix}
\DeclareRobustCommand{\VAN}[3]{#2}
\let\VANthebibliography\thebibliography
\def\thebibliography{\DeclareRobustCommand{\VAN}[3]{##3}\VANthebibliography}

\usepackage{gensymb}
\usepackage{float}

\usepackage{booktabs}
\usepackage{wrapfig}
\usepackage{subcaption}

\usepackage{hyperref}
\usepackage{fixltx2e}
\usepackage{graphicx}	% Including figure files
\usepackage{amsmath}	% Advanced maths commands
\usepackage{subcaption}

\usepackage{cleveref}

\graphicspath{{Plots/}}
    \title[Environmental vs. AGN Feedback Quenching]{The role of environment and AGN feedback in quenching local galaxies: Comparing cosmological hydrodynamical simulations to the SDSS}

\author[P. Goubert et al.]{
Paul H. Goubert,$^{1}$\thanks{E-mail: pgoubert@fiu.edu}
Asa F. L. Bluck ,$^{1}$\thanks{E-mail: abluck@fiu.edu}
Joanna M. Piotrowska$^{2}$
and Roberto Maiolino$^{3,4,5}$
\\
\\
$^{1}$Stocker AstroScience Center, Dept. of Physics, Florida International University,11200 SW 8th Street, Miami, 33199, Florida, USA\\
$^{2}$Department of Astronomy, California Institute of Technology, 1200 East California Boulevard, Pasadena, California 91125, USA\\
$^{3}$Kavli Institute for Cosmology, University of Cambridge, Madingley Road, Cambridge, CB3 0HA, UK\\
$^{4}$Cavendish Laboratory, University of Cambridge, 19 J. J. Thomson Avenue, Cambridge CB3 0HE, United Kingdom\\
$^{5}$Department of Physics and Astronomy, University College London, Gower Street, London WC1E 6BT, UK
}

\date{Accepted XXX. Received YYY; in original form ZZZ}

\pubyear{2023}

\begin{document}
\label{firstpage}
\pagerange{\pageref{firstpage}--\pageref{lastpage}}
\maketitle

% Abstract of the paper
\begin{abstract}
We present an analysis of the quenching of local observed and simulated galaxies, including an investigation of the dependence of quiescence on both intrinsic and environmental parameters. We apply an advanced machine learning technique utilizing random forest classification to predict when galaxies are star forming or quenched. We perform separate classification analyses for three groups of galaxies: (a) central galaxies; (b) high-mass satellites ($M_{*} > 10^{10.5}{\rm M_{\odot}}$); and (c) low-mass satellites ($M_{*} < 10^{10}{\rm M_{\odot}}$) for three cosmological hydrodynamical simulations (EAGLE, Illustris, and IllustrisTNG), and observational data from the SDSS. The simulation results are unanimous and unambiguous: quiescence in centrals and high-mass satellites is best predicted by intrinsic parameters (specifically central black hole mass), whilst it is best predicted by environmental parameters (specifically halo mass) for low-mass satellites. In observations, we find black hole mass to best predict quiescence for centrals and high mass satellites, exactly as predicted by the simulations. However, local galaxy over-density is found to be most predictive parameter for low-mass satellites. Nonetheless, both simulations and observations do agree that it is environment which quenches low mass satellites. We provide evidence which suggests that the dominance of local over-density in classifying low mass systems may be due to the high uncertainty in halo mass estimation from abundance matching, rather than it being fundamentally a more predictive parameter. Finally, we establish that the qualitative trends with environment predicted in simulations are recoverable in the observation space. This has important implications for future wide-field galaxy surveys.

\end{abstract}

% Select between one and six entries from the list of approved keywords.
% Don't make up new ones.
\begin{keywords}
Galaxies: evolution -- Galaxies: formation -- Galaxies: star formation
\end{keywords}

%%%%%%%%%%%%%%%%%%%%%%%%%%%%%%%%%%%%%%%%%%%%%%%%%%

%%%%%%%%%%%%%%%%% BODY OF PAPER %%%%%%%%%%%%%%%%%%

\section{Introduction}

Local galaxies exhibit a clear bimodality in several properties, chief among them their specific star formation rate (sSFR = SFR$/ M_*$) and optical color \citep{Strateva_2001, Brinchmann_2004}. As such, they can be sorted into one of two groups, either "blue and star forming" or "red and dead". The extinguishing of star formation in galaxies is commonly referred to as quenching \citep[e.g.,][]{Peng_2010}. Understanding how galaxies quench is paramount to comprehending galaxy evolution. Whilst the mechanisms that drive quenching have yet to be fully established, correlation of sSFR with a host of parameters has previously been identified.

Specific star formation rate has been demonstrated to correlate strongly with both intrinsic and environmental properties of galaxies. This suggests that there may be multiple routes to quenching: those originating from within the galaxy, and those originating from the environment in which the galaxy is situated. \citet{Peng_2010, Peng_2012} demonstrate that these two avenues for quenching are largely separable, i.e. both forms of quenching may act independently. This suggests that we are searching for (at least) two quenching mechanisms, which do not strongly influence each other.

Intrinsic quenching dependencies have been found with stellar and (more recently) black hole mass \citep[see][]{Baldry_2006, Peng_2010, Bluck_2016, Terrazas_2016, Terrazas_2017, Bluck_2020a, Bluck_2020b, Bluck_2022, Bluck_2023, Piotrowska_2022}. Additionally, correlation between quenching and environmental parameters has also been found, including with halo mass \citep{Woo_2013, Wang_2018}, local galaxy over-density \citep{Baldry_2006, Peng_2010, Bluck_2020b}, and the location of galaxies within their dark matter haloes \citep[e.g.,][]{Bluck_2019, Bluck_2020b}. Hence, our contemporary understanding of galaxy quenching introduces a dichotomy between intrinsic (often mass correlating) or environmental (often density correlating) processes. This is often explored by splitting galaxies into two groups: centrals and satellites. A central galaxy is defined as the most massive galaxy in a given halo, whilst a satellite galaxy is defined as any other galaxy within a given group or cluster. Centrals are found to have their star formation regulated almost entirely by intrinsic parameters \citep[e.g.,][]{Peng_2010, Bluck_2016, Bluck_2020b}, whereas satellites exhibit both intrinsic and environmental quenching \citep[e.g.,][]{Peng_2012, {Woo_2015}, Bluck_2019, Bluck_2020b}.

Intrinsic quenching mechanisms of potential importance include active galactic nucleus (AGN) and supernova (SN) feedback. It has been shown in previous works \citep{Croton_2006, Bower_2006, Bower_2008, Smith_2018, Henriques_2019} that while supernova explosions do create gas outflows, they do not generate the necessary energy required to suppress star formation in massive galaxies. SN explosions are not strong enough to expel gas from massive galactic halos, and their induced heating of the circum-galactic medium (CGM) is not sufficient to keep the hot gas halo stable from collapse \citep[e.g.,][]{Henriques_2015, Somerville_2015}. This leaves AGN feedback as the most promising theoretical route for quenching high mass systems.

AGN feedback is theoretically proposed to come in one of two principle forms: (i) `quasar mode' \citep[e.g.,][]{Hopkins_2006, Hopkins_2008, Nesvadba_2008, Feruglio_2010, Maiolino_2012}  and (ii)  `preventative mode' \citep[e.g.,][]{Bower_2006,Bower_2008, Croton_2006, Fabian_2012}. Quasar mode feedback describes the processes by which high luminosity AGN trigger galactic winds which can lead to outflows, while preventative mode feedback refers to the heating of the CGM by low luminosity AGN, potentially via relativistic jets or kinetic feedback. The latter provides a route to suppression of star formation through the long-term stabilization of the hot gas halo surrounding massive galaxies, preventing it from cooling, and ultimately starving the galaxy of fuel. This is the primary mode of intrinsic galaxy quenching currently adopted in most contemporary simulations \citep[see][]{Sijacki_2007, Vogelsberger_2014a, Vogelsberger_2014b, Schaye_2015, Zinger_2020, Terrazas_2020, Piotrowska_2022, Bluck_2023}. 

Observational analysis into the quenching of central galaxies offers insight into the intrinsic mechanisms by which galaxies quench, enabling a test of the theoretical AGN feedback paradigm. The quenching of centrals has been shown to best correlate phenomenologically with stellar velocity dispersion, which is well known to correlate strongly with central black hole mass \citep[e.g.,][]{Ferrarese_2000, Saglia_2016}, potentially linking the process to AGN feedback \citep[see][]{Bluck_2016, Bluck_2020a, Bluck_2020b, Bluck_2022, Terrazas_2016, Terrazas_2017, Piotrowska_2022, Brownson_2022}.  Indeed, contemporary cosmological models agree that the key predictor of intrinsic galaxy quenching ought to be black hole mass \citep[see][]{Piotrowska_2022, Bluck_2023}, which is explained by black hole mass being directly proportional to the feedback energy injected into galaxy haloes \citep[see][for an analytical derivation]{Bluck_2020a}.

The quenching of satellites, however, is influenced by environmental factors in addition to AGN feedback \citep[see][]{Peng_2010, Peng_2012, Woo_2013, Woo_2015, Woo_2017, Bluck_2014, Bluck_2016}. In \citet{Bluck_2020b}, the quiescence of high mass satellites is shown to behave much like centrals, i.e. exhibiting a strong correlation with central velocity dispersion, although with the notable difference of a weak secondary dependence on environmental parameters. Conversely, low mass satellite quenching appears to depend almost exclusively on environmental parameters, with prior work revealing local density to be most predictive quenching parameter for local systems \citep{Bluck_2020b}. 

Proposed physical mechanisms for environmental quenching include ram pressure stripping, galaxy-galaxy harassment, and galaxies transitioning away from nodes in the cosmic web \citep[see][]{vandenBosch_2008, Cortese_2009, Henriques_2015}. Additionally, pre-processing of satellites in groups prior to cluster membership may lead to early quenching as a result of lower levels of the same environmental processes.

Ram pressure occurs when fluids interact at high relative bulk velocities. When this happens gas in the two components experience an effective pressure, which can be strong enough to counteract gravitational stability. More specifically, the ram pressure stripping of satellite galaxies describes the process by which intra-cluster medium (ICM) wind strips gas from galaxies as they move through clusters \citep[e.g.,][]{Gunn_1972}. 

As a galaxy enters, and moves through a cluster, it can  be stripped of its CGM when it encounters the ICM of the halo. Due to the CGM being diffuse at large radii it typically has a low binding energy, hence stripping a galaxy of its hot gas halo is relatively easy. The stripping of the CGM can then trigger strangulation \citep{Balogh_2000} for the affected galaxy, rendering the refueling of cold gas impossible, which eventually quenching the galaxy. 

The stripping of the ISM on the other hand is much more difficult as it is strongly gravitationally bound to its galaxy. This can occur as a satellite galaxy moves near the center of a massive cluster, where very high densities and fast relative velocities occur. This stripping of the ISM in some cases results in `jellyfish galaxies' \citep[see,][]{Cramer_2019}. If the pressure is high enough, eventually there will no longer be sufficient cold gas for sustained star formation, ultimately quenching the satellite. This form of ram-pressure stripping thus leads to rapid quenching, whereas the former leads to slow quenching.

Additionally, galaxy-galaxy harassment has been proposed as a process that could affect star formation in satellite galaxies \citep[e.g.,][]{Moore1996, Moore_1998}. As satellite galaxies move through the halo at high relative velocities, systems could experience gas stripping from their CGM due to tidal forces from close galaxy passes. In the most extreme cases of radial orbits, satellites may also be dynamically stripped of gas through interaction with the central as well. 

Environment quenching is a complex process incorporating gravitation, (magneto-)hydrodynamics, and the physics of star formation. Cosmological hydrodynamical simulations offer unprecedented insight into how these factors interact to yield environmental effects on satellite galaxies. Whilst intrinsic quenching from AGN feedback is simulated using subgrid models (due to the small physical scale and relativistic nature of this process), the environment is allowed to evolve according to the exact prescribed laws of physics (i.e., the Euler equations coupled to the Poisson equation). Therefore, we are able to study mechanisms like ram pressure stripping and galaxy-galaxy interactions exactly as observed in the Universe, within the resolution limit of the numerical approximation. Hence, cosmological hydrodynamical simulations offer a direct comparable to our Universe for how gas content behaves in galaxies, clusters, and the cosmic web overall.

The highly complex and interlinked processes of galaxy quenching are difficult to analyse successfully via simple statistical techniques (such as correlations), yet this is tractable with machine learning approaches, which can account for highly non-linear relationships within a multidimensional and inter-correlated parameter space \citep[see, e.g.,][]{Teimoorinia_2016, Bluck_2022}. To first order, galaxies separate into star forming and quenched types (as discussed above). Hence, the type of machine learning required is clearly that of classification, which aims to accurately segregate physical categories on the basis of input data. Indeed, the combination of sophisticated classification algorithms and cosmological hydrodynamical simulations provides a powerful tool to determine how galaxies quench in our Universe \citep[see][for various demonstrations of this approach]{Piotrowska_2022, Brownson_2022, Bluck_2022, Bluck_2023}. 

Random Forest classification is a machine learning tool that not only accurately classifies complex data, but also determines the most predictive parameter(s) when sorting objects into a class. Therefore, in our case, it is particularly useful as it allows us to determine which feature is best to predict the state of star formation in a galaxy. The use of machine learning through Random Forest classification is ideal to hone in on the most important parameters regulating quiescence in both models and the observed Universe. Determining the optimal parameter(s) allows us to establish causal links to the mechanisms that drive quenching in the local observable Universe and in z $\sim$ 0 simulations. This method was previously used in \citet{Piotrowska_2022} in a study of the quenching of central galaxies. Black hole mass was found to be by far the most predictive parameter in both hydrodynamical simulations and the observed local Universe (as seen in the SDSS). This has aided in establishing a plausible causal link between quenching and AGN feedback in nature. Moreover, the lack of a strong link between quenching and AGN luminosity \citep{Hickox_2009,Aird_2012,Trump_2015} was resolved by this work, which found that quenching depends critically on the integrated energy released by AGN over their lifetime, not their current output. 

In this work, we build upon our prior works \citep{Piotrowska_2022, Bluck_2023} by applying Random Forest classification to three categories of galaxies. We reveal the most predictive quenching parameters for centrals, high-mass satellites, and low-mass satellites in the EAGLE \citep{Schaye_2015}, Illustris \citep{Vogelsberger_2014a, Vogelsberger_2014b}, and IllustrisTNG \citep{Nelson_2018, Pillepich_2018a} simulations at z = 0, as well as in the largest spectroscopic survey of local galaxies, the Sloan Digital Sky Survey \citep[SDSS;][]{Abazajian_2007}. This paper is novel in the approach of applying machine learning classification to rigorously compare the predictions from cosmological models to the largest low-redshift galaxy survey in order to reveal the role of environment in quenching galaxies. 

This paper is structured as follows. In Section \ref{Data} we describe the three hydrodynamical suites and the SDSS observational data. Section \ref{Methods} first details the SFR - $M_*$ plane in the simulations and the SDSS, then the approach we take to model the simulations with observational realism, and the estimation of black hole mass for the SDSS. In Section \ref{Methods} we also include a description of our Random Forest classifier. In Section \ref{Results}, we present our results, starting with the simulations, followed by those for the SDSS. Section \ref{Results} additionally includes an analysis into the impact on specific star formation rate of the location of satellites within their parent haloes, and a comparison on the importance of nearest neighbor density to quenching for low mass satellite galaxies. This is followed by a discussion of the results in Section \ref{Discussion}, and finally a summary of the paper in Section \ref{Summary}.

\section{Data}\label{Data}

\subsection{Hydrodynamical Simulations}

\par
\begin{figure*}
    \centering
    \begin{subfigure}{.33\textwidth}
        \caption{EAGLE}
        \includegraphics[width = \textwidth]{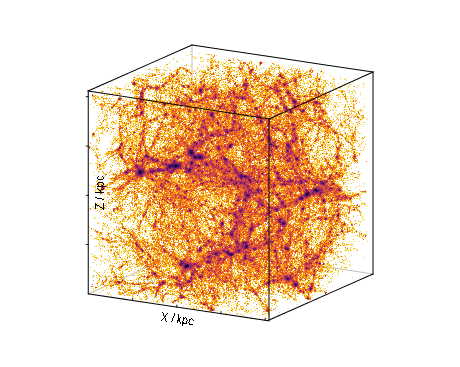}  
    \end{subfigure}
    \hfill
    \begin{subfigure}{.33\textwidth}
        \caption{IllustrisTNG}
        \includegraphics[width= \textwidth]{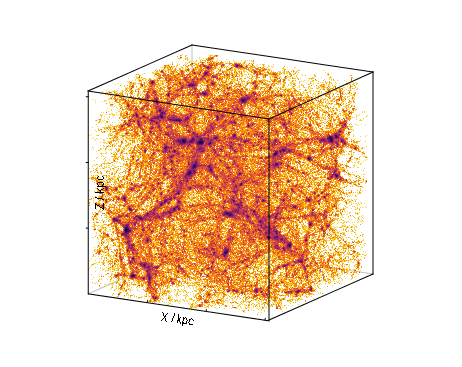} 
        
    \end{subfigure}
    \hfill
    \begin{subfigure}{.33\textwidth}
        \caption{Illustris}
        \includegraphics[width= \textwidth]{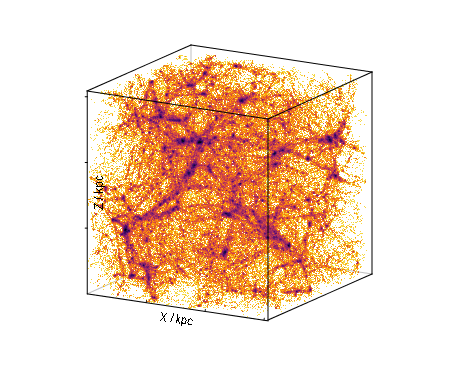}  
    \end{subfigure}

\vspace*{-3mm}
\caption{The topological configuration of each simulation suite presented in density of galaxies. The density of galaxy counts is indicated by the darkness of the filamentary structure. In each simulations over 3 orders of magnitude in density variation is found (from the field to massive clusters). These models provide an ideal laboratory for exploring the impact of environment on galaxy evolution, specifically in this work star formation and quenching.}\label{fig:SIMCubes}
\end{figure*}

 For the purposes of this study we make use of three hydrodynamical cosmological simulation suites: EAGLE \citep{Schaye_2015, Crain_2015}, Illustris \citep{Vogelsberger_2014a}, and IllustrisTNG \citep{Marinacci_2018, Naiman_2018, Nelson_2018, Springel_2018, Pillepich_2018b}. Each suite simulates an equal volume of $(\sim100 \, {\rm cMpc})^3$, although the Eagle and Illustris/ TNG simulations have significantly different seed density distributions, leading to different environmental distributions at $z = 0$ (see Fig.\ref{fig:SIMCubes}). Combined with the large range of local densities, which span over 3 orders of magnitude, we find that different simulations offer vastly different cosmic environments to study. 

\par The simulations occupy a volume of $\sim (100 \, {\rm cMpc})^3$, which is much smaller than the volume probed by the SDSS. This could result in the simulations lacking extreme environments, such as exceptionally massive clusters, which are present in the observations. However, this discrepancy should only affect a small fraction of the data, with the majority expected to be well represented by the simulations. To check the potential impact of this, we perform tests on our analysis while restricting the observational data to $M_{\rm halo} < 10^{14} M_{\odot}$ (an approximate upper limit in halo mass which is representative of the simulations). We find very little difference whether we impose this restriction or not, confirming the simulations and observational data are reasonably well comparable, with the observational data simply having larger sample sizes. Hence, we can investigate how environmental differences impact the quenching of galaxies in the simulations and compare these effectively to observations.

\par Due to difficulties that arise from resolving the physics of small spatial resolution the simulations make use of subgrid models, including for star formation, metal production, supernova feedback, accretion of gas onto black holes, and the ejection of energy from AGN feedback \citep[see][]{Schaye_2015, Crain_2015}. 

From the EAGLE simulations we chose to use the L100N1504 reference model, specifically the z = 0 snapshot; for Illustris the z = 0 snapshot of the Illustris-1 run is used; and the TNG100-1 run, snapshot z = 0, for IllustrisTNG. From all of these snapshots we extract the relevant parameters of halo mass, stellar mass, black hole mass, and star formation rate. Table.\ref{tab:SIM_ParentSample} shows both the size of the total and parent samples for each simulation run. This data can be extracted from a docker provided in \citet{Piotrowska_2022}.

\begin{table}
    \centering
        \caption{Simulation Total and Parent Sample Size. Total sample describes the entire simulation sample, while the parent sample are solely galaxies which meet our criteria for selection.}
    \begin{tabular}{ccc}
      \textbf{Simulation} & \textbf{Total Sample Size} & \textbf{Parent Sample Size} \\
      \hline
      EAGLE  & 325561 & 12931\\
      \hline
      Illustris  & 307786 & 23907\\
      \hline
      IllustrisTNG  & 337261 & 22029\\
      \hline
    \end{tabular}
    \label{tab:SIM_ParentSample}
\end{table}

\begin{figure*}
    \centering
    \includegraphics[width = 0.6\textwidth]{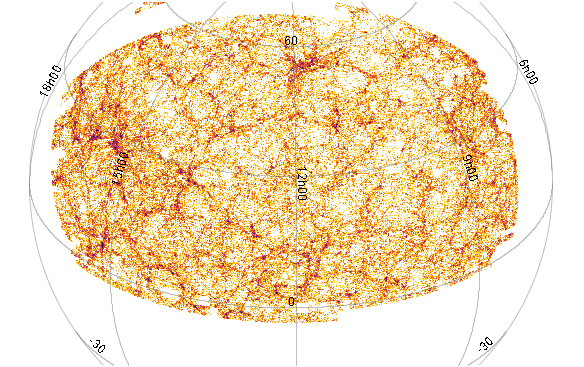}  
\vspace*{-3mm}
\caption{The sky plane configuration of the SDSS, presented in density of galaxies, for redshift $0.05 < z < 0.09$. The density of galaxy counts is indicated by the darkness of the filamentary structure. We see extensive variety in galactic and cluster density present in the SDSS. As in the simulations, over 3 orders of magnitude in density variation is seen. }\label{fig:SDSS_CosmicWeb}
\end{figure*}

\subsubsection{EAGLE}

EAGLE\footnote {Eagle Data Access: \url{http://icc.dur.ac.uk/Eagle/}} (Evolution and Assembly of GaLaxies and their Environments) \citep{Schaye_2015, Crain_2015, MCALPINE_2016} is a set of hydrodynamical cosmological simulations of a Dark Energy - Cold Dark Matter ($\Lambda$CDM) Universe, whose parameters are estimated from the \citet{Plank_2014}, defined in Table.\ref{tab:SimCosmoParams}.

 \par The simulation suites in EAGLE are run with a modified version of GADGET-3, an N-Body Tree-PM, smoothed particle hydrodynamics (SPH) solver \citep[described in][]{Springel_2005}. By simultaneously solving for gravity and hydrodynamics, EAGLE enables an accurate probe of the physics of galaxy evolution on scales of 1\,kpc - 100\,Mpc. The development of groups and clusters of galaxies, galaxy morphology, and gas accretion into halos take place organically, i.e. emerging from the fundamental equations for particle dynamics as large-scale phenomena. As such, mechanisms like tidal stripping, ram pressure stripping, and galaxy-galaxy interactions can transpire within the simulated Universe. While the spatial resolution achieved in EAGLE allows for the simulation of events and systems such as the ISM and CGM, certain processes must be represented through subgrid models \citep[a detailed list can be found in][]{Schaye_2015}. These models must then be calibrated with respect to observational data to properly represent our Universe. The self-consistent formation and evolution of structures in the EAGLE simulation constitute an avenue for insightful analysis into both intrinsic and  environmental effects present in the Universe.

\par The EAGLE simulation seeds black holes at a mass of $M_{\rm BH} = 10^{5} {\rm M_{\odot}} h^{-1}$ once a halo reaches $M_{\rm Halo} =  10^{10} {\rm M_{\odot}} h^{-1}$. Black holes then accrete matter according to a modified Bondi–Hoyle-Lyttleton accretion model, that takes into account the angular momentum of the infalling gas \citep[see][]{Bondi_1944, Hoyle_Lyttleton}, limited by the Eddington rate. Unlike Illustris and IllustrisTNG, EAGLE makes use of a sole AGN feedback mode with an energy injection rate of $\dot{E}_{\rm feedback} =\epsilon_{f}\epsilon_{r} \, \dot{M}_{BH}c^{2}$, where $\epsilon_{f}$ is the fraction of energy coupled to the ISM and $\epsilon_{r}$ is the radiative efficiency of the accretion disk. Each black hole then has an energy reservoir, increased by $\dot{E}_{\rm feedback} \delta t$ for each time-step, $\delta t$. Once a black hole has enough stored energy to heat its surrounding neighbors by a given amount, $\Delta T_{\rm AGN}$, it increases their temperature. This mimics the effect of a hot disk radiating energy, of which a small portion thermally influences its environment. For the specific simulation of EAGLE we focus on, $\Delta T_{\rm AGN}$ is set to $10^{8.5}K$. For a complete description of the black hole model we refer the reader to \citet{Booth_Schaye, Schaye_2015}.

\subsubsection{Illustris}

Illustris\footnote{Illustris Data Access: \url{https://www.illustris-project.org/}} \citep{Genel_2014, Vogelsberger_2014a,Vogelsberger_2014b} is a cosmological simulation that used the AREPO moving mesh code \citep[see][]{Springel_2010}. As with the EAGLE simulations, Illustris assumes a $\Lambda$CDM universe, but with cosmological parameters obtained through the nine year Wilkinson Microwave Anisotropy Probe Observations (WMAP 9, \citet{Hinshaw_2013}), listed in Table.\ref{tab:SimCosmoParams}. 

\par Using the AREPO code, the Illustris simulation solves the gravitational and fluid equations separately for dark matter and stars (which are still treated as particles) and for gas in all phases (which are solved in adaptive Voronoi bins). This approach has the advantage of adaptive resolution whereby denser regions are sampled to higher resolution than less dense regions \citep[see][]{Springel_2010}. The simulations' initial conditions take place only 300,000 years after the Big Bang and continue until they reach our local Universe. It is then possible to observe how the baryonic matter changes, modeled by solving the hydrodynamics of gas and the gravitational interaction. This is especially fruitful in analysing the formation and evolution of galaxies, stars, and halos (as well as the connection between these processes). We can then extract information on a plethora of mechanisms and how they affect a galaxies throughout cosmic time. The Illustris simulations differ from the EAGLE simulation suites as they employ separate code to model hydrodynamics, use different cosmological parameters, and adopt very different subgrid recipes for feedback.

\par The black holes in Illustris are seeded with $M_{\rm BH} = 10^{5} {\rm M_{\odot}} h^{-1}$, once unoccupied halos reach a threshold mass of $M_{\rm Halo} = 5 \times 10^{10} {\rm M_{\odot}} h^{-1}$, and similarly to EAGLE they grow according to a Bondi-Hoyle-Lyttleton based, and Eddington limited, accretion model. The accretion rate is then linked to the AGN feedback model, consisting of three regimes: (i) `quasar'; (ii) `radio'; and (iii) `radiative' \citep[see][]{Sijacki_2007, Sijacki_2015}.

\par The high accretion rate `quasar mode' requires that $\chi \equiv \dot{M}_{BH} / \dot{M}_{EDD} > 0.05$. This mode is similar to the AGN feedback in EAGLE, except here the energy is injected continuously. For values of $\chi$ lower than 0.05, the feedback switches to radio mode, the lower accretion rate mode. To reproduce the effects of relativistic jets on the CGM, a bubble with specific energy is created. The bubble is formed once the black hole increases its mass by a factor of $\delta_{\rm BH}=1.15$, and is given an energy E\textsubscript{Bubble},
where the efficiency of its mechanical heating is described by $\epsilon_{m} = 0.35$. The bubble can then heat the CGM as jets would and has a radius proportional to $\left(\frac{E_{Bubble}}{\rho_{CGM}}\right)^{\frac{1}{5}}$. Finally, the radiative mode, which is active at all accretion rates, allows for the ionising radiation from accreting black holes to modify the net cooling rate of nearby gas.

\subsubsection{IllustrisTNG}

Following in the footsteps of Illustris, IllustrisTNG\footnote{IllustrisTNG Data Access: \url{www.tng-project.org/}} \citep[The Next Generation, referred to as TNG hereafter,][]{Marinacci_2018, Naiman_2018, Nelson_2018,Springel_2018, Pillepich_2018b} makes use of an updated version of the AREPO moving mesh code (now including magnetic fields) applied to a $\Lambda$CDM universe. The new cosmological parameters are obtained from the Planck Collaboration \citep[see,][]{Plank_2016}, and are presented in Table.\ref{tab:SimCosmoParams}. 

\par The TNG simulation suites make use of the updated AREPO code, different cosmological parameters, and furthermore improves upon on the original Illustris simulation by radically updating its AGN feedback prescription. As with the other simulations discussed above, TNG enables a full gravitational and hydrodynamical solution to cosmic evolution on scales from 1\,kpc to 100\,Mpc. This enables the detailed evolution of galaxies, groups and clusters. Importantly for this work, these simulations model satellite - halo, satellite - central, and satellite - satellite interactions (both gravitational and hydrodynamical) self-consistently. Coupled with subgrid star formation models \citep[see][]{Pillepich_2018a}, this enables us to extract the testable predictions for environmental processes on galactic star formation.

\par The black holes are seeded with $M_{\rm BH} = 8 \times 10^{5} {\rm M_{\odot}} h^{-1}$, once unoccupied halos reach a threshold mass of $M_{\rm Halo} = 5 \times 10^{10} {{\rm M_{\odot}}} h^{-1}$, and then increase their mass through a Bondi-Hoyle-Lyttleton, Eddington limited, accretion rate model. TNG implements three different modes of AGN feedback. The high accretion rate mode is the same `quasar' mode as for Illustris, as is the radiative mode, whereas the low accretion rate mode is updated to a `kinetic' feedback mode \citep{Weinberger_2017}. The kinetic feedback is modelled as a stochastic injection of pure kinetic energy into neighboring gas cells around the black hole, applied as a momentum kick released in a random direction to preserve isotropy \citep[see][]{Weinberger_2017, Weinberger_2018}. The change in kinetic energy of a neighboring gas cell due to the feedback is given by $\dot{E}_{kinetic} = \epsilon_{k} \dot{M}_{BH}c^{2}$, where $\epsilon_{k}$ is the efficiency of energy transfer. The feedback energy is then transported through the ISM (adding turbulence) and eventually percolates to the CGM, providing long-term heating of the hot gas halo. The change in low accretion mode was motivated by the radio prescription being overzealous in its removal of the CGM around massive galaxies.

\begin{table}
    \centering
        \caption{Simulation Cosmological Parameters}
    \begin{tabular}{cccc}
      \textbf{Parameter} & \textbf{EAGLE} & \textbf{Illustris} & \textbf{TNG} \\
      \hline
      $\Omega_{\Delta}$  & 0.693 & 0.727 & 0.6911 \\
      \hline
      $\Omega_{m}$  & 0.307 & 0.2726 & 0.3089 \\
      \hline
      $\Omega_{b}$  & 0.04825 & 0.0456 & 0.0486 \\
      \hline
      h  & 0.6777 & 0.704 & 0.6774 \\
      \hline      
      $\sigma_{8}$  & 0.8288 & 0.809 & 0.8159 \\
      \hline
      n\textsubscript{s}  & 0.9611 & 0.963 & 0.9667 \\
      \hline
    \end{tabular}
    \label{tab:SimCosmoParams}
\end{table}

\subsection{SDSS}

The Sloan Digital Sky Survey Data Release 7\footnote{SDSS Data Access: \url{https://classic.sdss.org/dr7/access/}} \citep[SDSS DR7;][]{Abazajian_2009} functions as our parent observational sample. It constitutes the largest sample of spectroscopic and photometric data for local galaxies (at $z < 0.2)$. 

\subsubsection{SDSS Parameters}

We use public value added catalogs to extract parameters, including stellar mass and star formation rates \citep[from][]{Brinchmann_2004}, halo masses and central satellite classifications \citep[from][]{Yang_2009,Yang2011}, and stellar velocity dispersions \citep[from][]{Blanton_2005}. We take velocity dispersions ($\sigma_{*}$) from the NYU value-added galaxy catalog \citet{Blanton_2005}, allowing us to compute values for black hole mass via the $M_{\rm BH} - \sigma$ relation (see Section 3.4). We also use spectroscopic redshifts, galaxy magnitudes, and astrometry from the SDSS DR7 release \citep[see][]{Abazajian_2009}.

\par The star formation rates are taken from \citet{Brinchmann_2004}, determined by one of two techniques. For strong emission lines, without AGN signature, the SFRs are determined through dust corrected emission lines. In the absence of strong emission lines, and for AGN dominated galaxies, a relationship between specific star formation rate (sSFR) and the strength of the 4000{\AA} break (D4000) is used. Given the 3'' diameter of the SDSS fibres, the entire galaxy is rarely seen. Therefore, calculating the total SFR is done through an analysis of the galaxy color, using a relationship between the in-fibre SFR and in-fibre colours, as done in \citet{Salim_2007}. 

\par From \citet{Mendel_2014} we extract bulge-disk decompositions, the bulge and disk masses, along with the total mass, determined via SED fitting. We also extract the S\'ersic mass and z\textsubscript{max}, where z\textsubscript{max} is the maximum observable redshift for each object. The latter is used to perform inverse volume weighting by V\textsubscript{max}, used to correct Malmquist bias. We can use either $M_{\rm Sers}$, the S\'ersic mass, or the sum of the bulge and disk masses as the total mass for a galaxy. In this work we chose to use $M_{\rm Sers}$. The reason for this is due to issues with combining the bulge and disk masses, discussed in detail in \citet{Bluck_2014}.

\par We use lower halo masses from \citet{Yang_2009}, while higher halo masses are extracted from \citet{Yang_2007}. Specifically, the total stellar mass of galaxies in groups is used to ascertain halo masses through abundance matching. Group members are determined through a linking length algorithm. A galaxy is first linked to all galaxies within a certain distance. It is then further linked to all of its neighbors' neighbors. The process is iterated until we find ourselves far beyond the virial radius, at which point the process can be repeated to form a new group. Estimating the total dark matter masses is done by rank ordering the group halo and stellar mass functions. To recover this data we refer the reader to \citet{Piotrowska_2022}.

\subsection{Sample Selection}

We apply the same selection criteria to all simulations and observations. We first demand that only $M_{\rm Halo} > 10^{11} \, {\rm M_{\odot}}$ haloes be considered and further restrict the galaxies to have $M_{*} > 10^{9} \, {\rm M_{\odot}}$. This ensures that galaxies are well resolved in simulations, and additionally permits sufficient galaxy counts in the SDSS to apply robust corrections to number densities. 

We subdivide the satellite population into low mass satellites (with $M_{*} < 10^{10} \, {\rm M_{\odot}}$) and a high mass satellites (with $M_{*} > 10^{10.5} \, {\rm M_{\odot}}$). This cut is motivated by the analysis in \citet{Bluck_2020b}, where we find that high and low mass satellites quench via very different mechanisms. We apply the same mass limitations to the SDSS, and further require the redshift of a galaxy to be contained within $0.02 < z < 0.2$. We then classify the central galaxy of a group as its most massive member, and any other galaxy within the virial radius a satellite. Table.\ref{tab:Samples} shows the total parent sample size, central sample size, and satellite sample size for each database. 

\par To ensure an accurate estimate of black hole mass from velocity dispersion \citep[e.g., via the][relation]{Saglia_2016}, for the SDSS we can apply an inclination selection criteria, to remove edge-on galaxies whose galactic rotation could dominate the fibre measurement of $\sigma_{*}$ \citep[see][]{Bluck_2016}. We employ a $b/a$ constraint where we require $b/a > 0.8$ for disky galaxies ($B/T \le 0/5$). In \citet{Piotrowska_2022} it was shown that a similar constraint does not affect results in random forest classification, so we choose to perform our analysis without applying these inclination cut in order to preserve a larger sample size of satellite (especially low mass satellite) galaxies. However, we do additionally test to see if these types of cuts impact out conclusions.

\begin{table}
    \centering
        \caption{Sample Group size for each simulation and the SDSS}
    \begin{tabular}{ccccc}
      \textbf{Sample Group} & \textbf{SDSS} & \textbf{EAGLE} & \textbf{TNG} & \textbf{Illustris} \\
      \hline
        {\bf Total} & 480913 & 12931 & 22029 & 23907\\
      \hline
        Quenched & 254709 & 3853 & 6466 & 3430\\
      \hline
        Star Forming & 226204 & 9078 & 15563 & 20477\\
      \hline
        {\bf Centrals} & 380474 & 7125 & 11726 & 14145\\
      \hline
        Quenched & 195865 & 894 & 1483 & 620\\
      \hline
        Star Forming & 184609 & 6231 & 10243 & 13525\\
      \hline
        {\bf Satellites} & 100439 & 5806 & 10303 & 9762 \\
      \hline
        Quenched & 58844 & 2959 & 4983 & 2810\\
      \hline
        Star Forming & 41595 & 2847 & 5320 & 6952\\
      \hline
    \end{tabular}
    \label{tab:Samples}
\end{table}

\section{Methods}\label{Methods}

\subsection{Classifying high mass satellites and low mass satellites }

\par We split the satellite galaxies into two groups, high mass satellites and low mass satellites. This is motivated by our work in \citet{Bluck_2020b} where we found that the mechanisms of quenching vary dramatically with mass for satellites. As mass quenching cannot operate at $M_{*}< 10^{10}{\rm M_{\odot}}$ \citep[see][]{Peng_2010,Peng_2012, Bluck_2016}, we limit low mass satellites to below this threshold. This effectively removes AGN effects from this population. For high mass satellites, we require $M_{*} > 10^{10.5}{\rm M_{\odot}}$. The gap in coverage is intentional since there is an overlap between quenching modes which runs continuously with mass. As such, this enables a clear distinction in properties of the two populations. Unlike for satellites, we analyse all centrals together (i.e. impose no mass cuts). However, we note that there are essentially no quenched centrals at low masses \citep[see, e.g.,][]{Piotrowska_2022, Bluck_2022}.

\subsection{Identifying Star Forming and Quiescent Galaxies}

\begin{figure*}
\includegraphics[width=\textwidth]{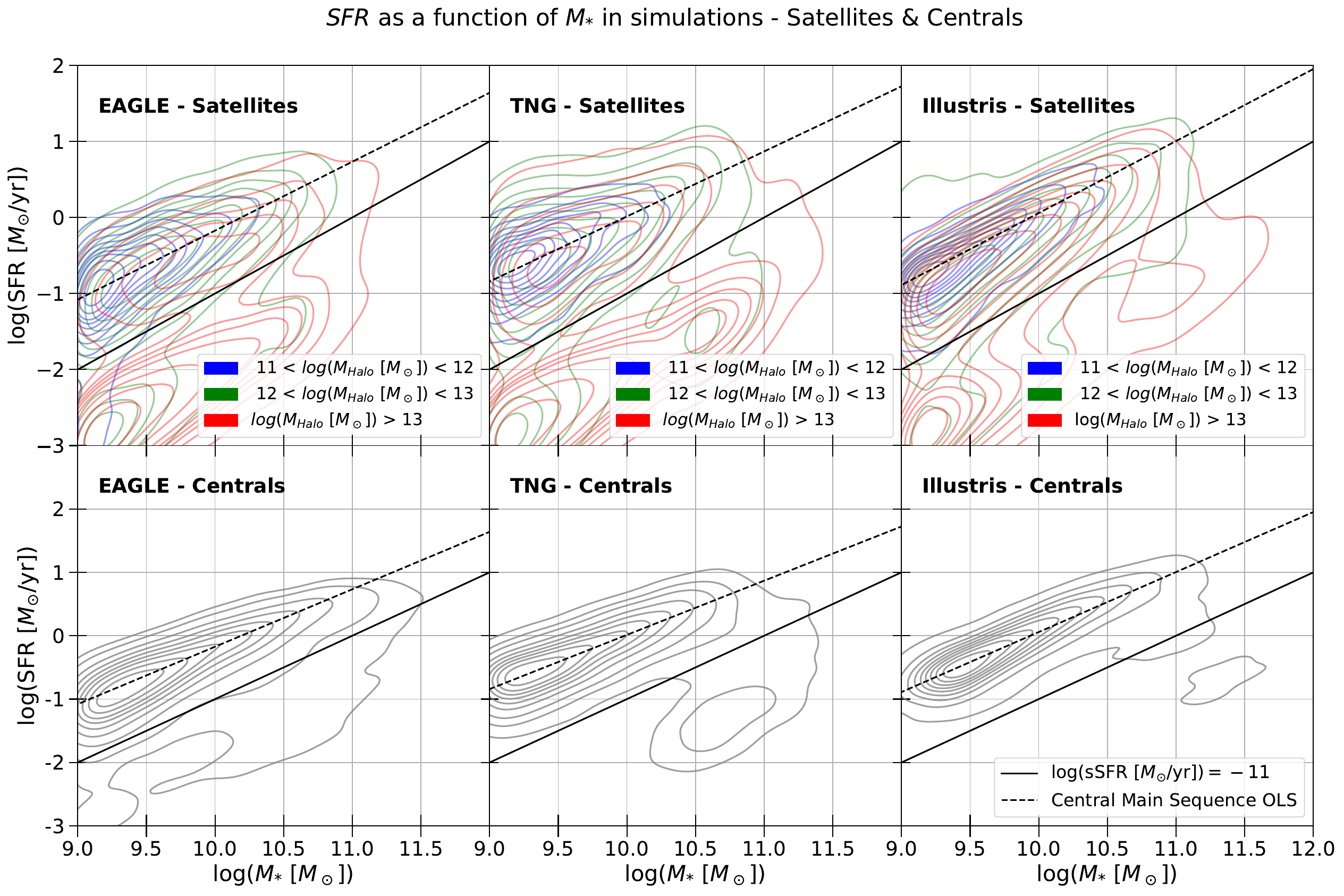}
\vspace*{-3mm}
\captionof{figure}{The SFR - $M_{*}$ relationship in simulations. The top panels present satellite galaxies, the bottom panels present central galaxies. For each row, the leftmost panel corresponds to the EAGLE suite, the middle panel to the TNG suite, and the rightmost panel to the Illustris suite. For sSFR < -11.5 yr$^{-1}$ galaxies we redistribute their sSFR's as a normal distribution centered around -12 with a dispersion of 0.33 dex, so all data may be visible when plotting. An OLS best fit to the central star forming main sequence is displayed in each plot (dashed lines). Additionally, the sSFR = -11yr$^{-1}$ line is shown as a solid line on each panel, which we use as our quenching threshold. For satellites, different halo masses are presented with different color density contours. It is clear that the central main sequence relationship well represents the main sequence for satellites in all masses of haloes. Additionally the sSFR quenching threshold clearly does a good job of separating the star forming and quiescent density peaks.
}\label{fig:SIMMainSequence}
\end{figure*}

\begin{figure*}
\includegraphics[width=\textwidth]{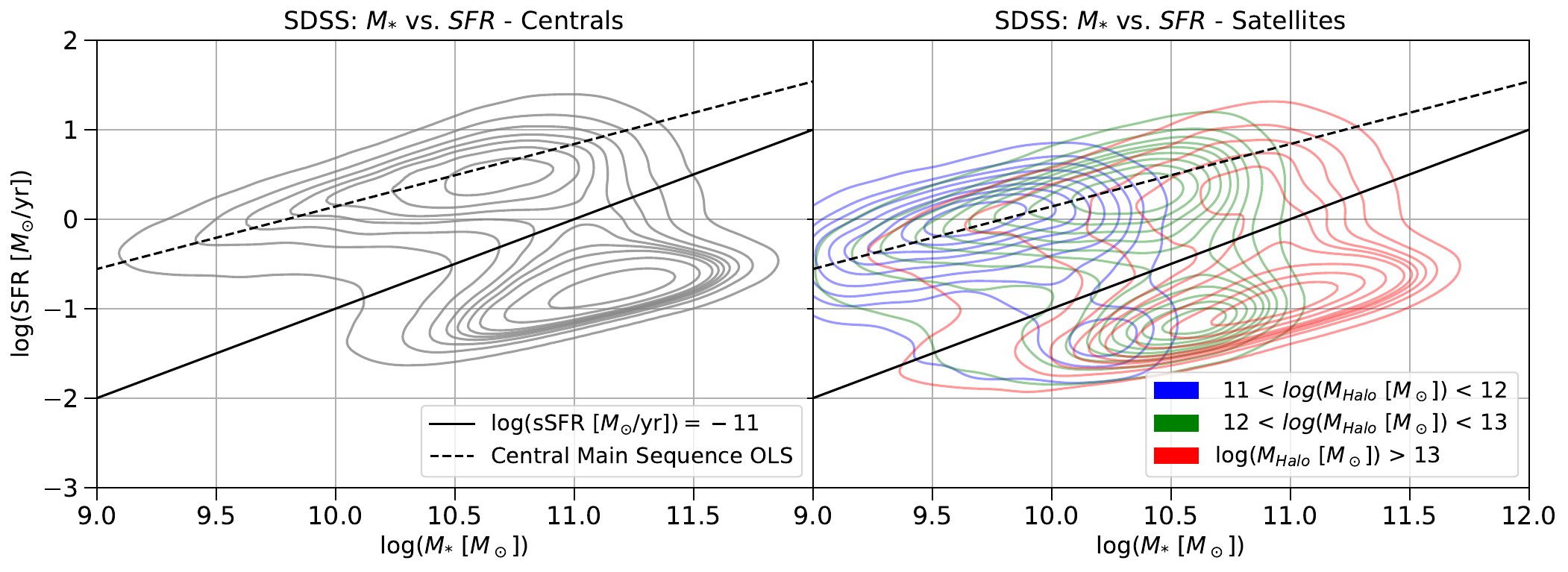}
\vspace*{-3mm}
\captionof{figure}{The SFR - $M_{*}$ relationship for observed galaxies in the SDSS. The structure of this plot is identical to Fig.\ref{fig:SIMMainSequence}, but here centrals are presented on the left panel and satellites on the right panel. Clearly, the central galaxy main sequence (dotted lines) is a good representation of the satellite galaxy main sequence at all halo masses. This indicates that star forming systems behave the same in different environments (although the probability of being quenched does vary, see Table. \ref{tab:f_Q_HaloMass}). Additionally, the sSFR quenching threshold (solid line) clearly does an excellent job of separating the star forming and quiescent density peaks in all types of local galaxies.}\label{fig:SDSSMainSequence}
\end{figure*}

\par We focus first on a comparison of the distribution of central and satellite galaxies in the SFR - $M_{*}$ plane for simulated and observed data. For simulations, in order to present the SFR in a finite range we redistribute extraordinarily low sSFR values to a randomly selected value in the distribution of quenched objects of the SDSS \citep[as in ][]{Bluck_2016}. This allows the distributions in the plane to be more comparable to the SDSS for visual comparison\footnote{But note that this has no impact on the designation of galaxies into star forming or quiescent classes.}. Fig.\ref{fig:SIMMainSequence} presents the satellite galaxies in the top row and centrals in the bottom row, while columns correspond to one of the simulation suites (EAGLE, TNG, and Illustris). The satellite plots are separated into three halo mass groups (as indicated by the contour colors). In all plots we draw two lines. The dotted line is an OLS (ordinary least squares) fit to the central main sequence (MS), defined as having an $\log(\mathrm{sSFR}/yr^{-1}) > -10.5$ \citep[as in][]{Bluck_2016}. Meanwhile, the solid line shows $\log(\mathrm{sSFR}/yr^{-1}) = -11$, which we use as our quenching threshold \citep[see][]{Piotrowska_2022}.

\par The OLS fit to the centrals Main Sequence fits well to the distribution for the Main Sequence in satellites as well. This indicates that the classification as either central or satellite, or even varying halo mass, does not influence where we find the MS. The group halo mass, however, does impact the fraction of quenched objects for satellites, which we display in Table.\ref{tab:f_Q_HaloMass}. The $11<\log(M_{\rm Halo}[{\rm M_{\odot}}])<12$ group has its entire distribution above $\log(\mathrm{sSFR}/yr^{-1}) = -11$, while $12<\log(M_{\rm Halo}[{\rm M_{\odot}}])<13$ has more contours beneath that line, and the $\log(M_{\rm Halo}[{\rm M_{\odot}}])>13$ group dips a few density contours well underneath it. We can then infer that, for satellites, halo mass must be correlated with quiescence.

\par In Fig. \ref{fig:SDSSMainSequence} we repeat this process for observed galaxies in the SDSS, separating centrals (left panel) and satellites (right panel). There is a similar effect as was seen for the simulations, where the centrals' MS OLS fit lays over the MS of the satellites, independent of the halo mass. Furthermore, increasingly larger fractions of objects are seen to be quiescent as the halo mass increases. This leads to the same conclusion as predicted by simulations: halo mass has a clear effect on the fraction of quiescent systems, which we investigate further through machine learning classification, but the main sequence of star forming systems is environment independent. That is, star forming galaxies are essentially identical in different environments, but the probability of ceasing star formation is a strong function of environment.

\par In summary, in Figs.\ref{fig:SIMMainSequence} \& \ref{fig:SDSSMainSequence} we show the SFR - $M_{*}$ plane for the satellites and centrals, in both simulations and observations, respectively. In both cases, the $\log(\mathrm{sSFR}/yr^{-1}) = -11$ line cleanly demarcates two groups, star forming and quenched. This is especially evident for the observed data, due simply to a much larger amount of data points than in the simulations (restricted by the box sizes). Hence, we adopt $\log(\mathrm{sSFR}/yr^{-1}) = -11$ as our quenching threshold in the local Universe. Galaxies with higher sSFR are considered star forming, and galaxies with lower sSFR are considered quenched. Clearly, this threshold is at least adequate for separating the density peaks in the SFR - $M_*$ plane in both simulations and observations, and moreover also works for both centrals and satellites, and in various mass haloes. This is a major advantage because it greatly simplifies the classification procedure (see Section 4).

\begin{table}
    \centering
        \caption{Quenched Fraction size of satellites in different halo mass ranges}
    \begin{tabular}{cccc}
      \textbf{Halo Mass} & \textbf{f\textsubscript{Q} EAGLE} & \textbf{f\textsubscript{Q} TNG} & \textbf{f\textsubscript{Q} Illustris} \\
      \hline
        $11<\log(M_{\rm Halo}[{\rm M_{\odot}}])<12$ & 0.055 & 0.008 & 0.016\\
      \hline
        $12<\log(M_{\rm Halo}[{\rm M_{\odot}}])<13$ & 0.251 & 0.234 & 0.168\\
      \hline
        $\log(M_{\rm Halo}[{\rm M_{\odot}}])>13$ & 0.723 & 0.689 & 0.472\\
      \hline
    \end{tabular}
    \label{tab:f_Q_HaloMass}
\end{table}

\subsection{Simulation Realism}

Since observational data is collected in a 2D plane, any measurement of distance conducted for observational data is two dimensional (plus redshift). This is in opposition to the 3D geometry of the box models of the simulations seen in Fig.\ref{fig:SIMCubes}, for which we would usually compute 3D distances. Consequently, we form two data sets, a `2D complete' sample set and an `SDSS-Like' data set, for which we chose a random 2D plane when computing any geometric features, such as the distance to the central for satellites or the local density of a galaxy. Additionally, we no longer use units of volume but rather units of area whenever relevant, particularly when computing density. While the `2D complete' data differs from the complete data solely in dimensionality, the `SDSS-Like' data is further subject to a weighing process, applied to reproduce the stellar mass function of the observational data.  

\par The distribution of stellar masses in the simulations differs greatly from that of the Sloan Digital Sky Survey. This is due to the strong mass incompletion of the SDSS, a result of a fixed magnitude detection threshold \citep[see][]{Abazajian_2009}. A common technique to correct for this is to weight by the inverse of the detection volume for each galaxy \citep[e.g.,][]{Thanjavur_2016}. However, our intent is to train a Random Forest classifier to identify quiescent galaxies. As such, the missing data is not recovered simply by weighting the SDSS stellar mass function. Moreover, in the SDSS this also has the deleterious effect of propagating uncertainties. As an alternative, in order to correct for the sample differences, we choose to weight the simulated stellar mass functions to match that of the observed data. We apply random sampling of the SDSS-weighted simulations {\it with return}. Explicitly, we assign a weight to each simulated galaxy mass bin with respect to the SDSS stellar mass function, computed as:

\[w_{i} = \bigg(\frac{H_{SDSS,i}}{H_{SIM,i}}\bigg)\]

\noindent where $H_{SDSS,i}$ is the height of the i\textsuperscript{th} bin in the stellar mass distribution for the observed data, and $H_{SIM,i}$ is the height of the i\textsuperscript{th} bin for the same range in simulated data. This allows us to simulate an SDSS-like view of the simulations (with approximately identical stellar mass functions) before making comparisons. Since the selection of data for simulations in the SDSS-like sample is performed with return, we form a sample with duplicated simulated galaxies, rather than removing galaxies from the sample. This is necessary since the very low detection rate of low mass galaxies in the SDSS would yield essentially zero galaxies in the simulations (which have a much smaller volume).

\begin{figure*}
\includegraphics[width=\textwidth]{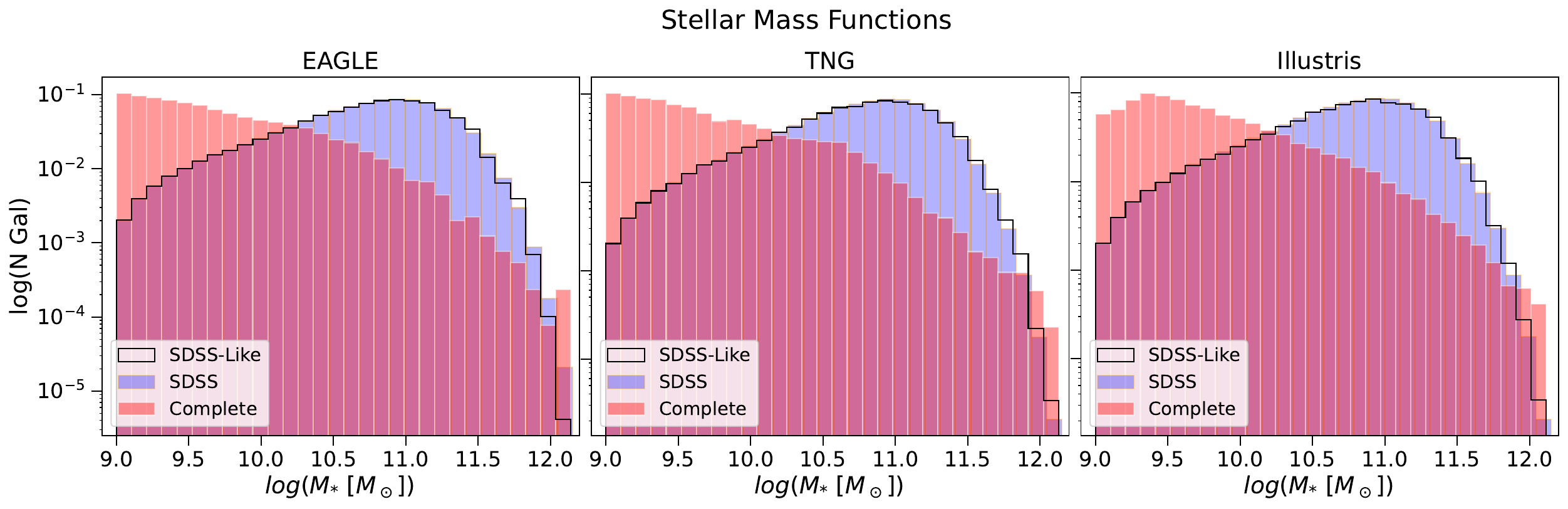}
\vspace*{-3mm}
\captionof{figure}{Stellar mass distributions for the simulations, shown before and after weighing to resemble the SDSS observed mass function. The SDSS stellar mass distribution is included for comparison. Post weighting, the mass distributions of the simulations are essentially identical, ensuring a fair comparison. \\}\label{fig:StellarMassFunction}
\end{figure*}

The stellar mass distribution of each sample set (simulation complete, simulation weighted, and SDSS) is presented in Fig.\ref{fig:StellarMassFunction}. We determine that this approach is better suited to the random forest classification than the usual $1/V_{\rm max}$ SDSS weighting, since it does not propagate errors in the observations. We ensure that a duplicated galaxy will never be placed in both the training and testing data sets (which would lead to spurious results), by generating the machine learning sub-samples prior to weighted sampling. This method does not propagate errors present in observed parameter values or over-weight lone objects, both of which could skew statistical results. Our weighted simulated data set is referred to as "SDSS-Like" in the future. Additionally, we perform more rigorous error analyses for the observational data (presented in the results sections).

\subsubsection{Nearest Neighbor Density and Topology}

We compute the galaxy density around each galaxy for the simulations and the SDSS as:

\begin{equation}
\delta_{n} = \frac{\rho_n}{\langle \rho_n \rangle}; \,\, \mathrm{where,} \,\, \rho_n = \left(\frac{n}{\frac{4}{3}\pi R^{3}}\right) \,\, \text{(for 3D)} 
\end{equation}

\begin{equation}
\delta_{n} = \frac{\Sigma_n}{\langle \Sigma_n \rangle}; \,\, \mathrm{where,} \,\, \Sigma_n = \left(\frac{n}{\pi R^{2}}\right) \,\, \text{(for 2D)} 
\end{equation}

\noindent where $n$ denotes the nearest neighbor (third, fifth, tenth, etc...) and R is the distance to the n\textsuperscript{th} nearest neighbor, computed in this work for $n$ = 3, 5, 10. Note that we normalise the densities by the mean density in each dataset, using as our final measurement an over-density, relative to the norm \citep[as in][]{Baldry_2006, Peng_2010}.

\par
\begin{figure*}
    \centering
    \begin{subfigure}{.33\textwidth}
        \caption{EAGLE}
        \includegraphics[width = \textwidth]{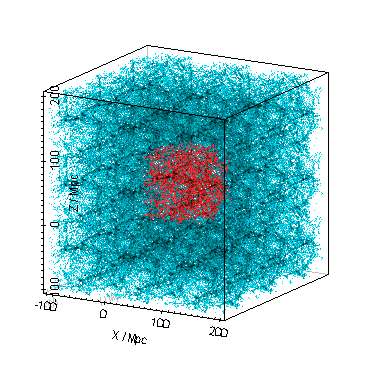}  
    \end{subfigure}
    \hfill
    \begin{subfigure}{.33\textwidth}
        \caption{TNG}
        \includegraphics[width= \textwidth]{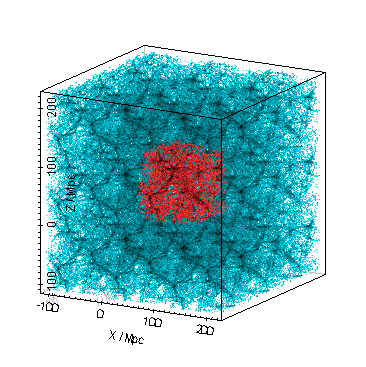} 
        
    \end{subfigure}
    \hfill
    \begin{subfigure}{.33\textwidth}
        \caption{Illustris}
        \includegraphics[width= \textwidth]{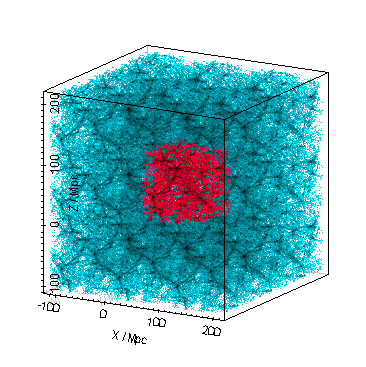}  
    \end{subfigure}

\vspace*{-3mm}
\caption{A 3D representation of the topology of the simulations. The teal regions mark additional cubes added to simulate the 3-torus topology, while the red region indicates the original cube in the center. This approach allows for the removal of edge effects in determining local densities and distances from central galaxies.}\label{fig:3torus}
\end{figure*}

 The simulations present an issue for density measurements due to their configurations as cubes. Edge effects impact the density of any object near one of the sides. In order to correct for this issue, we arrange a duplicated version of the original box to each of its sides (and corners). This allows for us to properly account for the periodic boundary conditions of the simulations. This representation allows for one side to experience the effects of the opposite side with respect to density. For example, if one were to cross the top plane of the cube, one would find oneself back at the bottom. This approach nullifies the edge effect issues in the simulations, while also conserving the topology and geometry of the simulations. In Fig. \ref{fig:3torus} we present the original cube in red, and the additional cubes in blue to illustrate this process.

\par Edges, however, are not unique to the simulations. Indeed, whether due to the physical limits of the survey, star diffraction spikes, or issues when recording data, we find edges in the SDSS. To test the impact of edge effects in observations we perform random forest classification on a smaller region of the survey, selected away from edges. The density parameters are still computed with respect to the rest of the survey. The results, presented in Appendix. \ref{appendix:EdgeSDSS}, demonstrate edge effects to have little impact on results. In fact, the absence of edges further affirms our result for low mass satellites and has no effect on central galaxies or high mass satellites. Therefore, we perform our analysis for the entire survey, rather than only a slice.

\par When dealing with a 2D configuration (2D complete and `SDSS-Like' datasets), we add cubes solely to the 2D plane in which we operate, preventing over counting. Additionally, we restrict a nearest neighbor to be within a sphere of radius 50\,Mpc, half the box size, from the pertinent galaxy. This is done to avoid duplicated objects in the replicated boxes, which would return false values of density. While the simulations operate in 6D phase space, 3 coordinate space (x, y, z) and 3 velocity space ($V_{x}, V{y}, V_{z}$), the SDSS is presented in a 2D + $z$ observational space. We know the position in right ascension and declination, and can estimate the line-of-sight velocity of each galaxy.

\par When computing density for observational data we must ensure that objects are truly near each other. In the SDSS, objects may appear nearby, if their right ascension and declination are similar, while at greatly different redshifts. Therefore, to assure objects are in the same physical region, we apply a redshift (or velocity) cut before evaluating density. Specifically, we require:

\begin{equation}
    \Delta V \equiv \frac{c \Delta z}{1 + \langle z \rangle} \, < \, 1500 \, \mathrm{km/s}    
\end{equation}

\noindent where, $\Delta z = z - z_i$, where $z_i$ is the redshift of any given neighbour of a galaxy with redshift, $z$; and $\langle z \rangle$ indicates the mean redshift of the two neighbouring galaxies. As such, we assure the galaxies exist in physically connected structures prior to any density, or halo location, calculations. For the `2D complete' and `SDSS-Like' data sets, we apply a similar velocity cut, computed with the peculiar velocity of each galaxy in the direction perpendicular to the (randomly selected) 2D plane. This aids in recreating observational conditions in the simulations. The relative velocity threshold is chosen to be just higher than the largest velocity dispersions observed in galaxy clusters.

\subsection{\texorpdfstring{$M_{\rm BH}$ in simulations and the SDSS}{MBH estimation in simulations and the SDSS}}

\subsubsection{In the simulations} \label{SimMBH}

In the simulations a notable portion of the galaxies appear to be stripped of their black hole, resulting in $M_{\rm BH} = 0$. While it is possible for galaxies to be stripped of their black holes for some physical reason (e.g. due to the 3-body interaction of merging black holes), this issue occurs much too frequently to be due to a true process. Rather, this appears to be an artifact of the simulations due to the errors in the Subfind Sublink algorithm. For centrals, this is almost never an issue ($\ll 1\%$ of systems are affected). For high-mass satellites this is a small issue ($<10\%$ of systems are affected), whereas for low mass satellites this issue becomes much more serious ($\sim 20 - 40\%$ of systems are affected). Therefore, we find it is necessary to combat this issue in our analysis, especially in cases where completeness is important. 

To address this issue we simply assign a `typical' $M_{BH}$ value to a galaxy with a missing black hole. More specifically, we assign a black hole mass value randomly drawn from a Gaussian distribution centered around the mean of a given class (central, high mass satellite, low mass satellite), with dispersion, $\sigma$, set by the standard deviation in ${M_{\rm BH}}$ values for the relevant class. 

\par In order to show that this method does not influence the conclusions of this work, we perform a Random Forest classification comparing a sample for which we require $M_{\rm BH} > 0$ (hence losing artificial stripped systems from our analysis) to a sample for which we re-assign a black hole mass as described in the method above. We present these results in Appendix. \ref{appendix:MBH_Reassign}. We find that our redistribution method does not alter any of the qualitative results, and has the distinct advantage of allowing us to recover a more substantial sample of satellite galaxies, especially low mass satellites. This is especially important for our later analysis of quenching as a function of location within the halo, controlled by black hole mass. Accordingly, we choose to adopt the correction method for our main analysis, but note again that this has no impact on our conclusions as to which parameter is most effective at predicting quiescence in simulations.

\subsubsection{In the SDSS}
Due to the difficulty in obtaining dynamical measurements of black hole masses in observational data, only $\sim$100 galaxies have known black hole masses in the local Universe \citep[e.g.,][]{Terrazas_2016, Terrazas_2017, Saglia_2016}. As an alternative, we turn to the relations between $M_{\rm BH}$ and the central velocity dispersion to estimate it. While black hole mass possesses a correlation with stellar mass, halo mass, and bulge mass, it is best correlated with $\sigma_{*}$ \citep{Saglia_2016}. That is, the RMS scatter in the $M_{\rm BH}$ - $\sigma_{*}$ plane is tightest. In \citet{Piotrowska_2022} multiple methods to estimate black hole mass are implemented to test how this could impact the results of random forest classification. There was no discernible change in the outcome of the classifications, including from using distinct calibrations for early and late galaxy types, pseudo or classical bulges, or from using bulge mass or other alternatives. We therefore choose in this work to determine $M_{\rm BH}$ via a single approach \citep[the optimal according to ][]{Piotrowska_2022}, using the following formula \citep[from][]{Saglia_2016}:

\begin{equation}
\log(M_{\rm BH}) = 5.246 \times \log(\sigma_{*}) - 3.77\,; \\ 
\sigma_{\rm rms} = 0.417
\end{equation}

\noindent This approach enables us to estimate black hole masses for $\sim$0.5 million galaxies with accurate central velocity dispersions measured\footnote{Note that in the SDSS the resolution limit is $\sigma_* = 70 \, {\rm km/s}$, hence values lower than this are unconstrained. However, we do know at least that they are lower than this limit. When translated to black hole mass via eq. 4, these systems are estimated to have $M_{BH} < 10^6 M_{\odot}$, i.e. lower than any dynamically measured supermassive black holes in the local Universe, and much lower than the quenched threshold of $M_{BH} \sim 10^{7.5} M_{\odot}$, see \cite{Bluck_2016, Bluck_2022}. As such this limitation is not significantly problematic for this work: i.e., all systems with unresolved velocity dispersions are predicted to be unaffected by historic AGN feedback. }. The price for this three order of magnitude increase in sample size is a reduction in accuracy of each measured black hole mass to $\sim$ 0.5 dex (from adding in quadrature the scatter on the $M_{\rm BH} - \sigma_*$ relationship to the typical uncertainty on $\sigma_*$). We note that this is comparable to the accuracy of halo masses \citep[$\sim$0.5 dex;][]{Yang_2007, Yang_2009}, and not drastically larger than that of stellar masses \citep[$\sim$0.2 - 0.3 dex;][]{Mendel_2014}. Moreover, we rigorously test the potential impact of uncertainties on the results for observations in Section 4.2.3.

\subsection{Random Forest Classification}

In this work we use a machine learning classifier, specifically a random forest classifier from the {\small SCIKIT-LEARN PYTHON} package \citep{pedregosa11}. Random forest classification uses decision trees to evaluate the relative importance of parameters when determining the class to which they belong. In our case, we consider two classes of galaxies - (i) actively star forming; and (ii) quenched (based on the cut in sSFR displayed in Figs 3 - 4). 

The input features for training the random forest classifier are chosen to be black hole mass ($M_{\rm BH}$), halo mass ($M_{\rm Halo}$), stellar mass ($M_{*}$), nearest neighbor density ($\delta$, evaluated for the third, fifth, and tenth nearest neighbor), the distance to the central galaxy ($D_{\rm cen}$), and a random number in the range 0 - 1. The ability of the random forest classifier to evaluate and separate parameters as either causal or simply nuisance (inter-correlated without causation) is greatly useful to us \citep[see][for a detailed discussion and demonstration using mock and simulated data]{Bluck_2022}. 

\par To correctly evaluate the input features we create two sample sets, a training set and a testing set. These separate data sets are generated by splitting the total data set prior to any analysis, and is done individually for each sample (`complete', `2D complete', `SDSS-Like' in the simulations, and the SDSS data). As such, each sample set will have its own training and testing sets generated, independent from the others. The testing data set remains unseen to the classifier during training, enabling a test of the stability of the classifier to novel data. For the weighted simulation samples (e.g., `SDSS-like') we split the sample prior to weighting to assure there are no repeated galaxies in the training and validations sample sets. To ensure that neither set is dominated by either star forming or quenched objects we balance each set to ensure 50\% are star forming and 50\% are quenched. 

During training, the classifier learns which input features (and thresholds) are best at sorting the two classes. The performance is quantified via the Gini impurity \citep{pedregosa11}:

\begin{equation}
    G(n) = 1 - \sum_{i}^{c = 2} p_{i}(n^{2})
\end{equation}

\noindent where we sum over all classes ($c=2$; star forming and quenched) and $p(n)$ indicates the probability of randomly selecting class, $i$, at node, $n$. Hence high values of the Gini parameter indicate poor separation of the classes. Each variable will be evaluated in their ability to separate the two classes at each branch, the most accurate will then be used. The Gini impurity will further be used to determine the optimal threshold for the best variable. The random forest classifier then iterates this procedure to fill out the entire decision tree. The full random forest is propagated through training on different bootstrapped random samples, which is proven to yield higher fidelity classifications.

\par From the trained data set, the relative importance of each input feature is determined through the following formula \citep{Bluck_2022, scikit-learn}:

\begin{equation}
        I_{R}(k) = \frac{1}{N_{\rm trees}} \sum_{\rm trees}\Biggl\{ \frac{\sum_{nk}N(n_{k})\Delta G(n_{k})}{\sum_{n}N(n)\Delta G(n)}\Biggl\}
\end{equation}

\noindent where $I_R(k)$ is the relative importance of input feature $k$. $\Delta G$ corresponds to the improvement of the Gini impurity from a parent node to a daughter node, and is additionally weighted by the number of features in the parent node. The numerator is summed over all nodes that use the relevant feature $k$ to split, while the denominator is summed over all nodes in the random forest. This ratio then gives the importance of the feature $k$ in classifying a galaxy as star forming or quenched, within the training sample. This is then averaged over all the trees in the random forest, which returns the relative importance, or quenching importance, that we use as our primary statistic when presenting classification results. 

\par To ensure statistically meaningful results we perform ten runs of both training and testing. We record the mean of the relative importance for each feature and use the standard deviation from the mean over the ten runs as the statistical error. 

\par We control for over-fitting by comparing the performance of the random forest classifier on the testing sample set, which is unseen by the classifier in training, to the training sample. We then plot the true positive - false positive curve for each and measure the area under the curve (AUC) for each set \citep[see][]{Teimoorinia_2016}. We require a very small difference in performance of the two classification runs, specifically requiring $\Delta AUC < 0.02$ \citep[as in][]{Bluck_2022}. This ensures that the random forest classifier does not learn pathological features from the training data, but only those which are effective on novel data. For a more in depth look into the RF classification architecture we refer readers to the appendix of \citet{Bluck_2022}, where a detailed mathematical discussion is provided, as well as numerous tests on the accuracy of this approach.

\section{Results}\label{Results}

\subsection{Cosmological Hydrodynamical Simulations}\label{sssec: SimResults}

\subsubsection{Random Forest Classification}

\begin{figure*}

    \includegraphics[width=\textwidth]{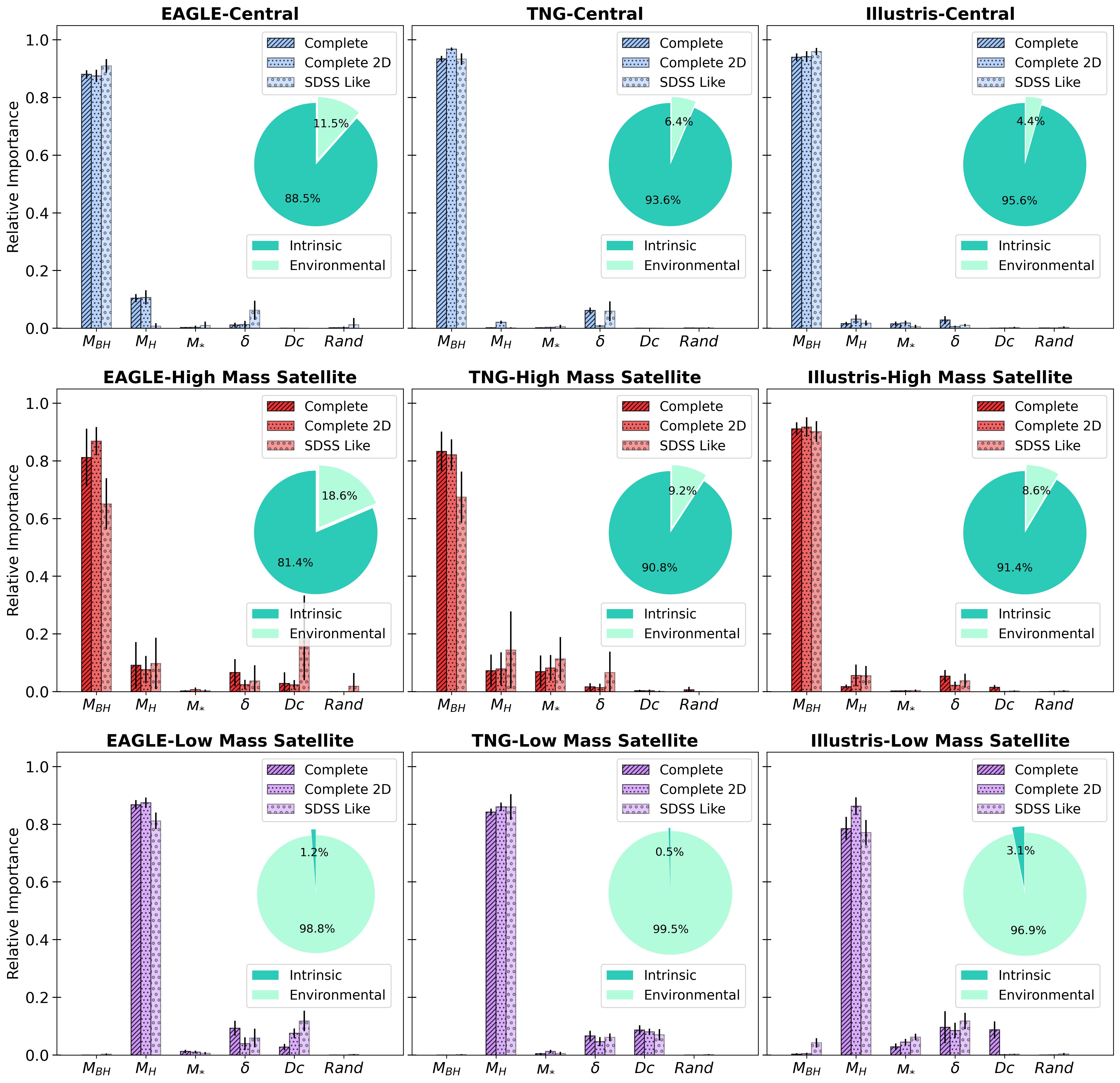}
    \vspace*{-3mm}
    \caption{Random forest quenching classification analyses for the cosmological simulations. Each row corresponds to a galaxy class (centrals, high mass satellites, low mass satellites), while each columns presents one of the simulation suites (EAGLE, TNG, Illustris). The parameters used for training the random forest are listed along the $x$-axis, with their relative quenching importances indicated by the $y$-axis bar heights. Uncertainties are given by the variance of 10 independent classification runs. Separate bars are shown for the complete simulated, the 2D complete simulated, and `SDSS-like' datasets. The pie charts on each panel indicate the total importance of intrinsic and environmental parameters for the complete sample. For all simulations, centrals and high mass satellites quench predominantly by intrinsic processes (with $M_{\rm BH}$ found to be by far the most dominant single parameter), whereas low mass satellites quench through environmental processes (with $M_{\rm halo}$ found to be by far the most dominant single parameter). These results are stable to the 2D complete data set and the SDSS sample selection. This indicates that both centrals and high mass satellites quench in these simulations predominantly via AGN feedback, yet low mass satellites quench predominantly as a function of environment in high mass groups and clusters.
    \label{fig:Simulations_RandomForest}}
\end{figure*}

\par Fig. \ref{fig:Simulations_RandomForest} shows the random forest classification results for the cosmological simulations. The height of each bar corresponds to the relative importance to quenching of its associated parameter. The input features tested are $\mathrm{M_{\rm BH}}$, $\mathrm{M_{\rm Halo}}$, $\mathrm{M_{*}}$, $\mathrm{\delta}$, $\mathrm{D_{cen}}$, and a randomly selected number between 0 and 1. The $\mathrm{\delta}$ parameter's relative importance represents the sum of the relative importance of the third, fifth and, tenth nearest neighbor. We carry out this classification test for centrals, high mass satellites, and low mass satellites, in the `complete', `2D complete' and `SDSS-Like' data sets, denoted by different hatches on the bars. The panels are each accompanied by a pie plot displaying the total importance of intrinsic and environmental parameters, defined as $\mathrm{M_{\rm BH}}$ and $\mathrm{M_{*}}$, and $\mathrm{M_{\rm Halo}, \delta, D_{\rm cen}}$, respectively. The pie charts correspond to the results of the random forest classification of the complete data set. This allows for a comparison in their respective importance to quenching.

\begin{figure*}

	\includegraphics[width=\textwidth]{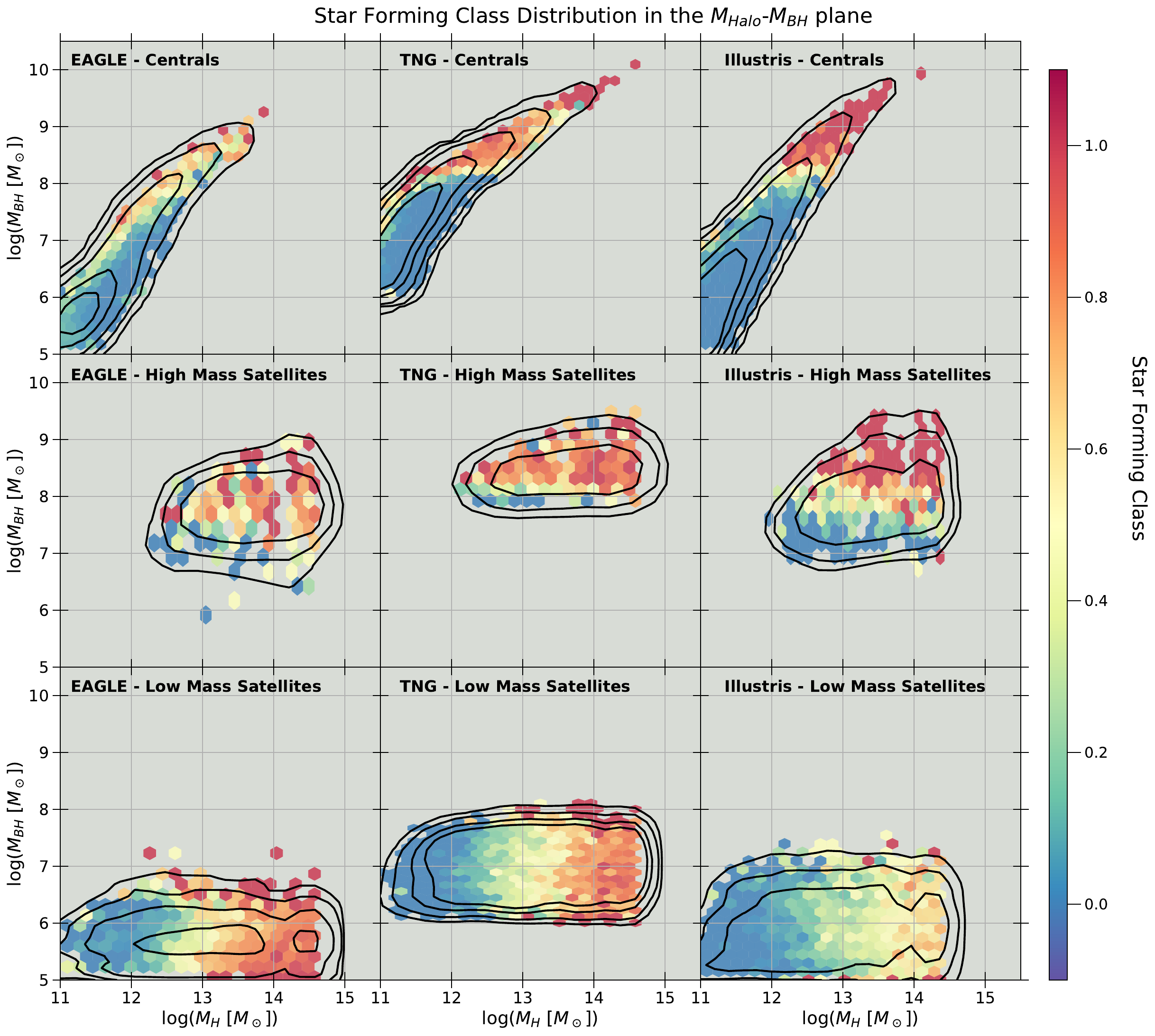}
    \vspace*{-3mm}
    \caption{The Halo Mass - Black Hole Mass relationship for each class of galaxy in every cosmological simulation suite, color coded by the average star forming class in each hexagonal bin. Each row corresponds to a galaxy class (central, high mass satellite, low mass satellite), while the columns represents each of the simulation suites in turn. TNG spawns black hole seeds at $M_{\rm BH}$ = $8 \times 10^{5} {\rm M_{\odot}} h^{-1}$, which leads to the raised values in low mass satellites compared to EAGLE and Illustris.}
    \label{fig:SimHexbin}
\end{figure*}

\begin{figure*}

	\includegraphics[width=\textwidth]{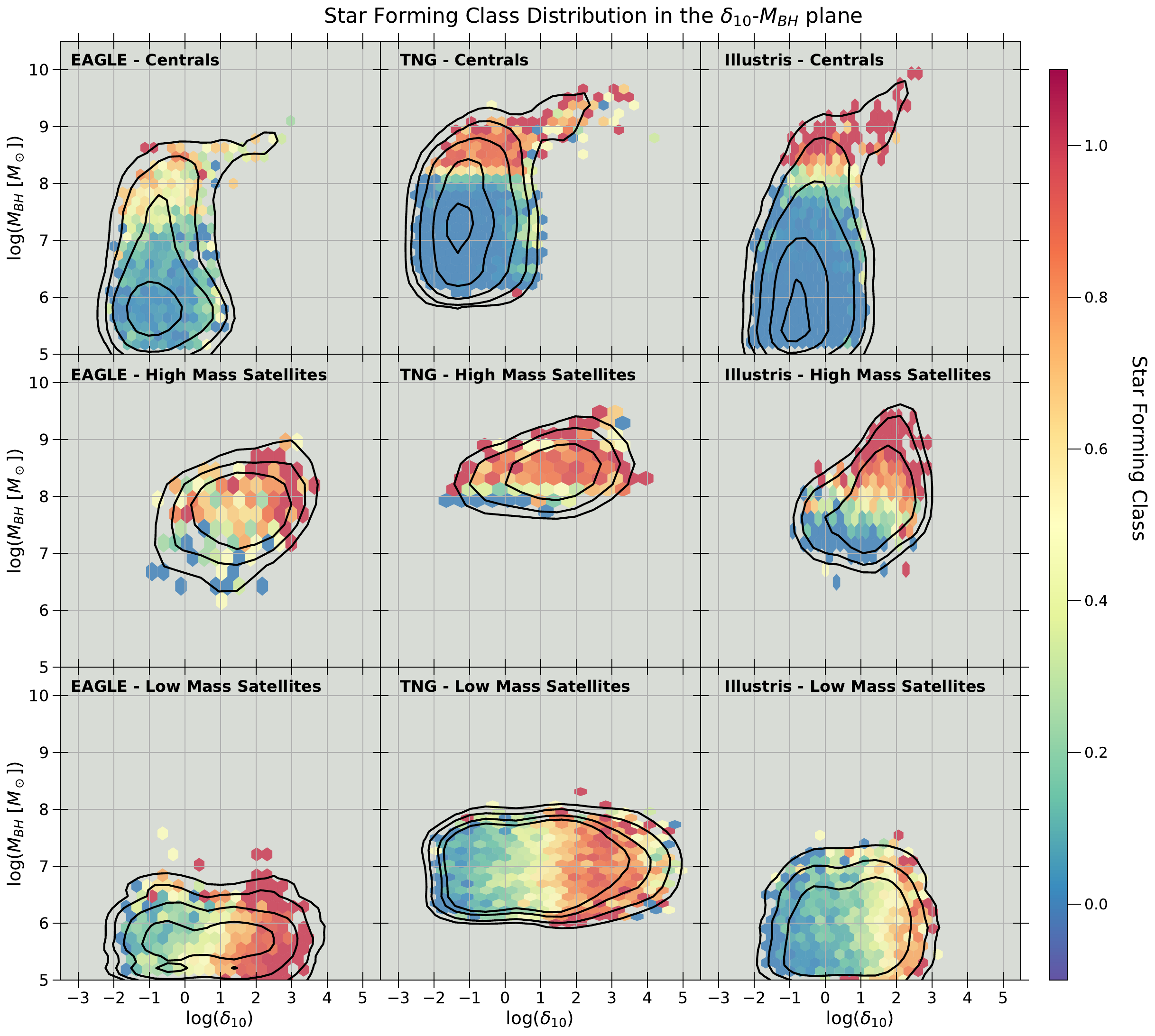}
    \vspace*{-3mm}
    \caption{The $\delta_{10}$ - Black Hole Mass relationship for each class of galaxy in every cosmological simulation suite. Each bin is color coded to its average star forming class. Each row corresponds to a galaxy class (central, high mass satellite, low mass satellite), while the columns represents each of the simulation suites in turn. }
    \label{fig:SimHexbinDen}
\end{figure*}

\par The top three panels of Fig. \ref{fig:Simulations_RandomForest} correspond to central galaxies. The most predictive parameter for quenching centrals is clearly black hole mass for all three of the simulations, in all data samples. The pie charts in these plots further make clear that the quenching of centrals is dominated by intrinsic parameters. These results are in close accord with our prior results at these epochs \citep[in][]{Piotrowska_2022}, as well as at higher redshifts \citep[see][]{Bluck_2023}. We repeat the analysis of central galaxies here in order to establish a baseline, environment-free quenching scenario in which to compare to satellite quenching at both high and low masses. Additionally, we add local densities as features input to the random forest classifier for the first time. This establishes what one may intuit, i.e. that central galaxies have little-to-no dependence of their quiescence on environment \citep[e.g.,][]{Peng_2012}.

\par The second row of plots shows the relative importance of the same parameters but now for high mass ($M_* > 10^{10.5} {\rm M_{\odot}}$) satellites. The complete data sets show the clearest results, where $M_{\rm BH}$ is the key parameter to determine quiescence, exactly as with central galaxies. The `2D complete' and `SDSS-Like' data sets agree with the raw sample, although a rise in the importance of other parameters is noticeable. In EAGLE, $M_{\rm Halo}$ and $\delta$ pick up some limited importance, as do $M_{\rm Halo}$ and $M_{*}$ in TNG, although they still do not begin to challenge $M_{\rm BH}$. Interestingly, while $M_{\rm BH}$ remains most dominant, the pie charts show that there is a slight gain for the environmental factors when comparing to the central galaxies. EAGLE, TNG, and Illustris have a growth of 7.1\%, 2.8\%, and 4.2\% in environmental importance, respectively. While the shift in favor of environmental factors may seem small for high mass satellites, it is much more evident for low mass satellites (discussed below).

\par The final row of panels presents classification results for the low mass ($M_* < 10^{10} {\rm M_{\odot}}$) satellites. In all simulations and data sets, halo mass is now the most predictive feature in determining quiescence of a galaxy. In all classification runs we also find a small dependence on $\delta$ and $D_{\rm cen}$. On the other hand, intrinsic parameters have essentially no importance. The pie charts convey this effect clearly, as the total importance in intrinsic factors fall from 81.4\% to 1.2\% for EAGLE, 90.8\% to 0.5\% for TNG, and from 91.4\% to 3.1\% for Illustris, when comparing high and low mass satellites.

\par Clearly, centrals and high mass satellites are subject to similar quenching mechanisms, driven by their supermassive black holes \citep[see also][for consistent prior results with centrals]{Piotrowska_2022, Bluck_2023}. In the simulations this is explicitly a result of AGN feedback, especially in the low luminosity preventative mode \citep[see, e.g.,][]{Terrazas_2020, Zinger_2020, Piotrowska_2022, Bluck_2023}. Conversely, low mass satellites achieve quiescence through environmental processes, rather than intrinsic factors, especially as a function of halo mass. We do find a secondary dependence on the location of galaxies within their haloes (as traced by $D_{\rm cen}$) and the local density on all scales (as traced by $\delta$). This small, albeit non-negligible, importance on $D_{\rm cen}$ suggests that the ram pressure stripping of ISM, which occurs most at the center of clusters, affects solely a small fraction of satellites. A more likely culprit for the quenching of low mass satellites is then the stripping of their CGM, or starvation, associated with the amount of ICM gas present in a given halo. We investigate the relation between quenching of low mass satellites and their location within a halo more in Section.\ref{sssec:HaloLoc}. This strongly suggests that the principal mode of environmental quenching in the simulations is related to the halo as a whole (e.g., via ram pressure stripping of CGM) rather than the interactions of satellites with each other (e.g., via dynamical harassment).

\subsubsection{The Quenching Angle}

\begin{figure*}
    \centering
    \begin{subfigure}{\textwidth}
        \includegraphics[width = \textwidth]{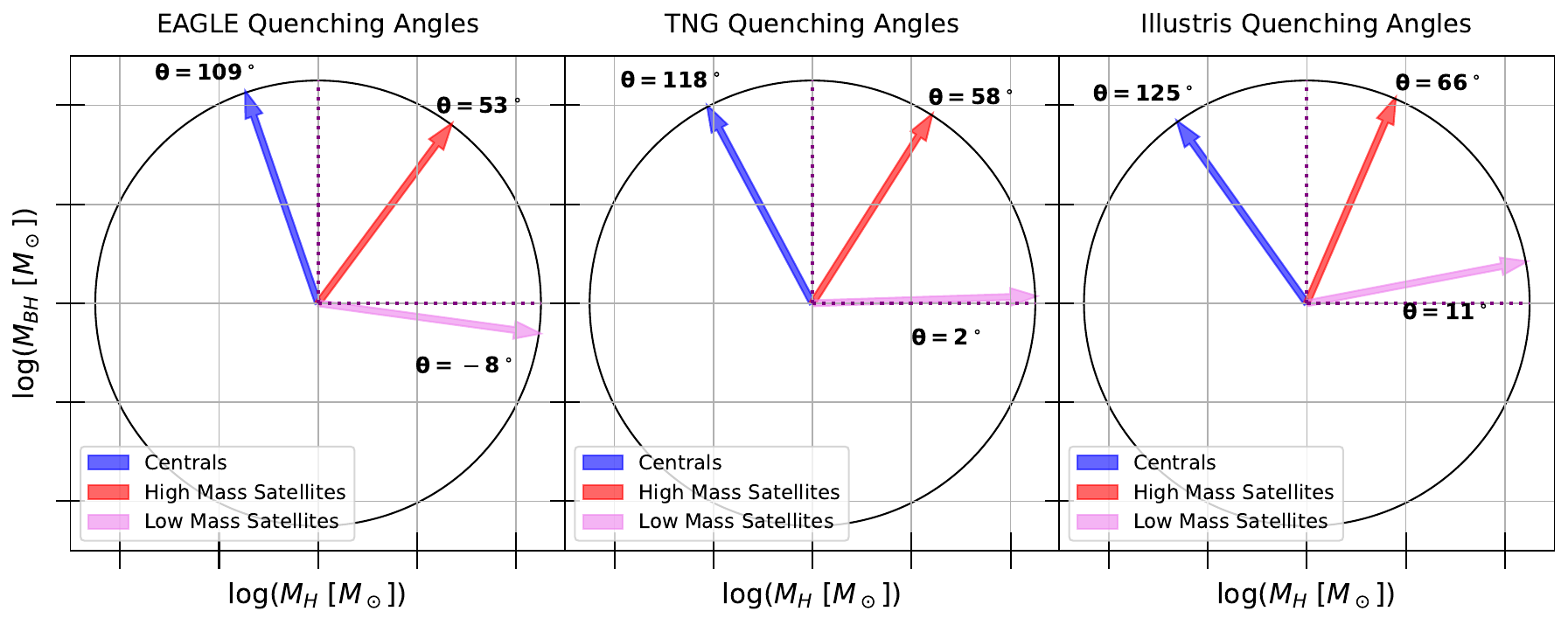}  
        \caption{}
        \label{fig:Sim_Quench_Halo}
    \end{subfigure}
    \hfill
    \begin{subfigure}{\textwidth}
        \includegraphics[width= \textwidth]{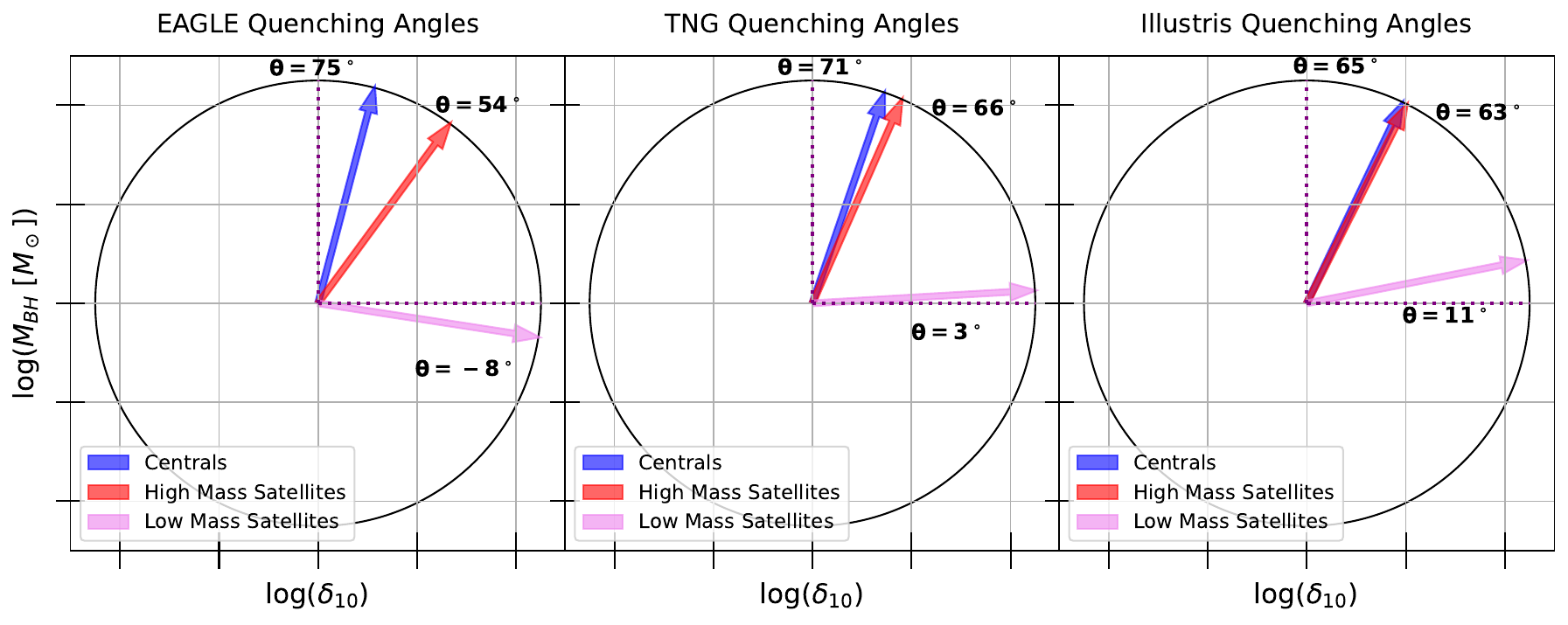} 
        \caption{}
        \label{fig:Sim_Quenc_Den}
    \end{subfigure}

\vspace*{-3mm}
\caption{{\it Upper panels (a):} Presentation of the quenching angles in the $M_{\rm Halo}$ - $M_{\rm BH}$ plane (computed via the ratio of partial correlations, see eq. \ref{QuenchAngEq}). For all simulations (displayed separately from left to right) a clear trend appears: Quiescence correlates predominantly with black hole mass for centrals and high mass satellites, while it correlates predominantly with halo mass for low mass satellites. Interestingly, centrals have a weak negative correlation with halo mass, whereas high-mass satellites have a weak positive correlation with halo mass. Low mass satellites are correlated almost entirely with halo mass. \\{\it Lower panels (b):}  Presentation of the quenching angles in the $\delta_{10}$ - $M_{\rm BH}$ plane, with the same format as in (a). We find the quenching vectors for central and high mass satellite galaxies to be extremely similar for all simulations, indicating strong dependence on black hole mass and weak dependence on galaxy density. However, low-mass satellite quenching strongly correlates with $\delta_{10}$, clearly favoring environmental over intrinsic processes for this population.}\label{fig:SIM_QuenchAng}
\end{figure*}

\par Fig. \ref{fig:SimHexbin} shows the $M_{\rm Halo}$ - $M_{\rm BH}$ relation for each galaxy class in simulations, color coded by the average star-forming class in each small hexagonal region. The top row of the plot presents central galaxies, the middle row corresponds to high mass satellites, and the bottom row shows low mass satellites. We note a change in the color gradient as one moves down from row to row, whereby quenching is regulated by $M_{\rm BH}$ for centrals and high mass satellites, but becomes clearly dependent only on halo mass for low mass satellites. This is in precise agreement to the evolution of the pie charts in Fig. \ref{fig:Simulations_RandomForest}. 

\par The central galaxy plots all show a positive correlation of star forming state with $M_{\rm BH}$, and a negative correlation with $M_{\rm Halo}$. This is interpreted as quenching for centrals to be more likely as the black hole mass grows, and less likely as the halo mass grows. Thus, more massive haloes require more massive black holes in the simulations in order to reach quiescence. Nonetheless, the dominant correlation is with black hole mass, not halo mass. As we go from centrals to high mass satellites, we find the color gradient of high mass satellites to be somewhat split. A dependence on both $M_{\rm BH}$ and  $M_{\rm Halo}$ is clear, though the former is noticeably stronger. While this is present for each of the simulations, Illustris displays a greater dependence on black hole mass than either TNG or EAGLE. In the final row, for low mass satellites, halo mass clearly dominates. The color gradient is seemingly fully horizontal for all simulations. Black hole mass has no impact in the quenching of the low mass satellites. 

\par Fig. \ref{fig:SimHexbinDen} takes the same approach as Fig. \ref{fig:SimHexbin}, but applies it to the $\delta_{10} - M_{\rm BH}$ plane. The importance of demonstrating both sets of plots will become clear when we present the SDSS data, where $\delta$ is found to be the most predictive parameter of quiescence for low mass satellites. Unanimously, the color gradient of central galaxies shows quenching to be positively correlated with black hole mass and density, albeit with a clear preference for black hole mass. The high mass satellites plots show a similar gradient as for centrals. On the other hand, the color gradient for low mass satellites evolves nearly horizontally for all simulations, as with halo mass. Consequently, we again find quiescence to be correlated to an environmental parameter for low mass satellites, and more heavily correlated to an intrinsic parameter for both centrals and high mass satellites. 

\par The growth of dependence in environmental factors for quenching as we descend from centrals to low mass satellites observed Fig.\ref{fig:SimHexbin} is parallel to that of the pie charts in Fig.\ref{fig:Simulations_RandomForest}. This is further shown in Fig.\ref{fig:SIM_QuenchAng}, where we quantify the color gradient as a quenching angle, or quenching vector, introduced in \citet{Bluck_2020a}. To determine the direction of the quenching angle, we make use of partial correlation analysis and the following formula:

\begin{equation}\label{QuenchAngEq}
    \theta_{Q} = \tan^{-1}\left(\frac{\rho_{yz\mid x}}{\rho_{xz\mid y}}\right)   
\end{equation}

\noindent where $x$ is $M_{\rm Halo}$, $y$ is $M_{\rm BH}$, and $z$ is sSFR (specific star formation rate). The numerator ($\rho_{yz\mid x}$) indicates the correlation of sSFR with $M_{\rm BH}$, at fixed $M_{\rm Halo}$; whereas the denominator ($\rho_{xz\mid y}$) represents the correlation of sSFR with $M_{\rm Halo}$, at fixed $M_{\rm BH}$. The angle, $\theta_{Q}$, therefore quantifies the most effective route through the 2D plane to lower sSFR (and therefore induce quenching). Should the quenching vector be along the vertical, $\theta_{Q} = 90\degree$, then the $y$ parameter dominates. Conversely, if the quenching vector is along the horizontal, $\theta_{Q} = 0\degree$, quenching depends solely upon the $x$ parameter. As such, in our case, the vertical projection of the quenching vector is proportional to dependence upon black hole mass, while its horizontal projection is proportional to dependence upon halo mass. A similar analysis is performed later for local over-density, where one need only switch $\delta$ for $M_{\rm Halo}$ in the above equation to get the new quenching angle definition.

\par In Fig. \ref{fig:SIM_QuenchAng} we represent the quenching vectors as arrows, displayed alongside their associated $\theta_{Q}$ values. The upper row of panels corresponds to the $M_{\rm Halo} - M_{\rm BH}$ plane, while the lower row presents the results for the $\delta_{10} - M_{\rm BH}$ plane. The plots follow the same order as the columns in Fig. \ref{fig:Simulations_RandomForest} and Fig. \ref{fig:SimHexbin}. 

\par Focusing first on the $M_{\rm Halo} - M_{\rm BH}$ plane (Fig. \ref{fig:Sim_Quench_Halo}), in all simulations central galaxies have angles exceeding 90\degree: 109\degree, 118\degree, and 125\degree \,\, for the EAGLE, TNG and Illustris simulations (respectively), indicating that quenching depends on increasing $\mathrm{M_{\rm BH}}$ and (more weakly) decreasing halo mass. Looking now at the quenching angles for high mass satellites, a rotation is noticeable, as the angles decrease to 53\degree, 58\degree, and 66\degree \,\, (in the same order as before). High mass satellites show a shared dependence on black hole mass and halo mass, albeit leaning significantly more towards $\mathrm{M_{\rm BH}}$, as indicated by all quenching angles having $\theta_{Q} > 45\degree$. For high mass satellites halo mass aids quenching (presumably through environmental routes), whereas in centrals increasing halo mass actually resists quenching (due to increasing the mass, temperature and density of the CGM, which all lead to accelerated cooling). Finally, low mass satellites have essentially no correlation with black hole mass for quenching, depending almost exclusively on halo mass. The quenching vectors are nearly horizontal at -8\degree, 2\degree, and 11\degree \,\, (in the same order as above), exemplifying a clear dominant correlation with halo mass. 

\par Now focusing on the $\delta_{10} - M_{\rm BH}$ plane (Fig. \ref{fig:Sim_Quenc_Den}), the quenching angles of central galaxies show a dominant correlation with black hole mass, pointing along 75\degree, 71\degree, and 65\degree, respectively. As with the $M_{\rm Halo} - M_{\rm BH}$ plane, the quenching vectors of the high mass satellites for all simulations begin to pivot (slightly) toward the horizontal. But the change in quenching angle between centrals and high mass satellites in the $\delta_{10} - M_{\rm BH}$ plane is not as severe as in Fig. \ref{fig:Sim_Quench_Halo}, and is in fact remarkably similar for TNG and Illustris. The quenching angles of low mass satellites approach 0\degree: -8\degree, 3\degree, and 11\degree \,\, (respectively), displaying a very strong correlation with galaxy over-density, and essentially no intrinsic dependence on black hole mass.

\par Viewing the change in the quenching vectors from centrals to low mass satellites, a complete 90\degree \,\, rotation is apparent. The driving factors for predicting quiescence change systematically from intrinsic properties in centrals to environmental effects in low mass satellites. High mass satellites, however, seem to be under the influence of both effects, while experiencing a greater dependence on intrinsic properties than environment. Thus, high mass satellites are more like centrals than low mass satellites in terms of their quenching. This result is consistent with our findings from the RF classification in Fig. \ref{fig:Simulations_RandomForest}, and quantifies what we have inferred visually based upon the color gradients in Fig. \ref{fig:SimHexbin}.

\begin{figure*}

	\includegraphics[width=\textwidth]{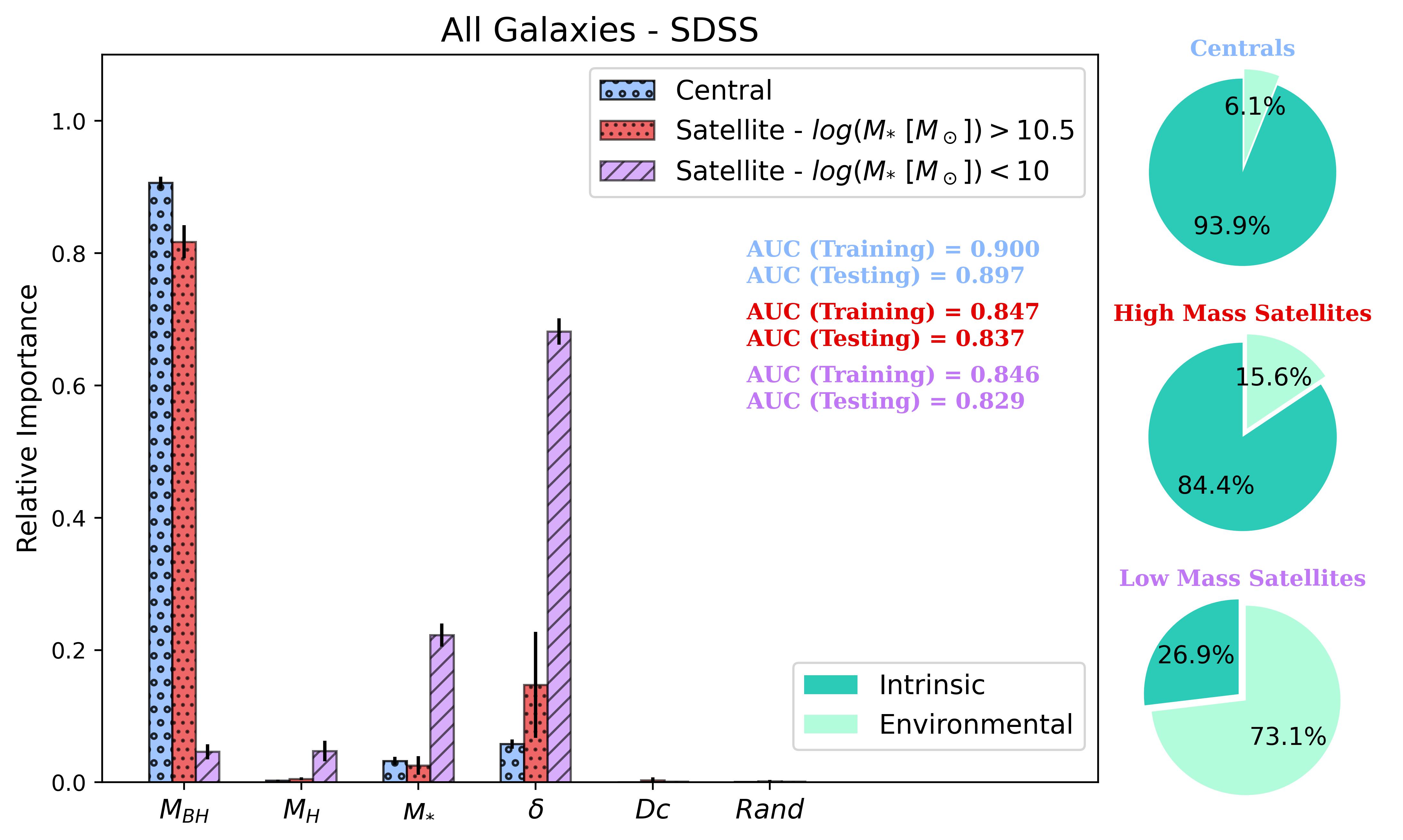}
    \vspace*{-3mm}
    \caption{Random forest quenching classification analyses for the SDSS. Classifications are performed separately for each galaxy class (centrals, high mass satellites, and low mass satellites). The parameters used for training the random forest are listed along the $x$-axis, with their relative quenching importances indicated by the $y$-axis bar heights. Uncertainties are given by the variance of 10 independent classification runs. The pie charts on the right of the main panel indicate the total importance of intrinsic and environmental parameters for each galaxy class. Intrinsic parameters clearly dominate the classification predictions for both centrals and high mass satellites, with $M_{\rm BH}$ (here estimated from the $M_{\rm BH} - \sigma$ relationship) being the clear best parameter. This is in complete agreement with the unanimous predictions from simulations in Section \ref{sssec: SimResults}. Yet, whilst low mass satellites are clearly quenched through environmental effects (as predicted by simulations), the dominant observable parameter is local over-density ($\delta$, indicating the sum of importances for $\delta_3, \delta_5, \delta_{10}$) {\it not} halo mass, which was predicted by the simulations. We also note a significant secondary dependence of low mass satellite galaxies on stellar mass, which is also not predicted by simulations. Hence, this figure represents excellent agreement with simulations for centrals and high mass satellites, but only partial agreement for low mass satellites. These results also hold for a black hole mass estimation using the H{\"a}ring \& Rix black hole mass - bulge mass relation, defined in \citet{Haring_Rix}.}
    \label{fig:SDSS_RandomForest}
\end{figure*}

\subsection{SDSS Results}

\subsubsection{Random Forest Classification}

\begin{figure*}

    \begin{subfigure}{\textwidth}
        \includegraphics[width=\textwidth]{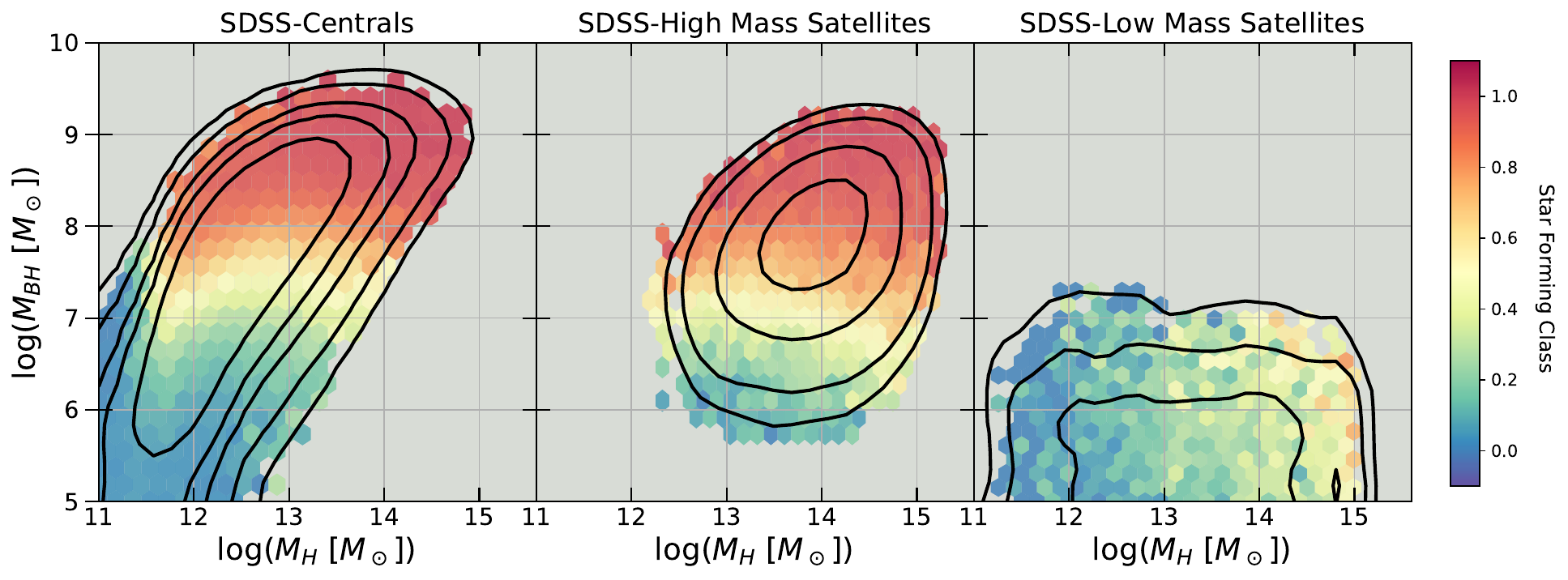}
            \caption{}
            \label{fig:SDSS_HaloHexbin}
    \end{subfigure}
    \begin{subfigure}{\textwidth}
        \includegraphics[width=\textwidth]{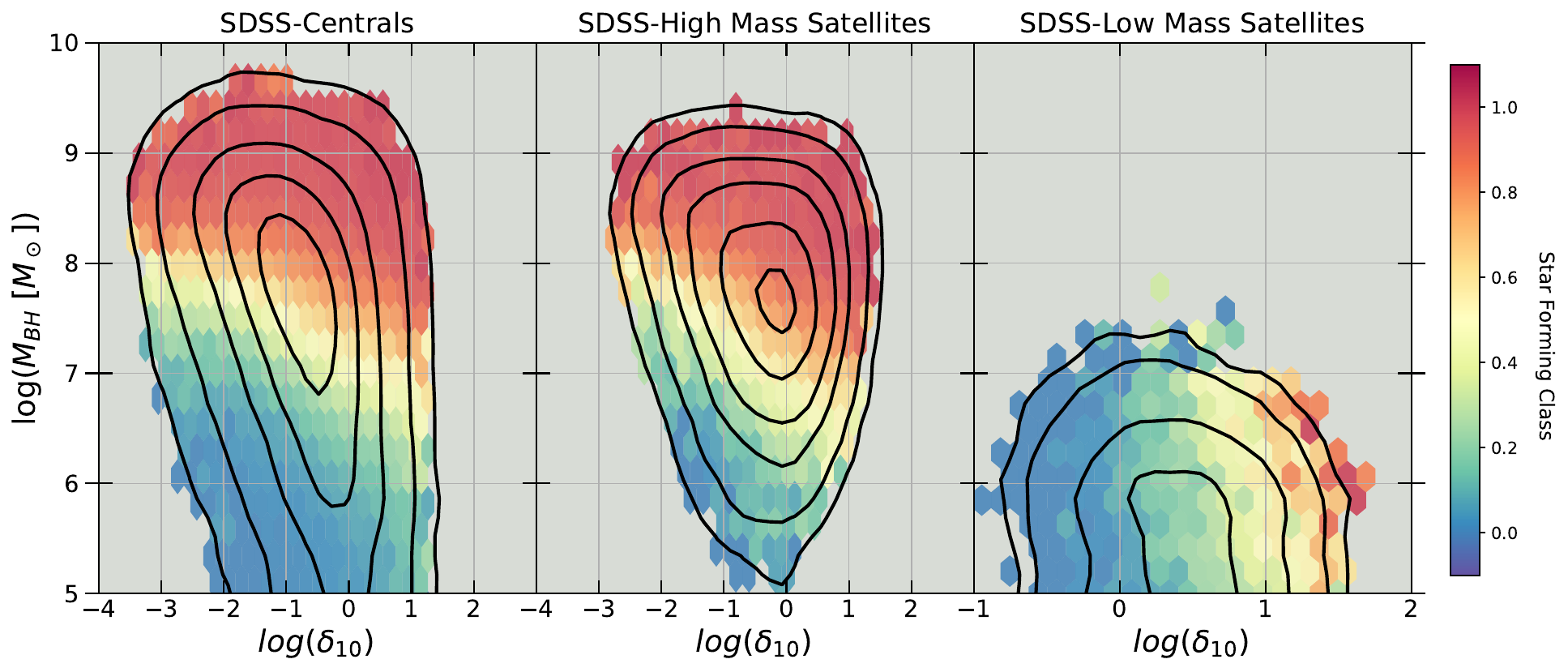}
            \caption{}
            \label{fig:SDSS_HexbinDen}
    \end{subfigure}
    \vspace*{-3mm}
    \caption{{\it Top row:} The halo mass - black hole mass relationship for SDSS galaxies, color coded by the average star forming class in each bin. Different classes of galaxies are presented from left to right. {\it Bottom row:} the $\delta_{10}$ - black hole mass relations for the SDSS data, color coded and separated into groups as above. Clearly, quenching progresses primarily as a function of black hole mass for centrals and high-mass satellites. Yet, for low-mass satellites quenching proceeds as a strong function of environment.}
    \label{fig:SDSS_Hexbin}

\end{figure*}

\par As was done previously for each of the simulation suites, we now execute random forest classification analyses for the SDSS. In Fig. \ref{fig:SDSS_RandomForest} we present the results of these machine learning runs. The bars are color coded to differentiate between the different galaxy types, with central galaxies in blue, high-mass satellites in red, and low-mass satellites in violet. The heights of each bar indicate the relative quenching importance for each parameter, as labelled by the $x$-axis. Uncertainties are given from the variance across 10 independent training and testing runs. The pie charts on the right of the plot show the total relative importance for intrinsic and environmental parameters. This plot also displays the AUC (the area under the ROC curve) of the training and testing sample, color coded to the pertinent category.

\par Identically to the classification analyses for the cosmological simulations, black hole mass is revealed to be by far the most important parameter for determining the quenching of central galaxies. This is also in agreement with previous work done by our group in \citet{Piotrowska_2022, Bluck_2016, Bluck_2020a, Bluck_2020b, Bluck_2022, Bluck_2023}. The pie chart for centrals displays intrinsic processes to be the dominant mechanism to engender quiescence, with essentially no importance given to environment.  

\par The high-mass satellite population appears extremely similar to that of central galaxies, with $M_{\rm BH}$ as the crucial quenching predictor. However, we find a weak secondary parameter, $\delta$, has climbed in importance. Environmental parameters in high-mass satellites account for a larger slice of the quenching pie than for centrals, very similar to the random forest classification for the simulations.

\par Focusing on the low mass satellite results, black hole mass has become entirely irrelevant. Here, $\delta$ is unambiguously found to be the strongest predictive feature, although importance to intrinsic attributes remains, with stellar mass accounting for nearly 25\% of the total importance. This clearly indicates a dominant environmental quenching mechanism for low mass satellites. While the pie charts of simulations and observational data for low-mass satellites agree (both are environment dominated), the classification analysis reveals a conflict -- the SDSS classification reveals $\delta$ to be most predictive environmental parameter, whereas the simulations all agree that $M_{\rm Halo}$ ought to be the most predictive. Hence, there is excellent agreement between observations and simulations for centrals and high-mass satellites, but only partial agreement for low-mass satellites.

\begin{figure*}
    \centering
    \begin{subfigure}{0.45\textwidth}
        \includegraphics[width = \textwidth]{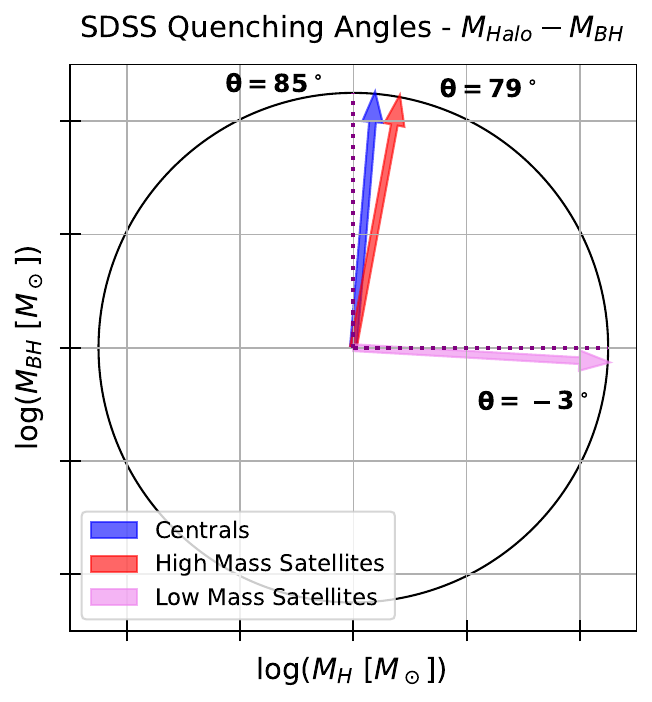}  
        \caption{}
        \label{fig:SDSSQuenchAngHalo}
    \end{subfigure}
    \hfill
    \centering
    \begin{subfigure}{0.45\textwidth}
        \includegraphics[width = \textwidth]{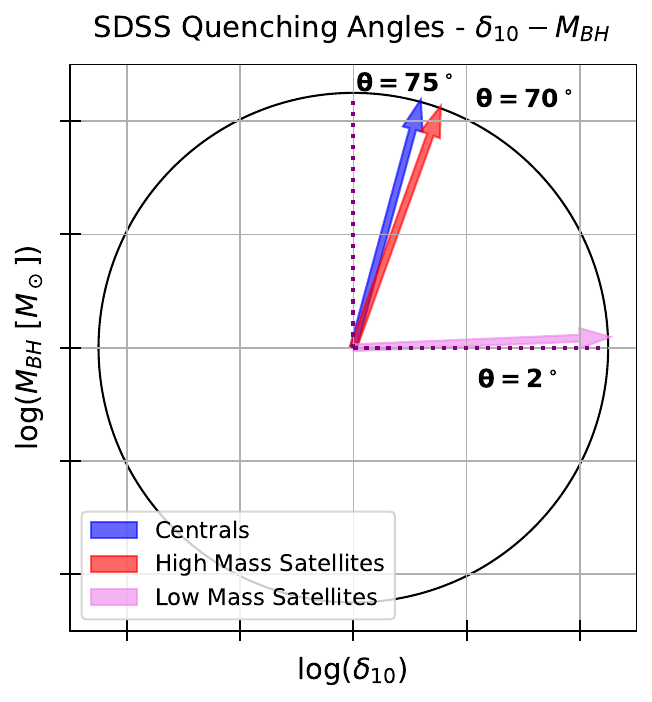} 
        \caption{} 
        \label{fig:SDSSQuenchAngDen}
    \end{subfigure}
    \vspace*{-3mm}
    \caption{The quenching angles for the SDSS. In the left panel (a), we present quenching angles for centrals, high-mass satellites, and low mass satellites in the $M_{\rm Halo} - M_{\rm BH}$ plane. As with simulations, there is clear rotation from intrinsic quenching dependence to environmental quenching dependence as we move through the galaxy classes. In the right panel (b), we present the same information as for (a) but now in the $\delta_{10}$ - $M_{\rm BH}$ plane. We find the quenching vectors for central and high-mass satellite galaxies to be extremely similar, with clear black hole mass dependence. However, for low-mass satellites, sSFR best correlates with $\delta_{10}$, favoring environmental factors. This confirms the general trend seen in the Random Forest classifications above.}\label{fig:SDSS_QuenchingAngles}
\end{figure*}

\subsubsection{The Quenching Angle}

\par Fig. \ref{fig:SDSS_HaloHexbin} shows the $M_{\rm Halo} - M_{\rm BH}$ plane for galaxies in the SDSS. The left panel corresponds to centrals, the middle to high-mass satellites, and the right panel to low-mass satellites. They are color coded by the average star forming class in each hexagonal bin. The central and high mass satellite planes show similar gradients from star forming to quenched, with both nearly vertical (indicating pure black hole mass dependence). This is in stark opposition to low mass satellites, which show a nearly horizontal gradient, indicating halo mass as the dominant feature (out of these two parameters). 

\par In Fig. \ref{fig:SDSS_HexbinDen} we repeat the method applied for Fig. \ref{fig:SDSS_HaloHexbin}, but now in the $\delta_{10} - M_{\rm BH}$ plane. The color progression for central galaxies reveals quiescence as being primarily correlated to black hole mass, albeit retaining a lesser, positive, correlation to local galaxy over-density. We observe this again for high mass satellites, although the correlation to over-density has visibly strengthened. On the other hand, the color progression of the low mass satellite plots is most correlated to over-density, while black hole mass is reduced to a non-factor. For low-mass satellites, increasing galaxy over-density leads to an increased likelihood for the galaxy to be quenched.

\par In both the $M_{\rm Halo} - M_{\rm BH}$ plane and the $\delta_{10} - M_{\rm BH}$ plane, it is clear that whilst quiescence is most strongly correlated to intrinsic parameters in centrals and high-mass satellites, it is most strongly correlated to environmental features for low-mass satellites. The switch from intrinsic to environmental is completely consistent with the pie charts in Fig. \ref{fig:SDSS_RandomForest}, and is exactly as predicted by simulations in Section 4.1. This effect is even clearer when illustrated in the quenching angle plots, presented in Fig. \ref{fig:SDSS_QuenchingAngles}, where the left panel corresponds to the $M_{\rm Halo} - M_{\rm BH}$ plane, and the right panel to the $\delta_{10} - M_{\rm BH}$ plane.

\par Focusing first on the $M_{\rm Halo} - M_{\rm BH}$ plane, the quenching vectors of centrals is nearly vertical, at 85\degree, the high-mass satellite vector is slightly less so at 79\degree, and finally the quenching angle rotates to -3\degree \,\, for low-mass satellites. These results indicate that quenching is strongly correlated to $M_{\rm BH}$ for centrals and high mass satellites, and $M_{\rm Halo}$ in the case of low mass satellites. Except for the quenching angle for centrals not crossing the 90\degree \, threshold, this plot presents a consistent narrative as for the simulations. As we move from centrals, to high mass satellites, and finally low mass satellites, a clear rotation from intrinsic to environment quenching dependence occurs. We note that whilst a similar analysis performed in \citet{Bluck_2022} for central galaxies finds an anti-correlation with halo mass, this is due to the use of $\Delta$sSFR as its `z' value when determining the quenching angles rather than sSFR. This result in fact further agrees with those found for simulations in this study.

\par We now hone in on the right panel of Fig. \ref{fig:SDSS_QuenchingAngles} to study quiescence in the $\delta_{10} - M_{\rm BH}$ plane. The quenching angle of central galaxies, 75\degree, represent a strong correlation to $M_{\rm BH}$, and a weaker correlation  with $\delta_{10}$. We find a very similar result in high-mass satellites, following a slight pivot towards the horizontal. Finally, the quenching vector of low-mass satellites points along a 2\degree \,\, angle, demonstrating a strong correlation with density, independent of black hole mass. 

\par When viewing both panels of quenching angles in Fig. \ref{fig:SDSS_QuenchingAngles}, we find a clear trend in the quenching of different galaxy classes. The quiescence of centrals is strongly correlated to intrinsic parameters, specifically black hole mass. The specific star formation rate of high-mass satellites is best correlated with black hole mass, but to a lesser extent is also correlated with environmental parameters. Finally, we find the quenching of low-mass satellites to be nearly solely correlated to environmental features, whether halo mass or local density. These general results are precisely as predicted by contemporary simulations in Section \ref{sssec: SimResults}. However, the main discrepancy remains from the random forest analysis: halo mass is the best environmental quenching parameter in simulations, but local galaxy over-density is the best parameter in observations.

\begin{figure*}
    \centering
    \begin{subfigure}{\textwidth}
        \includegraphics[width = \textwidth]{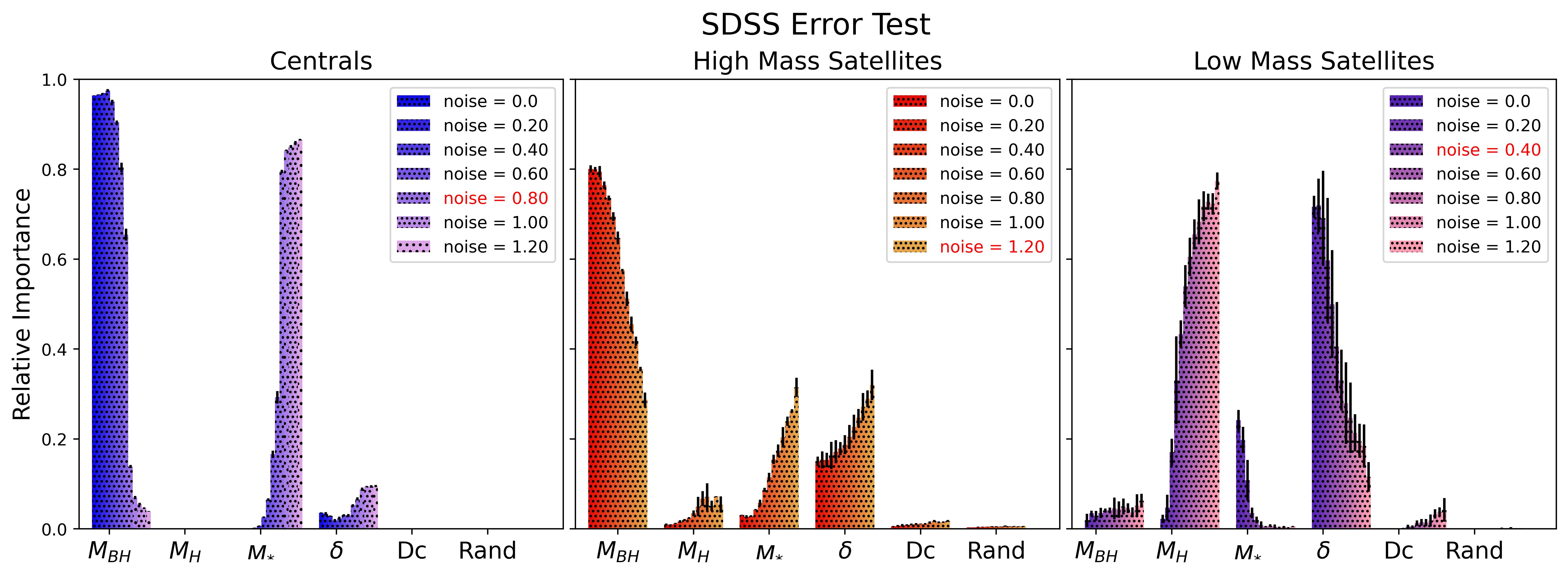}  
        \caption{}
        \label{fig:ErrRF}
    \end{subfigure}

    \begin{subfigure}{\textwidth}
        \includegraphics[width= \textwidth]{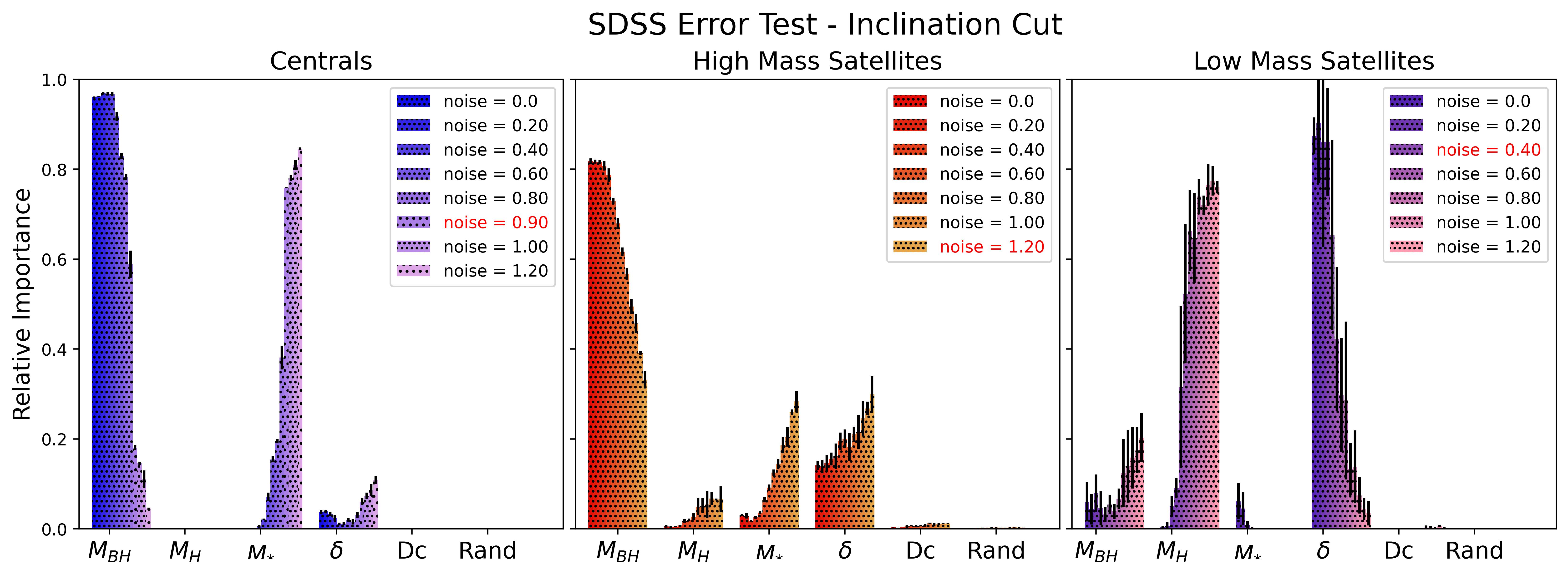} 
        \caption{}
        \label{fig:ErrCutRF}
    \end{subfigure}
    \caption{Differential measurement uncertainty tests. The top row of panels shows an error test on the random forest classification through the gradual application of noise to the dominant parameter of each class (and to the secondary parameter $M_{*}$ for low mass satellites). Centrals and high-mass satellites can withstand a large amount of noise to black hole mass, while the low-mass satellites exhibit a weaker resistance to noise added to local density. The bottom row of panels shows the same error test as above, but here applying an axial ratio cut to galaxies. Centrals and high mass satellites can now withstand an even larger amount of noise, while the low mass satellites still exhibit weak resistance. The threshold noise level at which the dominant parameter of each class yields to the second most valuable parameter is highlighted in the legends in red. This clearly demonstrates the sensitivity to noise for low-mass satellites, compared to the relative insensitivity for centrals and high-mass satellites.}
\label{fig:ErrorTest}
\end{figure*}

There are two possibilities to explain this discrepancy. First, the results could be real in the sense that they point to a difference in the mechanisms of environmental quenching in the simulations compared to the observed Universe. If this is the case, the move from halo mass to local over-density points towards a greater importance of satellite - satellite interactions in quenching low mass systems in nature, and a weaker dependence on halo properties, compared to the unanimous predictions from three contemporary cosmological simulations. Furthermore, this likely suggests a greater importance of dynamical stripping of satellites from interactions (which must correlate primarily with local density of galaxies) in nature over ram pressure stripping from halo crossing (which must correlate primarily with halo mass, with a secondary dependence on location within the halo). 

However, unlike in simulations, in the SDSS data there are uncertainties on the measured parameters. In \citet{Bluck_2022} we find that global uncertainty (applied equally to all parameters) does not strongly impact the RF classification results. However, differential uncertainty can impact the classification importances, potentially leading to erroneous conclusions. Hence, there is a possibility that the difference in the dominant environmental quenching parameter in the SDSS compared to the simulations could be a result of differential uncertainty in the SDSS parameter measurements. We explore this possibility in detail in the sect sub-section.

\subsubsection{Differential Uncertainty Tests}

\par To test how observational uncertainty could affect the Random Forest classification results in the SDSS, we re-execute RF classifications whilst incrementally adding Gaussian random noise to the dominant parameter of each class. This tests how stable the dominant quenching parameter for each galaxy class is to uncertainty. Specifically, since noise is added only to the dominant parameter, this tests the impact of {\it differential} measurement uncertainty on the RF results. We can the find the critical point at which noise changes the random forest rankings, and compare this with the known (or estimated) differential uncertainty on the parameters to establish whether measurement error could in principle lead to erroneous conclusions in the machine learning. This methodology is explained in detail in \citet{Brownson_2022} and in \citet{Piotrowska_2022}.

In Fig. \ref{fig:ErrorTest} we present these differential uncertainty experiments with and without an inclination cut (Fig. \ref{fig:ErrRF} and Fig. \ref{fig:ErrCutRF}, respectively). In the main results above we have not applied an inclination cut, favoring completeness over accuracy. But for the most accurate estimation of black hole masses through the $M_{\rm BH} - \sigma$ relation, one must remove the impact of differential disk rotation by requiring disk-like galaxies to present face-on \citep[see][for a discussion]{Bluck_2016}. By comparing both approaches, we establish that our results for centrals and high mass satellites are the same whether we opt to maximize completion or accuracy in the black hole measurements.

\par In Fig. \ref{fig:ErrRF} we systematically add random noise to the dominant quenching parameter for centrals (left panel), high-mass satellites (middle panel), and low-mass satellites (right panel). The dominant parameter for central and high-mass satellites, $M_{\rm BH}$, can handle a large amount of noise before a new feature gains the top spot. Explicitly, this occurs at 0.8 dex (centrals) and 1.2 dex (satellites). For centrals, stellar mass slowly rises in importance as noise is added to black hole mass, until it overtakes $M_{\rm BH}$ completely at the limit. For high-mass satellites, stellar mass and density both increase in importance as noise is added to black hole mass. 

The uncertainties on black hole mass are estimated to be $\sim$0.5 dex, accounting for uncertainty in $\sigma_*$ and in the scatter of the $M_{\rm BH} - \sigma_*$ relation \citep[see][]{Bluck_2022}. On the other hand, the typical uncertainty on stellar mass is lower, at $\sim$0.3 dex, accounting for uncertainty in the initial mass function (IMF), SED code, and stellar libraries \citep[see][]{Mendel_2014}. Hence, black hole masses are measured less accurately than stellar masses, yet the former clearly outperform the latter. Therefore, no amount of differential uncertainty between black hole mass and stellar mass could explain the observational results. From the error analysis, we see that the uncertainties in stellar mass would have to underestimated by $\sim$0.8 dex (yielding a total error of $\sim$1.2dex) in order for these results to be spurious. We note that this is four times that which is quoted conservatively in the literature. Moreover, the results for centrals and high-mass satellites are in precise accord with the predictions from three hydrodynamical simulations, further indicating reliability in these results. Hence, we conclude that differential measurement uncertainty cannot explain the importance of black hole mass for quenching centrals and high-mass satellites in observations \citep[see][for a similar result for centrals alone]{Piotrowska_2022}.

For low-mass satellites, the importance of the dominant parameter (local galaxy over-density) drops a lot faster with added noise than black hole mass does for either of the other two groups. At 0.4 dex of noise, halo mass begins to be as important as over-density, and the former dominates entirely at 0.5 dex of differential measurement error. This is a really interesting cutoff because this is the approximate uncertainty on the halo masses \citep[see][]{Yang_2007, Yang2009}. Since the halo masses are inferred from abundance matching, the primary uncertainty arises from scatter in the $M_{\rm Halo} - M_*$ relation. Other sources of error arise from the stellar mass estimates of the group members, and the group membership identification. For local densities, our only main source of error are from redshifts and detection limits. However, by normalizing the densities to over-densities the latter issue is largely mitigated. Consequently, we estimate that the differential measurement uncertainty between local over-density and halo mass is $\sim0.4$ dex, i.e. almost precisely the amount needed to give an erroneous result. Indeed, if the halo mass uncertainty is only slightly underestimated, one would expect to see local density as a better parameter than halo mass (provided both are equally good tracers of environmental quenching when measured perfectly). The above results suggest that the differential measurement uncertainty could be the reason for the discrepancy between simulations and observations for low mass satellites.

\begin{figure*}
    \centering
    \begin{subfigure}{.49\textwidth}
    \centering
        \includegraphics[width = \textwidth]{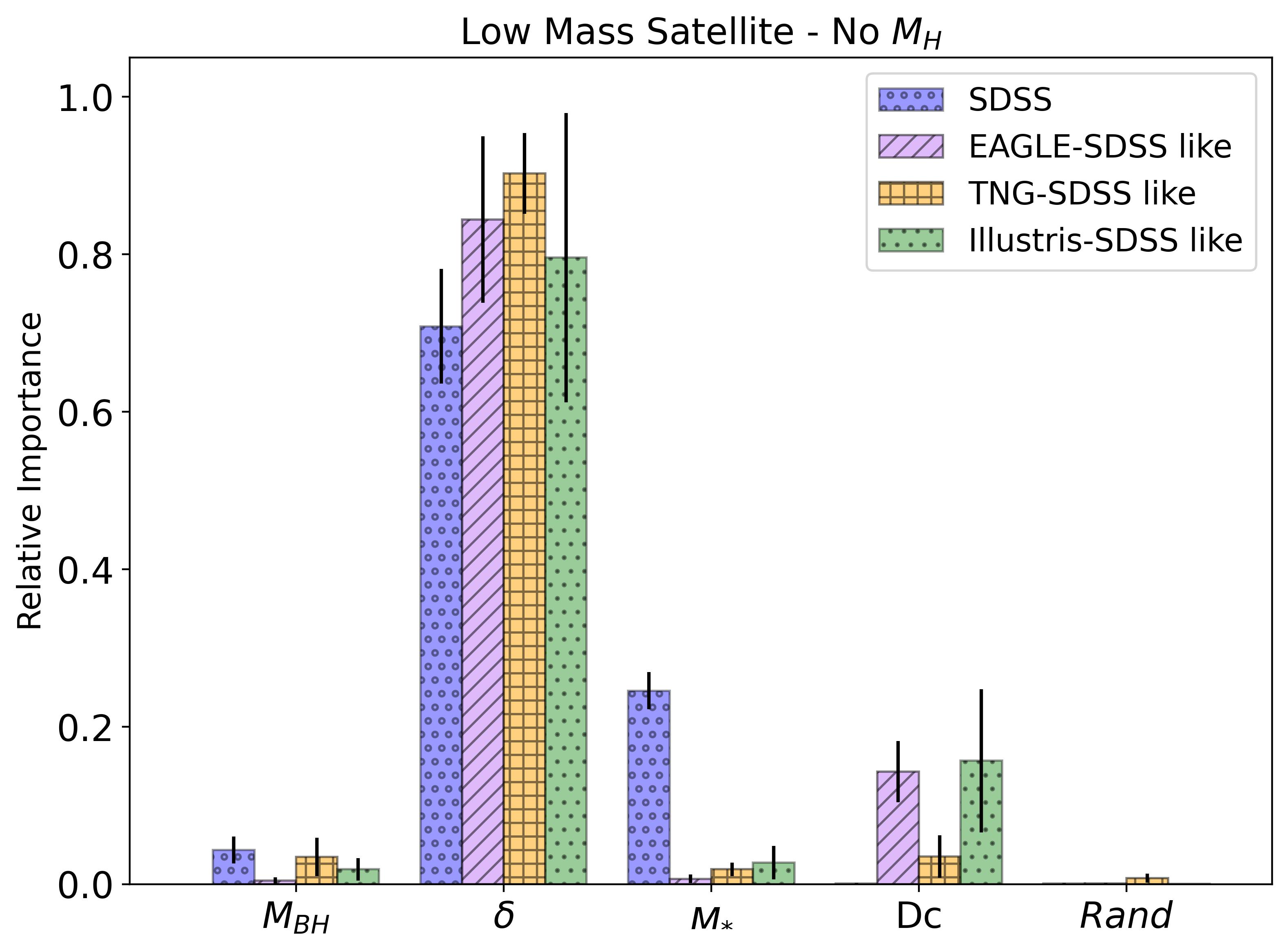} 
        \caption{}
        \label{SIM_RF_NOHalo}
    \end{subfigure}
    \hfill
    \begin{subfigure}{.49\textwidth}
    \centering
        \includegraphics[width= \textwidth]{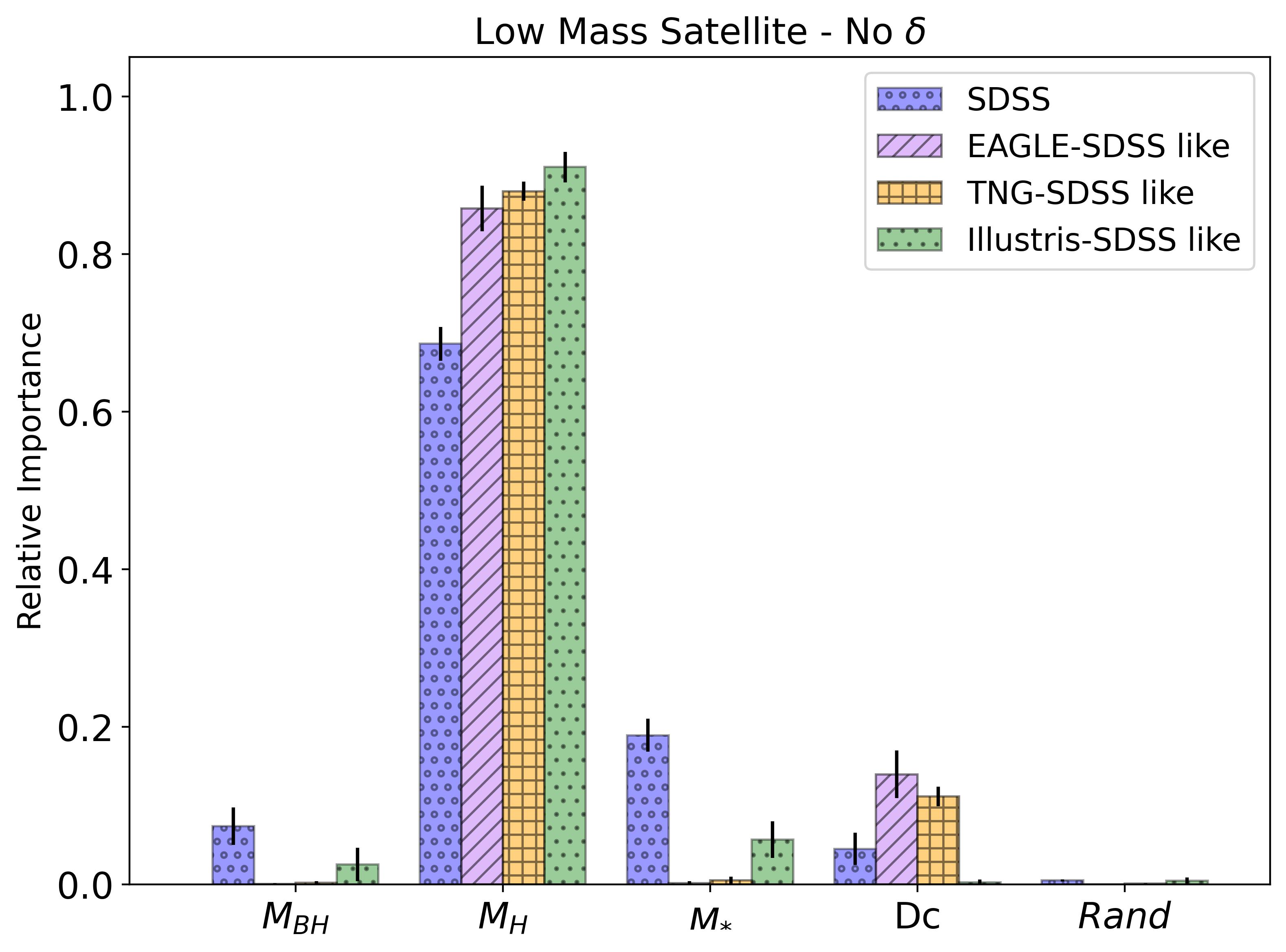} 
        \caption{}
        \label{SDSS_RF_NODEN}
    \end{subfigure}

\caption{Random forest classification analysis for low-mass satellites in the SDSS and simulations after removing the dominant parameter of each. In the left-hand panel (a), we remove halo mass as an input feature in the simulations, which results in local density becoming the best parameter to predict quiescence in simulations (as in observations in the full analysis). In the right-hand panel (b), we remove local density from the features, which results in halo mass becoming the most predictive parameter for observations (as in simulations in the full analysis). Hence, the simulations and observations agree on the top two parameters for predicting quiescence in low-mass galaxies, just not their order.}\label{fig:SDSS_SIM_ParamRemoved}
\end{figure*}

\par In Fig. \ref{fig:ErrCutRF} we apply inclination cuts to disky galaxies (with $n_S < 2$), restricting systems to presenting face-on. This cut allows central galaxies to resist even more artificial noise. $M_{\rm BH}$ withstands up to 0.9 dex of noise before stellar mass asserts itself as the most important feature. This same effect can be seen for high-mass satellites to a lesser degree. Low-mass satellites, however, still show a poor degree of resistance. Over-density loses out to halo mass after only about 0.4 dex, though the switch is slightly more gradual than it had been when no inclination cut was executed.

\par Whether the inclination cut is applied or not, the conclusion is essentially identical. Black hole mass remains the dominant quenching parameter for central and high-mass satellite galaxies, even after a large amount of noise is added. This strongly suggests that these results cannot reasonably be attributed to differential measurement uncertainty \citep[as also concluded for centrals in][]{Piotrowska_2022}. On the other hand, the most predictive parameter for quenching in low-mass satellites switches from local over-density to halo mass relatively quickly. It is therefore feasible that the dominance of $\delta$ may be due to an artifact of high error on the halo masses \citep[see][]{Yang_2007, Yang_2009, Behroozi_2010, Woo_2013}, rather than being inherent to the parameter itself.

\subsection{Halo Mass or Local Galaxy Density?}

To better understand the difference in the most predictive parameter for quenching of low-mass satellites in simulations and observations, we perform a random forest classification test after removing the dominant feature for each. This allows us to look for the second best quenching parameter for both simulations and observations. The results of these two experiments are shown in Fig. \ref{fig:SDSS_SIM_ParamRemoved}. We present this test in two panels: the first (left panel) is for classifications without halo mass as a training parameter; while the second (right panel) corresponds to a test without local galaxy over-density as a parameter. 

In the first panel, the most predictive parameter for both observed and simulated low mass satellites is clearly found to be $\delta$. That is, in lieu of halo mass, local over-density becomes the most important parameter in simulations, just as is observed in observations. In the second panel, we see that in the absence of local over-density, halo mass becomes the most important quenching parameter for low-mass satellites in observations, just as in simulations. Taken together, these tests show that the discrepancy between simulations and observations for low mass satellites is not as severe as it may have appeared initially. Simulations and observations agree on the top two parameters, they just disagree about their order of importance. This further suggests that the differential uncertainty explanation for the discrepancy (discussed above) is highly plausible.

\begin{figure}
    \includegraphics[width=0.49\textwidth]{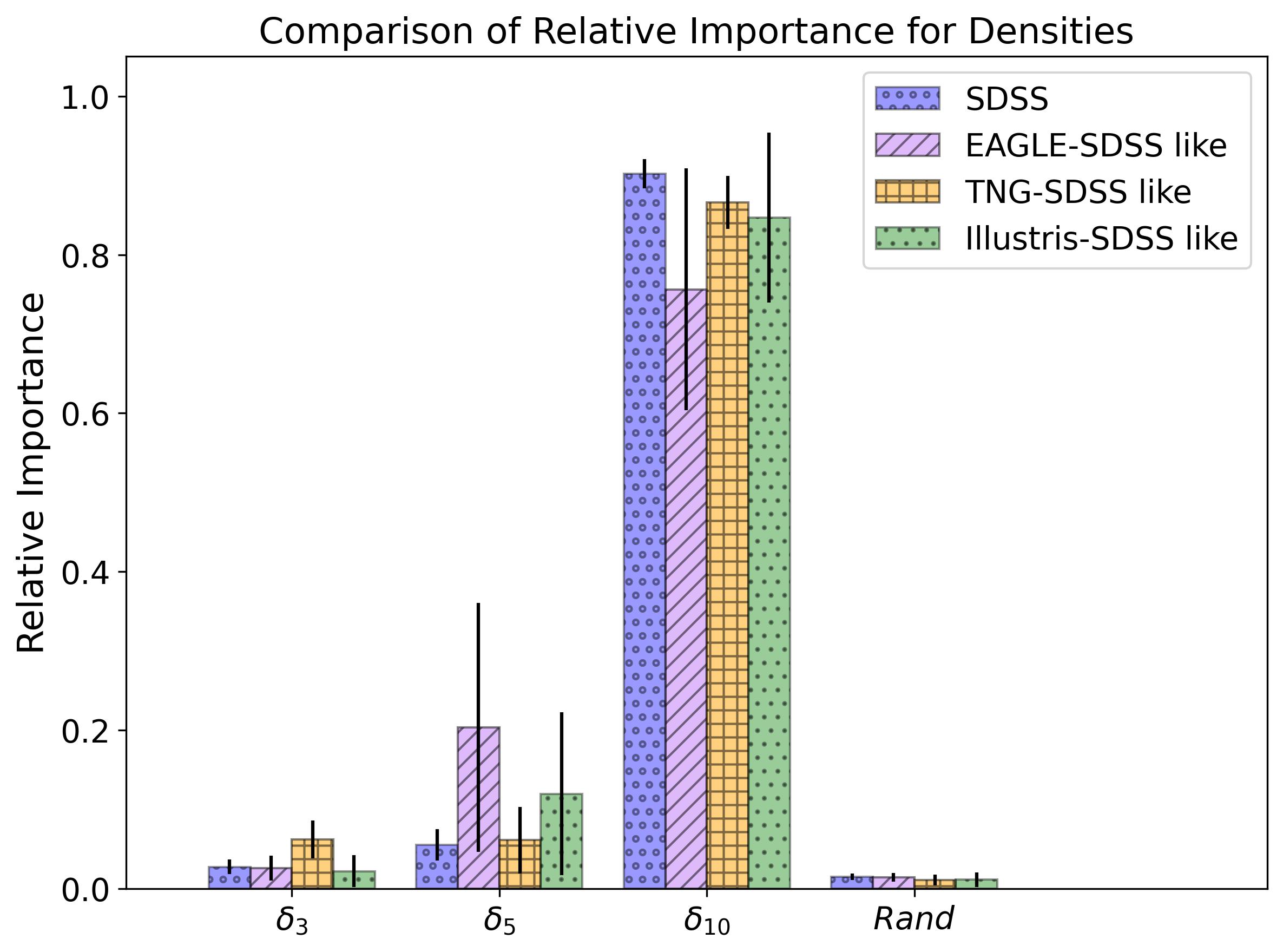}
    \vspace*{-3mm}
    \caption{Random Forest quenching classification for low-mass satellites in the SDSS and the simulations, comparing the different $\delta_{n}$ values (i.e., for $n$ = 3, 5, 10). $\delta_{10}$ is unanimously found to be the most important density scale for predicting quenching in this population of galaxies.}\label{fig:DensityComp}
\end{figure}

In all classification analyses so far, we have presented only the sum of importances for local galaxy over-density. However, important information is contained in the scale of the individual measurements. For example, $\delta_{3}$ best indicates local effects, e.g. galaxy-galaxy interactions; $\delta_{5}$ is a good tracer of mean halo density; and $\delta_{10}$ reveals information on the mean density of large haloes, as well as super-halo correlations. To explore this further, we compare the predictive power of $\delta_{3}, \delta_{5}, \delta_{10}$ over quenching in simulations and observations through a random forest classification (presented in Fig. \ref{fig:DensityComp}). This is performed for low mass satellites only, where environment becomes important.  

Interestingly, the results from Fig. \ref{fig:DensityComp} are unanimous: $\delta_{10}$ carries the most importance for determining whether a low mass satellite galaxy is star forming or quenched in both observations and simulations. However, very few groups exceed 10 satellite galaxies with $M_* > 10^9 M_\odot$ (see Table. \ref{tab:HaloSize}). Hence, the importance of $\delta_{10}$ could point to either a super-virial mechanism  (e.g., due to some form of pre-processing), or else a link between halo mass and super-virial scales, lost in the abundance matching method.

If the simulations are essentially correct in their modelling of satellite quenching, halo mass is the most important parameter. Differential measurement uncertainty could plausibly yield the observational result of local over-density being the best parameter, given the high uncertainty on halo mass relative to local densities. However, in order for this explanation to hold one more thing must be true: {\it Local density must be a very good proxy for halo mass.} If this were not the case, simply having a high uncertainty on halo mass would not yield importance to local over-density.

This is an important point because if the simulations are correct, local galaxy over-density may offer a more reliable estimator of halo mass than currently achieved via abundance matching. Leveraging the results from Fig. \ref{fig:DensityComp} and Table. \ref{tab:HaloSize}, we see that it is actually the over-density of galaxies on very large scales (mostly well beyond the typical virial radius of groups) which are most effective for predicting quiescence, and hence (if the simulations are correct) for constraining halo mass. Yet, the conventional abundance matching approach considers galaxies only within the virial radius \citep[e.g.][]{Behroozi_2010}).

Galaxy - galaxy correlations persist at distances up to the isotropy scale \citep[i.e., $\sim$100\,Mpc for a $\Lambda$CDM Universe; see, e.g.][]{Henriques_2015}. Hence, it is at least feasible that the information on these large scales is valuable for accurately estimating halo masses. Furthermore, the large spatial scales of the most important over-densities for predicting quenching suggest that satellite - satellite interactions are unlikely to be the reason for the importance of galaxy over-density (since galaxies beyond the 5th nearest neighbour are typically far too distant to have interacted directly). Taken together, these considerations support the possibility that local galaxy over-densities at large scales are an effective alternative to abundance matching for estimating halo masses.

In conclusion to this part of the results, one of the following two options must be true: (i) contemporary cosmological simulations fail to model low-mass satellite quenching accurately; or (ii) local galaxy over-density at the 10th nearest neighbour is a better tracer of the true underlying halo masses in observations than the currently adopted abundance matching approach, at least around the quenching threshold of $M_{\rm Halo} \sim 10^{13} M_{\odot}$ for low-mass satellites \citep[see][]{Yang_2007, Yang_2009}. 

Whilst the first option may well be true, there are several reasons to doubt its veracity. First, simulations do accurately predict the quenching characteristics of centrals and high-mass satellites. Second, halo masses are more poorly constrained than local galaxy over-densities, and hence differential measurement uncertainty may well be important here. Third, the simulations do get the top two parameters right for low mass satellites, and correctly predict the general environmental dominance. Fourth, the simulations all completely agree with each other on the importance of halo mass. Fifth, physical processes important to environmental quenching (e.g., ram pressure, tidal torques, etc.) are all modelled directly in these state of the art simulations. So, it is hard to accept that they would be missing classical environmental quenching mechanisms. There are no subgrid prescriptions to tweak for environment in these simulations, and the hydro and gravity solvers are already proven to be optimised (at least for the well resolved massive galaxies analysed in this work). Of course, resolution effects may still matter. Yet, environmental effects operate at scales of $\sim$ 10\,kpc - 10\,Mpc, which are very well resolved in these simulations, especially at high densities (which matter most for environmental quenching).

So, this leaves us with possibility (ii). In order for local galaxy over-density to appear as the most important quenching parameter for low mass satellites, even when halo mass is the underlying driver, local over-density on large scales must be an excellent predictor of halo mass (especially at the high group/ cluster masses where satellites quench). Indeed, it must be a better estimator of halo mass than the current status quo in abundance matching, especially when establishing a halo mass quenching threshold. This is an exciting possibility as it may point the way to improving the estimation halo masses in wide-field galaxy surveys moving forwards. We plan further work on this possibility in a series of papers (Bluck et al. in prep.; Goubert et al. in prep.).

\begin{table}
    \centering
        \caption{Halos occupation for N\textsubscript{Sat} $\le$ 10 and N\textsubscript{Sat} $\ge$ 10 in simulations and observations. We also include the mean distance to the tenth nearest neighbor for satellite galaxies, $\langle D\textsubscript{10} \rangle$, and the average virial radius, $\langle R_{\rm vir} \rangle$, both in units of Mpc.}
    \begin{tabular}{ccccc}
      \textbf{Halo Size} & \textbf{EAGLE} & \textbf{Illustris} & \textbf{TNG} & \textbf{SDSS} \\
      \hline
       N\textsubscript{Sat} $\le$ 10 & 1308 & 2305 & 1826 &  37680\\
      \hline
       N\textsubscript{Sat} $\le$ 10 & 94.8\% & 94.4\% & 94.6\% & 96.7\%\\
      \hline
       N\textsubscript{Sat} $\ge$ 10 & 72 & 137 & 105 & 1287\\
      \hline
       N\textsubscript{Sat} $\ge$ 10 & 5.2\% & 5.6\% & 5.4\% & 3.3\%\\
      \hline
       $\langle D\textsubscript{10} \rangle$ & 2.00 & 1.74 & 1.58 & 1.54\\
      \hline
       $\langle R_{\rm vir} \rangle$ & 0.174 & 0.143 & 0.166 & 0.487 \\
    \end{tabular}
    \label{tab:HaloSize}
\end{table}

\subsection{Satellite quenching as a function of location within the halo} \label{sssec:HaloLoc}

\begin{figure*}
	\includegraphics[width=\textwidth]{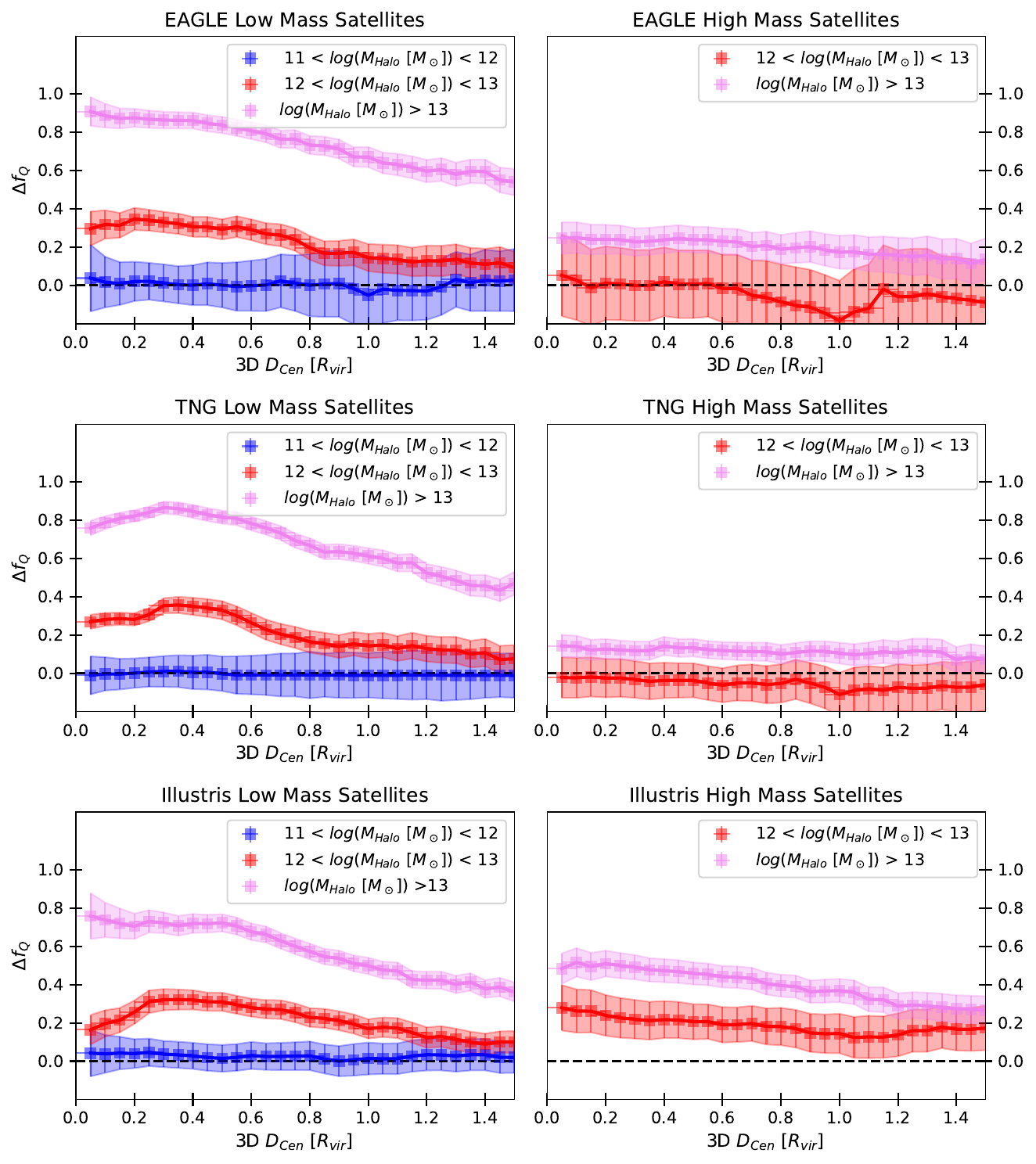}
    \vspace*{-3mm}
    \caption{The quenching of satellites as a function of location with their groups or clusters. We present the delta quenched fraction, $\Delta f_Q$, as a function of the distance to the nearest central galaxy, $D_{\rm cen}$ for the simulations (displayed separately along each row), separated into bins of halo mass (as indicated on the legends). The bins are formed from objects that are within a defined $D_{\rm cen}$ range of 0.15$R_{\rm vir}$, while the shaded regions are the margins of error for each bin, computed via Poison statistics. Low-mass satellites are displayed on the left-hand column, and high-mass satellites are displayed on the right-hand column. $\Delta f_Q$ measures the offset in quenched fraction for each satellite population relative to a control sample of central galaxies at the same black hole mass (which is found previously to be the dominant intrinsic driver of quenching). High-mass satellites exhibit weak environmental dependence at all locations with all masses of haloes. Conversely, low-mass satellites are strongly offset to higher quenched fractions, relative to centrals, at close distances to their centrals and within high mass groups/ clusters.}
    \label{fig:SIM_DelfQ}
\end{figure*}

\begin{figure*}
	\includegraphics[width=\textwidth]{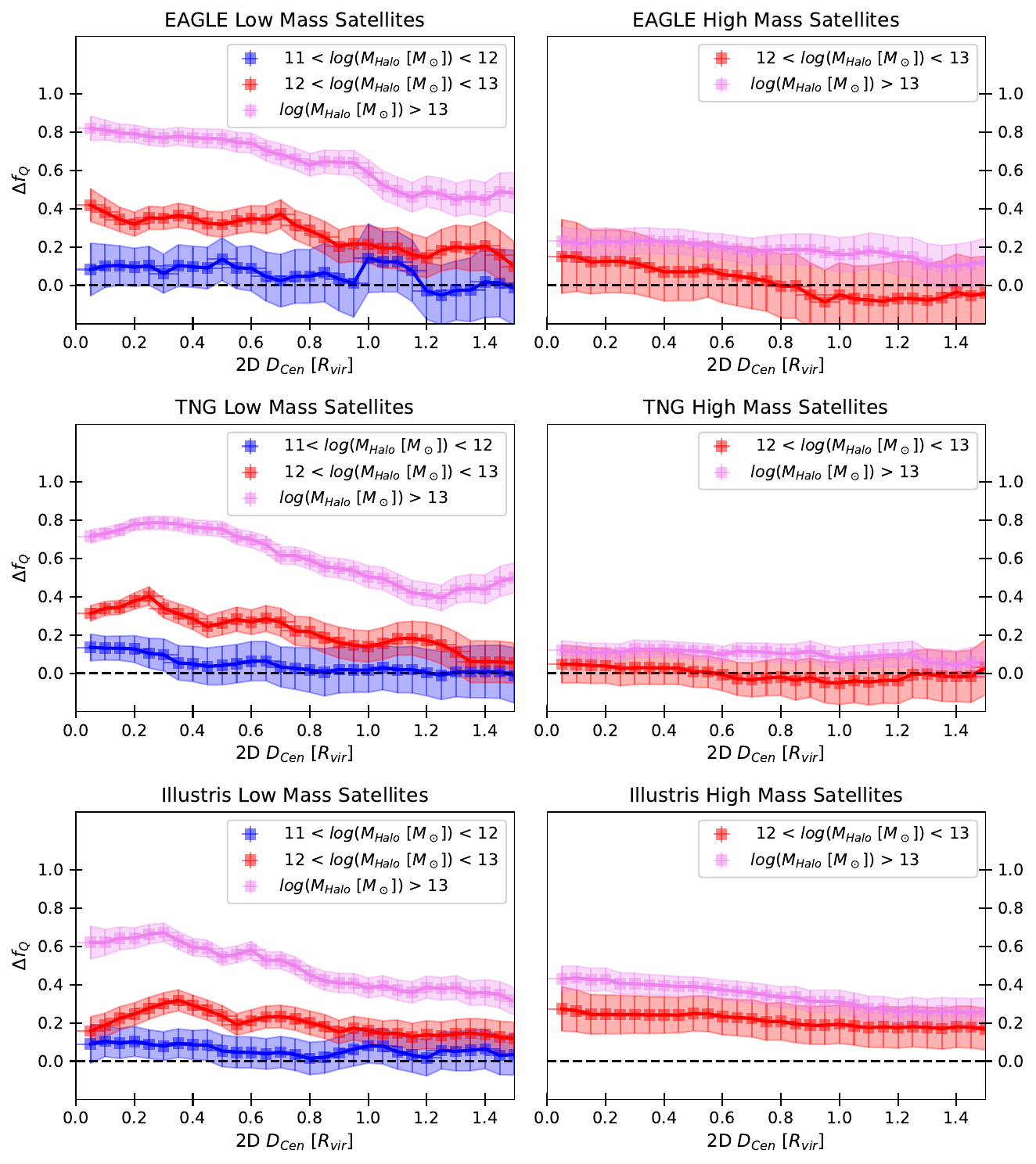}
    \vspace*{-3mm}
    \caption{This figure has an identical structure to Fig. \ref{fig:SIM_DelfQ}. However, here we analyse a random 2D projection of the groups and clusters, and add a 0.5 dex Gaussian random error to $M_{\rm Halo}$, in order to more effectively compare to the SDSS. Note that restriction to 2D, and even significant uncertainties on the halo masses, do not obscure the qualitative trends in simulations. This is very promising for the accurate determination of the role of environment in galactic star formation in contemporary observational data. }
    \label{fig:SIM_DelfQ_2D}
\end{figure*}

\begin{figure*}
	\includegraphics[width = \textwidth]{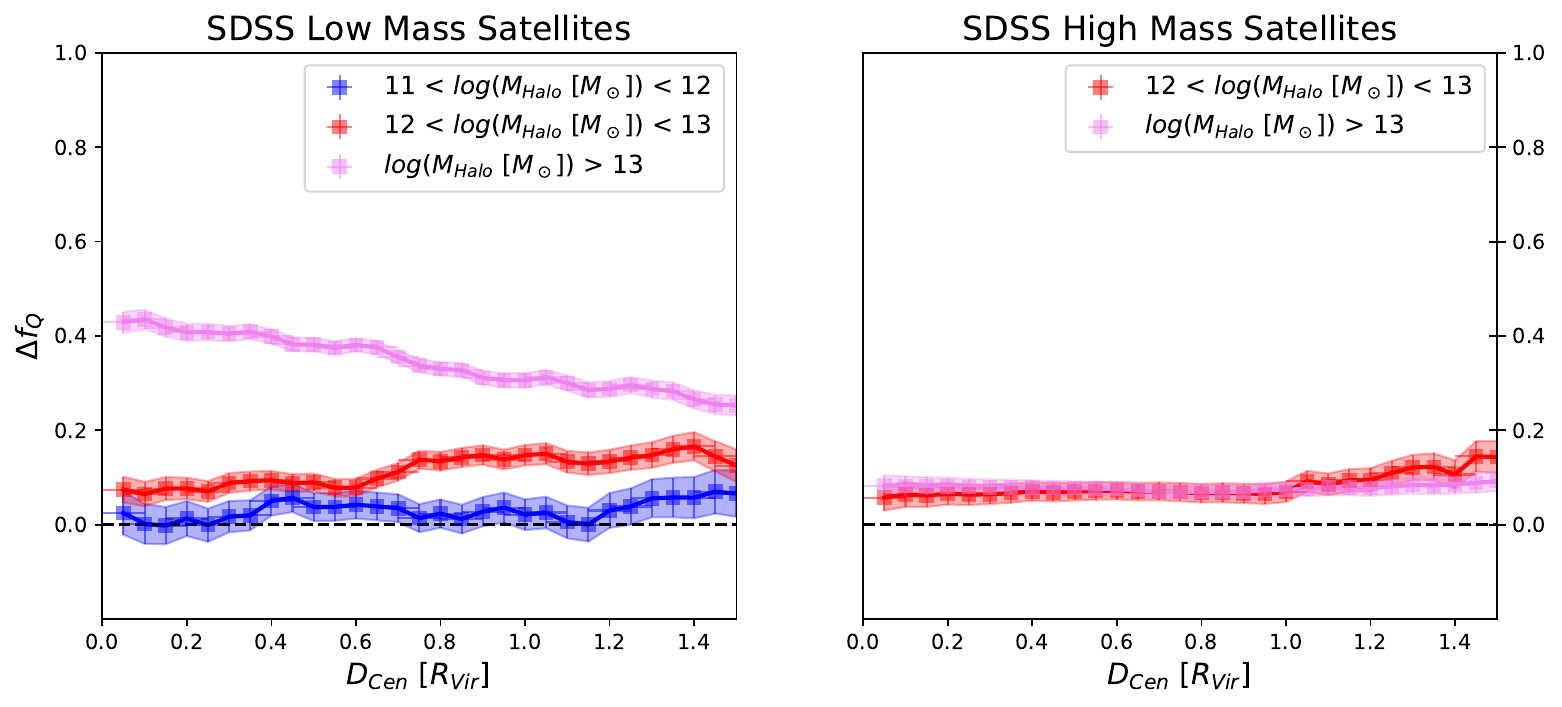}
    \vspace*{-3mm}
    \caption{Satellite quenching as a function of the location within dark matter haloes. We present the delta quenched fraction, $\Delta f_Q$, as a function of the distance to the nearest central galaxy, $D_{\rm cen}$, for the SDSS. The narrow shaded regions for the SDSS when compared to the simulations is due solely to the much larger sample size of observational data. This statistic measures the quenching enhancement of satellites relative to centrals matched at the same estimated black hole mass (which is known to be the dominant intrinsic quenching driver). Results are shown for different mass haloes (see legends), and for low-mass (left panel) and high-mass (right panel) galaxies. Note that there is a strong quenching dependence on the location of satellites within the halo for low-mass satellites in high-mass haloes, yet high-mass satellites are not strongly affected by environment in any location within any mass halo. }
    \label{fig:SDSS_DelfQ}
\end{figure*}

In Figs. \ref{fig:SIM_DelfQ}, \ref{fig:SIM_DelfQ_2D} \& \ref{fig:SDSS_DelfQ} we show how the satellite quenched fraction evolves with increasing distance from the central, and increasing halo mass. We define a new statistic, $\Delta f_{Q}$, which measures the offset in a satellite galaxy population's quenched fraction, relative to centrals within +/- 0.1 dex in $M_{\rm BH}$. This removes the effect of $M_{\rm BH}$ on quenching, allowing us to focus exclusively on environmental quenching. Note that $M_{\rm BH}$ is unambiguously identified as the dominant intrinsic quenching parameter, and hence this allows us to effectively remove the impact of intrinsic quenching mechanisms before assessing the role of environment.  We then explore how $\Delta f_{Q}$ evolves as a function of the distance of satellites to their respective central galaxies, at different halo mass intervals. These figures all follow the same structure, with low mass satellites in the left column and high mass satellites in the right column. 

\par The first two figures present this analysis for the simulations, where each row corresponds to a simulation suite. The first figure displays results for the complete simulation data utilizing a 3D distance to the central. Concentrating on the low-mass satellite panels, the simulations show that for $11<\log(M_{\rm Halo} / \rm M_\odot)<13$, $\Delta f_{Q}$ is low. It is near zero for $11<\log(M_{\rm Halo} / \rm M_\odot)<12$, and around 0.2 for $12<\log(M_{\rm Halo} / \rm M_\odot)<13$. However, at $\log(M_{\rm Halo} / \rm M_\odot)>13$ we find that we begin with a high $\Delta f_{Q}$ which gradually decreases as the distance to central increases. Using $\Delta f_{Q}$ tells us that while the quenched fraction of centrals in this population is low, the same population of satellites has an elevated quenched fraction. Therefore, it is reasonable to claim that what drives this quenching has to be an environmental effect, since we know that centrals quench via intrinsic mechanisms and we have controlled for the dominant parameter ($M_{\rm BH}$). 

Following the same method, we see that for high mass satellites (right panel of Fig. \ref{fig:SIM_DelfQ}), $\Delta f_{Q}$ is near 0 in all situations, irrespective of halo mass. Therefore we assert that high mass satellites quench in the same way as centrals, largely independent of environmental effects. There is a weak increase in the typical $\Delta f_{Q}$ values of satellites in the highest mass haloes, but this is only marginally significant and clearly far less than seen for low mass satellites. Importantly, these results show that environmental quenching is mass dependent. At low masses, the environment in which a satellite resides has a dramatic effect on its star forming state (even after controlling for intrinsic quenching). Whereas, for high mass satellites, environment has at most a very small impact on the star forming state, once black hole mass is controlled for.

\par In order to more directly compare to the SDSS data, we additionally derive a 2D version of $D_{cen}$ (obtained by randomly projecting the simulation cube into a plane), and apply 0.5 dex of Gaussian random noise to the halo mass parameter, to simulate the uncertain abundance matching data \citep{Yang_2009,Yang2011}. We present these results in Fig. \ref{fig:SIM_DelfQ_2D}. There is a relatively small shift towards lower quenched fractions for the $\log(M_{\rm Halo} / \rm M_\odot)>13$ group, but other than that all results are very similar to the full fidelity 3D analysis. For low mass satellites we still see lower $\Delta f_{Q}$ in the two lower halo mass bins, and higher, gradually decreasing, $\Delta f_{Q}$ at larger halo masses. Additionally, the high mass plots are nearly identical to the unaltered ones in the previous figure. Hence, even with the two major limitations of the SDSS accounted for (2D distances and uncertain halo masses), the main signature of environmental quenching within dark matter haloes ought to be clearly visible. 

\par In Fig. \ref{fig:SDSS_DelfQ} we show this analysis for the observed SDSS data, with the left panel corresponding to low mass satellites and the right panel to high mass satellites. The SDSS data is weighted by $1/V_{max}$ in this plot to remove statistical incompletion \citep[see, e.g.][]{Bluck_2014, Bluck_2016, Thanjavur_2016}. Hence, we can make a like-to-like comparison with Fig. \ref{fig:SIM_DelfQ_2D}. We elect to restrict the acceptable weights so as to not overwhelm the lower mass objects, as some will have extreme statistical effect that corrupts results. This only impacts the scatter in the trends, not the qualitative results.

The results on the role of halo location in quenching satellite galaxies is very similar in observations to the plots seen previously for simulations. For low-mass satellites (left panel), the two lower halo mass bins have low $\Delta f_{Q}$, while the highest halo mass range exhibits a much greater differential quenched fraction that decreases with $D_{cen}$. The high mass satellites (right panel) exhibit the same property as seen for simulations, i.e., low $\Delta f_{Q}$ values that remain approximately constant with $D_{\rm cen}$. Hence, there is excellent agreement between the simulation predictions and the observed Universe in the SDSS with respect to both high- and low-mass satellite quenching. High-mass satellites are not strongly impacted by environment in any halo mass or halo location, hence they behave similarly to centrals (as seen several times before in this paper). Conversely, low-mass satellites are more frequently quenched in high-mass haloes, and preferentially so when closer to their respective centrals.

\section{Discussion - Intrinsic vs. Environmental}\label{Discussion}

\subsection{Centrals and High Mass Satellites}

Our analysis into local centrals reveals the key importance of supermassive black hole mass when determining if they are quenched or star forming. Through Random Forest classification analyses we show that black hole mass dominates other properties (such as halo mass, stellar mass, or local density) in predicting the quiescence of galaxies in both simulations and observations of the local Universe. This is in direct agreement with prior work by \citet{Piotrowska_2022}, \citet{Bluck_2022}, and \citet{Bluck_2023}.

\par High-mass satellites share very similar quenching dependences to central galaxies. When performing the Random Forest classification experiments (as seen in Fig. \ref{fig:Simulations_RandomForest}) and the quenching vectors (in Fig. \ref{fig:SIM_QuenchAng}), it is clear that black hole mass is the most predictive parameter by a wide margin in the simulations. There is, however, the beginning of a pivot towards environmental factors as demonstrated by the pie charts in Fig. \ref{fig:Simulations_RandomForest} and the quenching angles in Fig. \ref{fig:SIM_QuenchAng}. The observational data returns nearly identical results to the simulations, with black hole mass dwarfing all other features in its power to determine the quenching of high-mass satellites. 

In the simulations studied in this work, the cause of central and high mass satellite quenching is explicitly known to be AGN feedback \citep[see][]{Vogelsberger_2014a,Vogelsberger_2014b, Schaye_2015, Weinberger_2017, Weinberger_2018}. Black hole mass is an excellent tracer of the total integrated energy of AGN feedback over the lifetime of a galaxy \citep[e.g.,][]{Bluck_2020a, Zinger_2020, Piotrowska_2022}. Therefore, the importance of black hole mass for predicting the quenching of all types of high-mass galaxies in the SDSS can be explained effectively by the AGN feedback paradigm. 

It is important to appreciate that the simulations used in this work do not predict a strong connection between quenching and the instantaneous AGN luminosity \citep[see][]{Piotrowska_2022, Ward_2022, Bluck_2023}. Hence, exploring the location of AGN on the SFR - $M_*$ plane is not an effective route to testing AGN feedback quenching (despite  being an extremely common method explored in the literature). Instead, a long term tracer of feedback energy is required, like black hole mass. Consequently, the approach used in this work to test AGN driven quenching is essentially the optimal strategy, despite being (necessarily) indirect. See \citet{Bluck_2023} for further discussion on this important point.

Whilst AGN feedback is a clearly a plausible explanation to the observational results, one can always speculate that a close dependence of central (and high-mass satellite) quenching on central velocity dispersion (used to estimate $M_{\rm BH}$) could be due to some other cause. Three possibilities are worth considering: (i) supernova feedback \citep[e.g.,][]{Matteucci_2006}; (ii) halo mass quenching from virial shocks \citep[e.g.,][]{Dekel_2006}; and (iii) dynamical stabilization \citep[e.g.,][]{Martig_2009}. We consider each in turn next.

As shown in \citet{Bluck_2020a}, the total energy released from supernovae must be directly proportional to stellar mass. Yet, in Fig. \ref{fig:SIM_QuenchAng} we find no importance on stellar mass once black hole mass is available to the classifier for both central and high-mass satellite galaxies. Hence, we reject the possibility of supernovae quenching massive galaxies. See also \citet{Somerville_2015} for a review of the theoretical limitations of supernovae as a high-mass quenching source.

Virial shock heating could provide a route to quenching galaxies through stabilizing the CGM \citep{Dekel_2006}. The energy released from virial shocks is shown to depend primarily on halo mass \citep{Bluck_2020b}. Yet, in Fig. \ref{fig:SIM_QuenchAng} no dependence of quenching on halo mass is found for centrals or high mass satellites. So, we reject this possibility as well. See also \citet{Bluck_2016,Bluck_2020b, Bluck_2022} for similar conclusions. Note that in this case the relative uncertainties on halo mass and black hole mass in observations are similar (both $\sim$0.5 dex), hence differential uncertainty is highly likely to not be relevant here \citep[see also][for further tests on this possibility]{Piotrowska_2022}.

Finally, we consider dynamical stabilization of the ISM \citep[e.g.,][]{Martig_2009, Gensior_2020}. In this scenario a dense central bulge (which is expected to have high central velocity dispersion) provides stabilizing torques, preventing giant molecular gas cloud (GMC) collapse in the disk. Such a scenario may help to reduce star formation in certain galaxies. However, this mode does not prevent gas inflow into galaxies (either from CGM cooling or cosmic inflows). Consequently, a stabilization mechanism alone would lead to higher gas fractions in quiescent systems than star forming ones, which is emphatically not seen. Indeed, it is well known that quenched galaxies have depleted gas reservoirs in both atomic and molecular Hydrogen \citep[see, e.g.,][]{Saintonge_2016, Saintonge_2017, Brownson_2020, Ellison_2021, Piotrowska_2020, Piotrowska_2022}. Hence, we reject this hypothesis as well.

As a result of the above discussion it seems that AGN feedback is by far the most plausible mechanism for driving all types of massive galaxy quenching in nature. Whether a system is a central or a satellite is irrelevant to whether it can form a massive enough black hole to quench the system. Furthermore, environmental effects do not lead to a significant increase in the probability of high mass satellites being quenched. From considering Figs. \ref{fig:SIM_DelfQ} - \ref{fig:SDSS_DelfQ}, it seems that environment is only effective at quenching relatively low mass galaxies. This can be explained by high-mass systems having deeper potential wells, and hence higher binding energies of both their ISM and CGM. Hence, environmental processes such as ram pressure or dynamical stripping become systematically less effective for higher satellite masses.

Interestingly, our results strongly imply that `mass' (intrinsic) and environmental quenching are not separable phenomena \citep[as was argued for in][]{Peng_2010, Peng_2012}. It is true that intrinsic quenching is independent of environment, but environmental quenching is {\it not} independent of intrinsic parameters. This can be seen very clearly by noting the utter lack of dependence of quenching on environment for high mass satellites, once the dominant intrinsic parameter ($M_{\rm BH}$) is accounted for (see Fig. \ref{fig:SDSS_DelfQ}, right panel). Moreover, we see an explicit dependence of low-mass satellite quenching on stellar mass in Fig. \ref{fig:SDSS_RandomForest}. 

For over a decade the received wisdom has been that centrals quench via intrinsic routes and satellites quench by environment. But this paper reveals that this is not true. All high-mass galaxies quench primarily as a result of intrinsic quenching, which is well explained by simulations utilizing AGN feedback. Whether a high mass galaxy is a central or a satellite is irrelevant to its quenching mechanism. Low-mass centrals are almost invariably star forming (they cannot quench by either AGN feedback or environment), but low-mass satellites may be environmentally quenched (see below). Hence, a better way to separate intrinsic and environmental quenching in galaxy populations is to cut in mass, not galaxy type.

\subsection{Low Mass Satellites}

\par In both cosmological simulations and observations from the SDSS, the quenching of local low-mass satellite galaxies is best predicted by environmental parameters. This is clearly apparent in Fig. \ref{fig:Simulations_RandomForest} and Fig. \ref{fig:SDSS_RandomForest}, wherein pie charts indicate a near total environmental importance to quenching of more than 95\% for all simulations, and nearly 75\% for the SDSS. We further present this through constructing quenching angles in Figs. \ref{fig:SIM_QuenchAng} \& \ref{fig:SDSS_QuenchingAngles}. The quenching angle for low mass satellites are all nearly horizontal, indicating a strong correlation with halo mass or local galaxy over-density, and a complete absence of correlation with black hole mass. 

However, our random forest classifications do reveal a discrepancy as to which parameter is most predictive of quenching in the simulations compared to the SDSS. The simulations (whether using the complete or SDSS-weighted samples) all share the same result of $M_{\rm Halo}$ being the best parameter. The SDSS on the other hand has a clear preference of $\delta$ (the sum in importances of $\delta_3$, $\delta_5$ and $\delta_{10}$), along with a significant secondary dependence upon $M_{*}$. This contrast between the most predictive parameters could indicate a real difference in the mechanisms by which low-mass satellites quench in simulations compared to in our Universe. 

\par Ram pressure stripping is proportional to the relative velocity squared of a satellite (compared to the CGM) multiplied by the density of the CGM through which it moves ($P_{\rm ram} = \rho v^2$). Increasing the mass of the halo increases both of these parameters \citep[e.g.,][]{Mo_2010}. Hence, ram pressure stripping is more effective in higher mass haloes. It is particularly easy for ram pressure effects to strip the CGM, as opposed to the ISM, since the former is much more weakly bound to the satellite. Hence, ISM stripping is likely only important at the very centre of high mass clusters. If ISM stripping were the mechanism that drives the quenching of satellites we would see a much greater concentration of quenched galaxies near the center of a halo, and nearly none as we approach the edges of the cluster. However, Figs. \ref{fig:SIM_DelfQ} - \ref{fig:SDSS_DelfQ} show that quenched objects exist throughout the halo, albeit those at the center are more frequent. Therefore, while ISM stripping may play a role for a fraction of galaxies, strangulation due to CGM stripping is the more likely culprit for the majority of systems. This explains why it is halo mass, not location within the halo, which best predicts quiescence in simulations from the random forest analysis. 

\par The random forest classification for the SDSS low-mass satellites reveals $\delta$ as most predictive parameter for quenching, not halo mass. This indicates that low mass satellites in over-dense regions are more likely to be quenched. Initially this may suggest that satellite - satellite interactions could be an important route to quenching. However, Fig. \ref{fig:DensityComp} shows that it is the tenth nearest neighbor which is the most important of the nearest neighbor parameters. Should quenching be due to events such as galaxy-galaxy harassment leading to tidal stripping, one would expect that the closest of the nearest neighbor densities should be most important. And yet, $\delta_{3}$ is actually found to be the least important, followed by $\delta_{5}$. Therefore, there must be some other reason for link between density and quenching in observational data.

\par Very few galaxy clusters contain ten or more satellite galaxies, whether in the simulations or observations, as seen in Table. \ref{tab:HaloSize}. The most important density parameter being $\delta_{10}$ then raises some pressing questions as to which physical processes could be involved, and hence what $\delta_{10}$ truly represents. Since the tenth nearest neighbor is generally a super-virial parameter, in combination with the unanimous agreement for halo mass in simulations, we posit that $\delta_{10}$ might represent a more accurate constraint on $M_{\rm Halo}$ than the value inferred from abundance matching, especially at the $M_{\rm Halo}$ quenching threshold. 

The large uncertainty in halo mass for observational data ($\sim 0.5$ dex) lends further credence to this hypothesis. $\delta_{10}$ could reveal information on the halo which is lost in the current abundance matching method, which truncates at the virial radii \citep[e.g.,][]{Yang_2007, Yang_2009}. Furthermore, as ram pressure stripping is closely linked to halo mass, the secondary importance of $M_{*}$ could be explained by its proportionality to the satellite's gravitational potential, and hence retention of both the satellite's ISM and, potentially, CGM as well. More massive galaxies would be more likely to hold on to their gas reservoirs during an event of ram pressure stripping.

\par This is further exemplified in Figs. \ref{fig:SDSS_SIM_ParamRemoved} and \ref{fig:ErrorTest}. Fig \ref{fig:SDSS_SIM_ParamRemoved} shows the result of random forest classification tests on low mass satellites for the simulations and the SDSS, following the removal of their, respective, most predictive parameters. As such, the left panel shows the result after withdrawing density as an input feature, while the left panel presents the analysis without halo mass. It is clear that, should density be removed, then in all cases halo mass prevails. Similarly, the left panel reveals that without the halo mass, $\delta$ (representative of the sum of relative importance of $\delta_3$, $\delta_5$, \& $\delta_{10}$) is most predictive for all sample sets. Hence, the simulations are clearly {\it nearly} right about the quenching of satellites. Therefore it becomes interesting to speculate as to whether observational limitations obscure the true importance of halo mass in the SDSS.

Referring back to the differential measurement uncertainty test in Fig. \ref{fig:ErrorTest}, the differential error required for halo mass to surpass density as the most predictive parameter is quite low ($\sim$ 0.4 dex). In fact, the required amount of noise is less than the total uncertainty in halo mass for observational data \citep[see][]{Yang_2007}. Therefore, it is conceivable that, for observations, density is most important in random forest classification due to the high uncertainty in halo mass, rather than it being inherently more predictive. If this is the case, it points towards a more effective way to estimate halo masses than abundance matching. Although above the scope of the present work, we plan to investigate this possibility in upcoming work (Goubert et al. in prep.).

In Figs. \ref{fig:SIM_DelfQ} - \ref{fig:SDSS_DelfQ}, we investigate how the location of satellites within their parent dark matter haloes impacts quenching. This analysis reveals a correlation between the relative quenched fraction (compared to centrals at the same $M_{\rm BH}$) of low-mass satellites and the distance to their central galaxy. For higher mass halos, as $D_{\rm cen}$ increases the relative quenched fraction decreases. This is essentially as expected since the inter-cluster medium (ICM) is denser at the center, and furthermore satellites on radial orbits reach their peak velocities at their pericenters (near the central). Hence, ram pressure is most effective at low $D_{\rm cen}$. Additionally, tidal effects from satellite - central interactions peak at the center of the group and cluster, and the number density of satellites also rises with decreasing $D_{\rm cen}$ leading to a greater potential for tidal stripping from satellite - satellite interactions as well.

It is important to note that the current value of $D_{\rm cen}$ does not reveal any information as to where in the halo a galaxy could have been previously. For example, a satellite observed on the edge of the virial radius now could have been near the center of the galaxy cluster in the past. Nonetheless, the simulations still predict a significant dependence of the quenching of low mass satellites in high mass haloes on the distance to the central. This is likely because many galaxies at high distances from their centrals are newly accreted systems, and essentially all galaxies at low distances from their centrals have been in the group/cluster for some time. Thus, there is predicted to be a real statistical difference as a function of group-centric distance. We confirm the general trends predicted by simulations in the SDSS, noting that the limitations in halo mass accuracy and 2D projection on the plane of the sky do not obscure these underlying secondary relationships.

Perhaps the most important result that arises from our $\Delta f_Q$ analysis is the difference in dependence on environment (both halo mass and location within the halo) between high- and low-mass satellites. Low-mass satellites in high mass haloes are preferentially quenched in the center of their groups/ clusters, but are much more likely to be star forming in their outskirts. Low-mass satellites in low mass haloes are not likely to be quenched. Moreover, high-mass satellites in any mass of halo are not significantly enhanced in their quenching, compared to centrals with the same $M_{\rm BH}$. This implies that environmental quenching is only important for satellites with relatively low masses, which reside in the ultra-dense environments near the centre of clusters or high-mass groups.

\subsection{How to Quench a Galaxy}

As a result of the above discussion, there are essentially two ways to quench a galaxy in the local Universe: 
\begin{enumerate}

\item  Host a high mass black hole, which leads to intrinsic quenching independent of environment. This form of quenching is available to both centrals and satellites, and is clearly a consequence of AGN feedback.\\

OR\\

\item Be a low-mass satellite close to the centre of a high-mass halo, which leads to environmental quenching. This form of quenching is clearly only available to satellites, but crucially not all satellites, and is most likely a consequence of ram pressure and/or dynamical stripping.
\end{enumerate}

Intrinsic quenching is completely decoupled from environment, yet environmental quenching is {\it not} decoupled from the intrinsic properties of galaxies. Only galaxies with shallow enough potential wells are vulnerable to environmental quenching. Interestingly, both of these quenching mechanisms effectively remove the possibility of gas accretion into the system, starving the galaxy of the fuel required for future star formation. This implies that galaxies may continue to form stars as they quench, which leads to enhanced stellar metallicities of quiescent centrals and satellites \citep[as observed in][]{Peng_2015, Trussler_2019, Bluck_2020b}.

Are there other quenching mechanisms? Certainly there is no evidence for them in this work. Yet, clearly the place to look would be low-mass centrals, which are unaffected by both environment and strong AGN feedback. In the SDSS, the number of quenched low-mass centrals is completely negligible \citep[see][]{Peng_2012, Bluck_2014}. However, deeper local Universe surveys do find evidence of quiescent isolated dwarfs \citep[e.g.,][]{Penny_2018, Polzin_2021}. Nevertheless, these systems may not be permanently quenched like higher mass galaxies, but rather experiencing dramatic oscillations around the main sequence (as a result of supernova or intermediate black hole feedback). Indeed, it is hard to envisage what could prevent these systems from re-accreting gas from the IGM and ultimately rejuvenating over long time-scales.

Finally, it remains to be seen whether the quenching dichotomy of the local Universe remains unchanged at higher redshifts, although see \citet[][]{Bluck_2022, Bluck_2023, Bluck_2024} for evidence that the intrinsic quenching mechanism is stable across cosmic time. In the coming few years the VLT-MOONRISE survey will provide rest-frame optical spectroscopy for $\sim$100\,k galaxies at Cosmic Noon, yielding unprecedented measurements of cosmic environment in the early Universe \citep[see][]{Cirasuolo_2020, Maiolino_2020}. This will enable an analogous analysis to this paper at the peak epoch of galaxy assembly for the first time.

\section{Summary}\label{Summary}

In this paper we determine which types of galaxies quench via intrinsic or external mechanisms. Additionally, we identify the key intrinsic and environmental parameters which regulate quenching. To do so, we first split galaxies into one of three classes: centrals, high-mass satellites, and low-mass satellites. The dominant feature(s) for predicting quiescence are then determined for each group via Random Forest classification, partial correlation coefficients, and visual assessment of quenching gradients. The random forest approach is chosen for its ability to determine causal relationships in complex, inter-correlated data \citep[see][]{Bluck_2022}. We first analyse three state-of-the-art cosmological hydrodynamic simulations (EAGLE, Illustris, IllustrisTNG), utilizing snapshots at $z\sim 0$, to extract testable quenching predictions for each of the above populations. Additionally, we provide a detailed comparison with the largest spectroscopic galaxy survey at low redshifts, the SDSS.

\par We define high-mass satellites as satellites with $M_{*} > 10^{10.5} {\rm M_{\odot}}$, and low-mass satellites as satellites with $M_{*} < 10^{10} {\rm M_{\odot}}$. A galaxy is defined to be quenched if it has an sSFR $< 10^{-11} {\rm yr}^{-1}$, and star forming if it lies above this limit. \\

\par Our primary results are as follows:

\begin{enumerate}

    \item All simulations predict that intrinsic parameters should be most predictive of quiescence for both centrals and high-mass satellites. Specifically, supermassive black hole mass is the most predictive quenching parameter for both classes of galaxies.\\

    \item From observations of galaxies in the local Universe with the SDSS, we confirm these predictions from simulations. \\
    
    \item All simulations predict that low-mass satellite quenching should be regulated by environmental processes, with halo mass being the optimal parameter for predicting quiescence.\\

    \item From SDSS observations we confirm that low-mass satellite galaxies are environmentally quenched. However, we find that the local galaxy over-density ($\delta$), summed over 3rd, 5th and 10th nearest neighbours, is the most predictive parameter (unlike in simulations).\\
    
    \item Hence, there is complete agreement between cosmological simulations and observations as to the quenching of centrals and high mass satellites, but there is a mild disagreement regarding the quenching of low mass satellites.\\

    \item We explore in detail the origin of this discrepancy, and highlight the following possibilities:

        \begin{enumerate}
        
            \item Low-mass satellites in the simulations may quench differently from those in observations. Ram pressure stripping is likely most connected to halo mass, whereas galaxy over-density lends itself more towards tracing galaxy - galaxy harassment, or tidal stripping. \\

            \item Local galaxy over-density could be most predictive in the observations due to the high uncertainty in halo mass from abundance matching. This is supported by a differential measurement uncertainty test, in which we find that a relatively small amount of noise applied to the $\delta$ parameters yields in halo mass prevailing as most predictive parameter in the SDSS.\\

            \item When comparing individual values of $\delta_{n}$ ($n$ = 3, 5, 10), the tenth nearest neighbor density is by far most predictive of quiescence in both simulations and observations. However, very few groups or clusters contain ten or more satellites (above our mass threshold). Therefore, $\delta_{10}$ is unlikely trace galaxy - galaxy interactions, potentially ruling out (a).\\
            
            \item Alternatively, it is possible that $\delta_{10}$ reflects a more global environmental tracer, which may strongly connect to halo mass. If this is true, this suggests that $\delta_{10}$ can be used to improve extant halo mass estimation from abundance matching, which do not currently incorporate information on scales beyond the virial radius \citep[e.g.,][]{Behroozi_2010, Moster_2010}.
            
        \end{enumerate}

        \item Finally, we note that the qualitative dependence of satellite quenching on environment in 6D phase space in simulations is recovered in observation-like 2D+$z$ space. This is encouraging as it strongly suggests that observational limitations are not seriously problematic for exposing the role of environment in galaxy evolution.

    \end{enumerate}

\par In summary, we find observations of our local Universe to be in remarkable agreement with cosmological simulations with respect to the quenching of centrals and high-mass satellites. Both populations have quenching best predicted by black hole mass, which clearly implies AGN feedback is likely the underlying quenching mechanism \citep[see][for an equivalent result for centrals]{Piotrowska_2022}. Hence, we discover that high-mass satellites behave very similarly to centrals in their quenching, and quite unlike low-mass satellites. 

On the other hand, low-mass satellites quench as a result of environment. However, the optimal parameter to trace environment is different in simulations and observations. Whilst simulations predict halo mass should be most predictive, the observations suggest local galaxy over-density is superior. From a variety of tests, we suggest that the most likely explanation for this discrepancy is the high uncertainties on the halo masses in the SDSS. If this is the correct explanation, our results further imply that halo mass estimation may be improved by including information on galaxy clustering beyond the virial radius.

Finally, we emphasize that the received wisdom regarding central and satellite galaxy quenching (i.e. that centrals intrinsically quench but satellites environmentally quench) is not entirely correct. All high-mass galaxies (both centrals and satellites) quench as a result of intrinsic AGN-feedback. Essentially no low-mass centrals are quenched in the SDSS. But, low-mass satellites may quench via environmental routes, closely connected to halo mass, halo location, and local density. Consequently, the most effective way to separate quenching mechanisms is to cut in mass, not galaxy class. Ultimately, a galaxy may quench by either: (a) hosting a high mass black hole (which implies strong historic AGN feedback); or (b) being a low mass system in a dense environment (which leads to effective ram pressure and dynamical stripping).

\section*{Acknowledgements}

We thank the referee, Dr Andrea Negri, for an insightful and highly positive report which greatly improved the presentation of this work. AFLB acknowledges a faculty start-up grant at the Florida International University (FIU). PG acknowledges support from a Teaching Assistantship at FIU. RM acknowledges a Royal Society Research Professorship, as well as support from the STFC and ERC Advanced Grant (695671) ‘QUENCH’.

%%%%%%%%%%%%%%%%%%%%%%%%%%%%%%%%%%%%%%%%%%%%%%%%%%
\section*{Data Availability}

 All data used in this study have been previously published and are
available at the following online locations:
\begin{itemize}
    \item EAGLE: \url{http://icc.dur.ac.uk/Eagle/}
    \item Illustris: \url{www.illustris-project.org/}
    \item TNG: \url{https://www.tng-project.org/}
    \item SDSS DR7: \url{https://classic.sdss.org/dr7/access/}
    \item MPA-JHU release of spectrum measurements:
\url{https://wwwmpa.mpa-garching.mpg.de/SDSS/DR7/}
    \item SDSS Group Catalogue: \url{https://gax.sjtu.edu.cn/data/Group.html}
    \item NYU Galaxy Value Added Catalogue:
\url{http://sdss.physics.nyu.edu/vagc/}
    \item morphological catalogues:
    \begin{itemize}
        \item \url{http://dx.doi.org/10.1088/0067-0049/196/1/11}
        \item \url{http://dx.doi.org//10.26093/cds/vizier.22100003}
    \end{itemize}

\end{itemize}

%%%%%%%%%%%%%%%%%%%% REFERENCES %%%%%%%%%%%%%%%%%%

% The best way to enter references is to use BibTeX:

\bibliographystyle{mnras}
\bibliography{Citations} % if your bibtex file is called example.bib

% Alternatively you could enter them by hand, like this:
% This method is tedious and prone to error if you have lots of references
%\begin{thebibliography}{99}
%\bibitem[\protect\citeauthoryear{Author}{2012}]{Author2012}
%Author A.~N., 2013, Journal of Improbable Astronomy, 1, 1
%\bibitem[\protect\citeauthoryear{Others}{2013}]{Others2013}
%Others S., 2012, Journal of Interesting Stuff, 17, 198
%\end{thebibliography}

%%%%%%%%%%%%%%%%%%%%%%%%%%%%%%%%%%%%%%%%%%%%%%%%%%

%%%%%%%%%%%%%%%%% APPENDICES %%%%%%%%%%%%%%%%%%%%%

\appendix

\section{SDSS Edge Effects}

\label{appendix:EdgeSDSS}
\begin{figure*}

	\includegraphics[width=\textwidth]{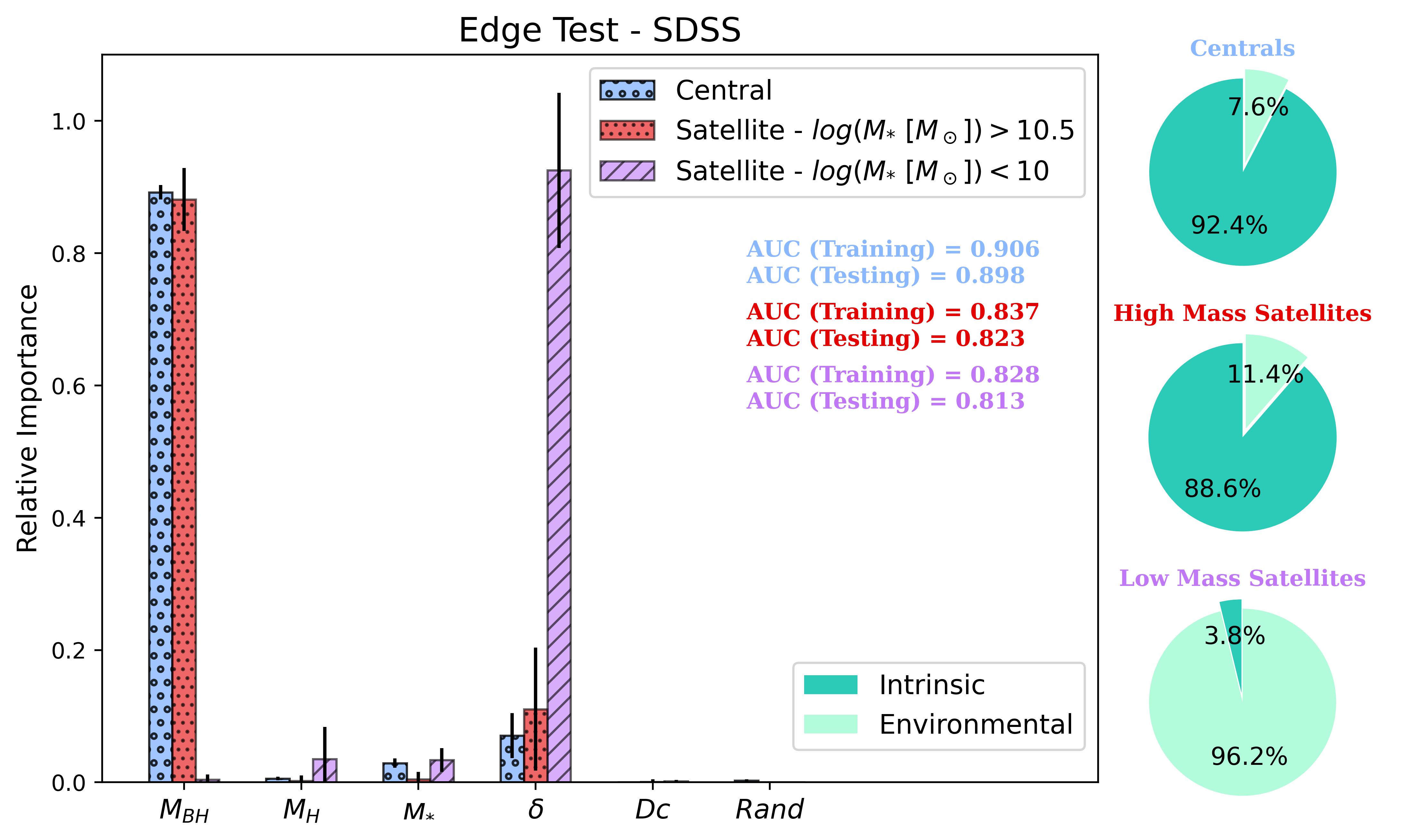}
    \vspace*{-3mm}
    \caption{Random Forest classification results for a region of the SDSS which is selected to be free of significant edge effects (see Fig. A2). The parameters used for training the random forest are listed along the $x$-axis, with their relative quenching importances indicated by the $y$-axis bar heights. Uncertainties are given by the variance of 10 independent classification runs. The pie charts on the right of the main panel indicate the total importance of intrinsic and environmental parameters for each galaxy class. We find results consistent with our earlier analysis on the entire parent sample of observational data. Black hole mass clearly dominates for centrals and high-mass satellites. Meanwhile, for low-mass satellites, local over-density remains the best predictor of quiescence. In fact, $\delta$ has gained slightly in importance when compared to Fig. \ref{fig:SDSS_RandomForest}. Therefore, edge effects have little impact on results of the random forest classification in this work. As such, we choose to study the entirety of the observational data in the main body of the paper.}
    \label{fig:SDSS_RandomForest_Edge}
\end{figure*}

To determine the potential impact edge effects could have on the analysis of SDSS environments, we perform a random forest classification test for a particular region of the survey. We choose a portion of the SDSS away from any survey edges, and additionally ensure as few as possible aberrations within the region. The results of this random forest classification are presented in Fig. \ref{fig:SDSS_RandomForest_Edge}. As expected, black hole mass remains the most predictive parameter for centrals and high-mass satellites. Similarly, the most predictive parameter for low-mass satellites stays as local over-density. As a matter of fact, $\delta$ is now even more important (as one might reasonably have expected). Therefore, we conclude that edge effects have little impact on the results of analysis conducted for the entire SDSS parent dataset in the main body of this paper.

\begin{figure*}
    \centering
    \textbf{\large Edge Effect Test Region}\par\medskip
    \includegraphics[width=\textwidth]{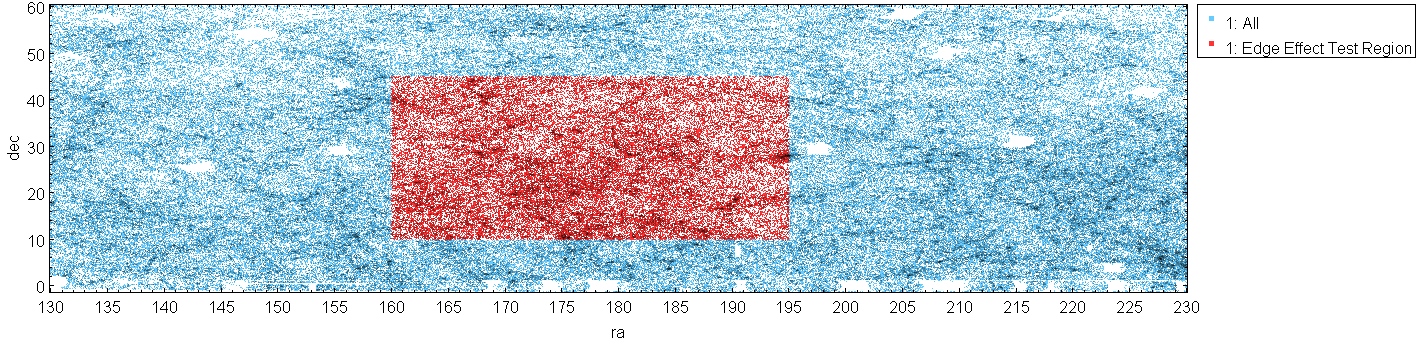}
    \vspace*{-3mm}
    \caption{The region selected to perform the edge effect test is presented as a red square within the blue of the total SDSS survey. We find the nearest edge to be 57.07 Mpc away from the limit of the region and, hence, likely free of any impact on gravitationally bound structures.}
    \label{fig:SDSS_Edge_Region}
\end{figure*}
%%%%%%%%%%%%%%%%%%%%%%%%%%%%%%%%%%%%%%%%%%%%%%%%%%

\section{Re-assigning Missing $M_{\rm BH}$ values in simulations} \label{appendix:MBH_Reassign}

A significant fraction of low mass satellites, and a much smaller fraction of high mass satellites and centrals, have missing central black holes in simulations. This is due to inaccuracies of the identification of black holes in the Subfind Sublink algorithm, especially for low mass satellites in high mass groups and clusters. In Section. \ref{SimMBH}, we detail the process by which we re-assign missing black hole mass values in the simulations. Specifically, we assign a black hole mass value from a Gaussian distribution centered around the mean black hole mass of the distinct galaxy class (central, high mass satellite, low mass satellite), and with $\sigma$ defined as the standard deviation of $M_{\rm BH}$ of said class. 

\begin{figure*}
    \centering
    \includegraphics[width=\textwidth]{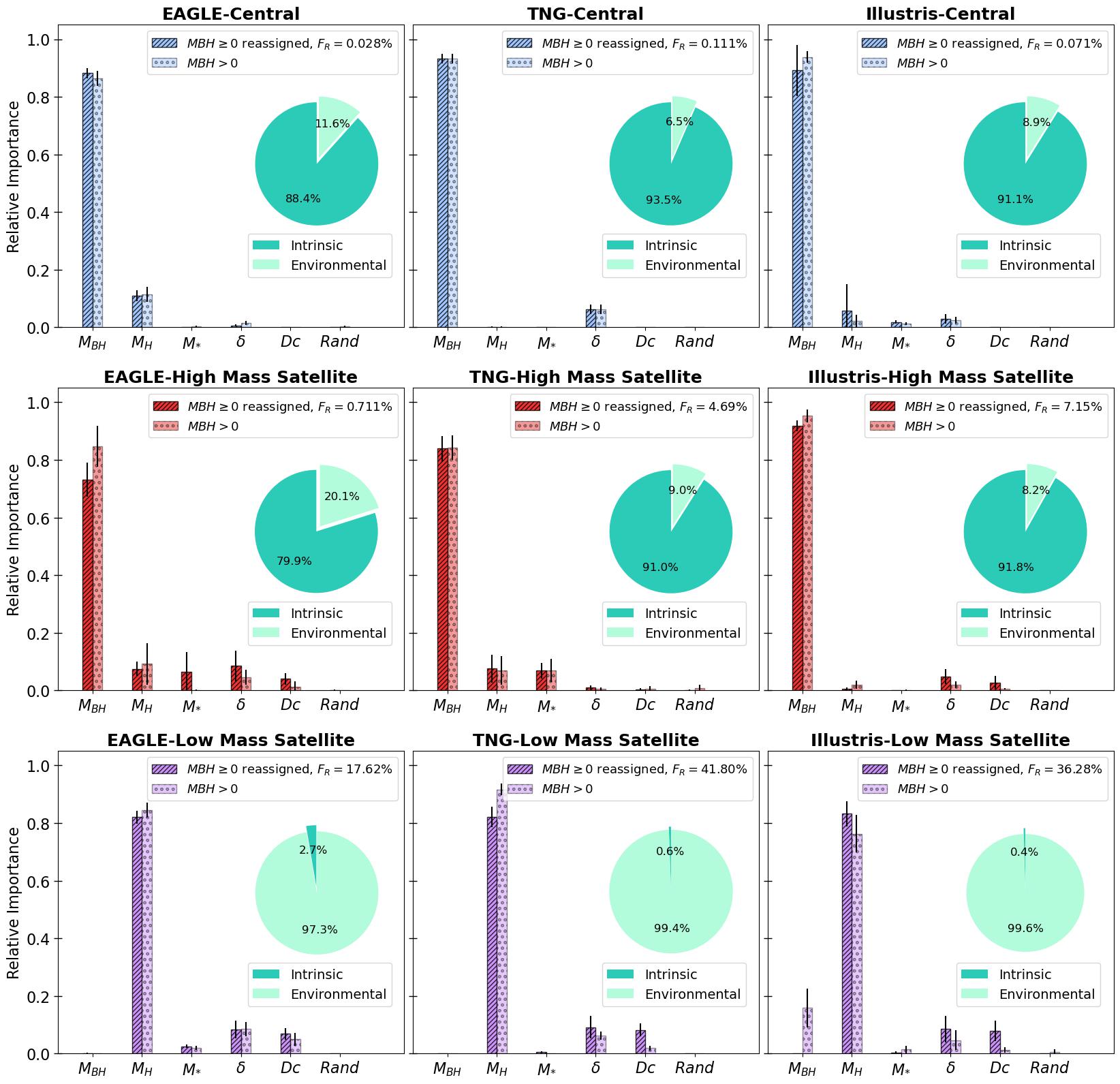}
    \vspace*{-3mm}
    \caption{Random Forest classification for the simulations results, before and after assigning black hole mass values for galaxies lacking black hole particles (shown as different hatched bars). Each row corresponds to a galaxy class, while each column corresponds to a simulation suite. We find that our method for assigning $M_{\rm BH}$ values does not alter the results of Random Forest classification. Furthermore, the change from black hole mass to halo mass as most predictive parameter as we move from class to class remains evident. Therefore, we determine that replacing absent black hole masses with reasonable values does not impact the results in any significant manner. The fraction of systems with re-assigned black hole masses is indicated on the legends for each simulation and galaxy population. }
    \label{fig:MBH_SIM_RF}
\end{figure*}

We show in Fig. \ref{fig:MBH_SIM_RF} that the results from the random forest do not significantly change whether we set a selection criteria of $M_{\rm BH} > 0$ (removing all galaxies for which this is an issue), or assign a new black hole mass value as previously detailed (ensuring completeness at the expense of accuracy). Therefore, for all analyses we elect to sacrifice sample purity to increase the completeness of the sample. This is especially important later in the paper where we assess offsets in quiescence as a function of location within the group or cluster, at fixed black hole masses.

\newpage
% Don't change these lines
\bsp	% typesetting comment
\label{lastpage}
\end{document}